\documentclass[11pt,a4paper]{article}
\usepackage[backend=bibtex, style=numeric, sorting=ynt]{biblatex}
\usepackage[utf8]{inputenc}
\usepackage{amsmath}
\usepackage{authblk}
\usepackage{graphicx}
\usepackage[margin=1in]{geometry}
\usepackage{textcomp}
\usepackage[T1]{fontenc}
\usepackage{authblk}
\usepackage{hyperref}
\usepackage{array}
\usepackage{multirow} 
\usepackage{tabularx} 
\usepackage{indentfirst}
\usepackage{subfigure}
\usepackage{diagbox}
\usepackage{booktabs}
\usepackage{float}
\usepackage{bm}
\usepackage{adjustbox}
\usepackage{amsfonts}
\usepackage[justification=centering]{caption}
\newcommand\keywords[1]{\textbf{Keywords}: #1}
\title{Autonomous Orbit Determination Analysis of a Conceptual Cislunar Navigation Constellation based on Inter-Satellite Range Measurement}

\author[1,2,3]{Haohan Li}

\author[1,2,3]{Yuxuan Miao}

\author[1,2,3]{Xiyun Hou\thanks{Corresponding author: email: houxiyun@nju.edu.cn}}

\author[1,2,3]{Bosheng Li}

\author[4]{Jinjun Zheng}

\author[5]{Hao Yu}

\author[6]{Kanglian Zhao}

\author[6]{Huan Yan}

\affil[1] {School of Astronomy and Space Science, Nanjing University, Nanjing 210023, China}
\affil[2]  {Institute of Space Environment and Astrodynamics, Nanjing University, Nanjing 210023, China}
\affil[3] {Key Laboratory of Modern Astronomy and Astrophysics, Ministry of Education, Nanjing 210023, China}
\affil[4]{Institute of Telecommunication and Navigation Satellites, China Academy of Space Technology, Beijing 100094, China}
\affil[5]{Shanghai Aerospace Control Technology Institute, Shanghai 201109, China}
\affil[6] {School of Electronic Science and Engineering, Nanjing University, Nanjing 210023, China}
\date{}

\begin{document}
\selectfont
\maketitle

\begin{abstract}
   With the community's increasing interest in the cislunar space, building a navigation constellation servicing the whole cislunar space has become a pressing need. Previous studies mainly focus on constellations using orbits close to the Moon, which limits the servicing volume of the constellation. In this work, a four-satellite constellation using one L3 orbit, one L4 orbit, one L5 orbit and an orbit close to the Moon is proposed. The orbit determination accuracy is an important factor to be considered when designing parameters of the constellation. In this study, the mode of autonomous orbit determination (AOD) based on inter-satellite range data is considered. With such a model, the out-of-plane design parameters are identified as the main parameters influencing the AOD accuracy. For the AOD based on short arcs, we find that the increase of the out-of-plane amplitude can improve the AOD accuracy, and the out-of-plane initial phases have a more complex influence. A novel relative planarity factor (RPF) $P_\text{r}$, which has negative correlation with the AOD accuracy, is proposed as the metric to evaluate the variation of AOD performance. Using $P_\text{r}$, we demonstrate that the coplanarity of the constellation can significantly reduce the AOD accuracy. For the long arc AOD, the influence of different parameters is insignificant.
\end{abstract}

\section{Introduction}
\par The cislunar space is becoming an important region for human space activities. Satellites in the cislunar space operate in orbits quite different from those of artificial Earth satellites. This space is dominated simultaneously by the gravity of both the Earth and the Moon. There are many types of orbits in the cislunar space. Among them, of particular interest are those around the collinear and the triangular libration points. In the lunar vicinity, the distant retrograde orbit (DRO) and the nearly rectilinear halo orbit (NRHO) are also attracting significant interest \autocite{browna2024,sun2024}. Currently, there are already spacecraft moving on some of these orbit types, and additional missions are planned to utilize them.
\par With more and more lunar and cislunar missions in the near future, the servicing ability of existing deep space network (DSN) is becoming insufficient. Furthermore, there are regions in the cislunar space which are invisible to the existing DSN \autocite{BAKERMCEVILLY2024}. A good way to solve these problems is to deploy a cislunar navigation constellation servicing the whole cislunar space. The first idea of deploying satellites in the cislunar space for navigation and communication was proposed by Farquhar in the 1960s \autocite{Farquhar1970}. His idea was to send a relay station at the Earth-Moon L2 point to communicate with the Moon's farside. Carpenter et al. mentioned the feasibility of an onboard navigation system utilizing an Earth-Moon L2 satellite and GPS, and also analyzed its navigation accuracy \autocite{Carpenter2004}. Bhasin and Hayden suggested placing communication satellites on the L1 and L2 halo orbits for supporting future missions \autocite{Bhasin2004}. Xu et al. developed a four-satellite constellation based on the periodic orbits near L2, L3, L4 and L5 \autocite{2013Xu}. Hill et al. proposed the well-known LiAISON (Linked Autonomous Interplanetary Satellite Orbit Navigation) autonomous orbit determination (AOD) strategy \autocite{Parker2012}. This strategy utilizes the three-body dynamics and inter-satellite range data between navigation satellites to autonomously determine their orbits. The orbit they used is the strongly unstable halo orbit. Liu et al. suggested using stable distant retrograde orbits (DROs) to replace halo orbits in the LiAISON strategy \autocite{Liu2014}. This substitution enables the acquisition of long-term inter-satellite range data, which is essential for accurate orbit determination and station-keeping. Wang et al. continued with this idea and tested the orbit determination accuracy in a more realistic mission scenario \autocite{Wang2019}. Liu et al. firstly suggested a navigation constellation using the dynamical substitutes around the triangular libration points (dynamical substitute orbits, DSOs) \autocite{Liu2020Master}. Besides the common orbits near the Moon or the libration points, Gao et al. proposed a type of special long-period orbit based on a homoclinic connection of L1 Halo and explored the capability of navigation in the lunar vicinity \autocite{2024GAO}. Li et al. extended the LiAISON method to simultaneously estimate satellite orbits and the lunar ephemeris using DRO-based inter-satellite ranging, achieving meter-level lunar ephemeris accuracy without ground support \autocite{Li_navi_2026}.
\par To enhance navigation and communication performance in cislunar space, particularly in terms of coverage and positioning accuracy, significant research efforts were devoted to optimizing satellite constellations. For instance, Patel et al. employed integer linear programming to minimize the number of satellites required to achieve a specified coverage requirement \autocite{Patel2024}. Separately, He et al. evaluated and compared the GDOP values for various orbital configurations of a four-satellite constellation \autocite{he2025}.
\par The orbit determination accuracy of navigation satellites is a critical factor in assessing the performance of a navigation constellation. Xu et al. analyzed the AOD accuracy of a cislunar navigation constellation based on inter-satellite link (ISL). They also pointed out the improvement of the orbit determination (OD) accuracy by incorporating the support from GNSS satellites \autocite{XU2024}. Turan et al. analyzed the performance of the AOD for cislunar small satellite formations using crosslink radiometric measurements \autocite{TURAN2023}. Iess et al. proposed a novel orbit determination and time synchronization architecture for a lunar radio navigation constellation based on multiple spacecraft per aperture (MSPA) and K-band spread-spectrum signals, achieving a signal-in-space error (SISE) better than 10 m \autocite{Iess_2025}.
\par Till now, influences of orbit parameters on the OD accuracy have not investigated in detail. A practical concern when constructing such a cislunar navigation constellation is how the orbit parameters influence the OD accuracy of the satellites. In our study, a four-satellite constellation consisting of the following orbits: one L3 Lissajous orbit, one L4 triangular libration point orbit (TLPO), one L5 TLPO and one DRO is proposed, and the influence of orbital parameters on the AOD accuracy is investigated. The reason we limit the satellite number to 4 is that this is the minimum number of satellites necessary for real-time navigation in the cislunar space. The reason we choose the L3, L4, and L5 points is that the three satellites have a fixed and uniformly distributed relative geometry with respect to each other and the Earth. Since the Moon is the most important target in the cislunar space, putting a navigation satellite in its proximity is reasonable. The reason we choose the DRO is that its stability property has irreplaceable advantages over other orbit types close to the Moon \autocite{Gao2020}. In our previous work, an indicator named the dynamic and geometric dilution of precision factor ($DAGDOP$) was proposed to efficiently characterize the OD accuracy of a navigation constellation. This indicator is somewhat similar to the GDOP indicator but differs by incorporating contributions from the dynamics, rather than the geometry. This indicator is valid even for a constellation of only two satellites \autocite{Gao2022}. In this work, we try to find the answers to the following two questions:
\begin{itemize}
    \item How does the AOD accuracy change with these orbit parameters?
    \item Which orbit parameters have the most significant effects on the AOD accuracy?
\end{itemize}
Besides these two questions, we will also discuss how the AOD accuracy changes if different orbit types are chosen. Our studies show that the AOD accuracy based on short arcs changes drastically with the orbit parameters, and the coplanarity of the satellite constellation is the most important factor influencing the short arc AOD accuracy. We propose the relative planarity factor (RPF) $P_\text{r}$ as an indicator of the AOD accuracy. For the AOD based on long arcs, its accuracy does not vary drastically with the change of orbit parameters.
\par This paper is organized as follows. Fundamentals of this work are introduced in Section~2. Properties of the orbits used are introduced in Section~3. Section~4 analyzes the AOD accuracy of the constellation in both the short-arc scenario and the long-arc scenario. Section~5 analyzes the influence of different orbit parameters and different AOD arcs. Section~6 discusses several problems and Section~7 concludes the whole work.

\section{Some Fundamentals}

\subsection{Time and Reference Frame}
\par In order to model the relativistic effects of the cislunar space more conveniently using the EIH equation, we choose the BCRS as the reference frame \autocite{petit_2010} and the TDB as the time standard. Furthermore, to clearly show the orbits in the cislunar space, we geometrically shift the origin of BCRS from solar system barycenter to the Earth's center. In this work, we call the BCRS frame centered at the Earth E-BCRS frame. Since only a simple geometrical transformation is involved, the relationship between the coordinates in the BCRS and the E-BCRS is
\begin{equation}
\mathbf{{r}_{E-B}}=\mathbf{{r}_{B}}-\mathbf{{r}_{E}},
\label{eq:E-BCRS}
\end{equation}
where $\mathbf{{r}_{E-B}}$ is the coordinate in the E-BCRS, $\mathbf{{r}_{B}}$ is the coordinate in the BCRS, and $\mathbf{{r}_{E}}$ is the coordinate of the Earth in the BCRS. In the following, to simplify the expression, the subscript ${E-B}$ is omitted.

\subsection{The Force Model}

\par We have developed an OD software based on the above reference frame and time standard. Satellites in the cislunar space follow the following motion equation:
\begin{equation}
\ddot{\mathbf{r}}=\mathbf{F}_{0}+\mathbf{F}_{M}+\mathbf{F}_{3rd}+\mathbf{F}_{ns}+\mathbf{F}_{srp}+\mathbf{F}_{QE}+\mathbf{F}_{PN}+\mathbf{F}_{tide},
\label{High-Fidelity_Force}
\end{equation}
where $\mathbf{F}_{0}$ is the monopole gravity of the Earth and $\mathbf{F_{M}}$ is the monopole gravity perturbation of the Moon. These two forces are the most important ones for space objects moving in the cislunar space. They can be expressed as: 
\begin{equation}
\begin{aligned}\left\{\begin{aligned}
&\mathbf{F}_{0}=-\frac{\mu_E}{r^3}\mathbf{r},\\
&\mathbf{F}_{M}=-\mu_M\left(\frac{\mathbf{\Delta}_{M}}{{\Delta}_{M}^3}+\frac{\mathbf{r}_{M}}{{r}_{M}^3}\right),\\
\end{aligned}\right.\end{aligned}
\end{equation}
where $\mu_E$ and $\mu_M$ are the gravitational constant of the Earth and the Moon, $\mathbf{r}$ is the vector from the Earth to the satellite defined by Eq.~\eqref{eq:E-BCRS}, $\mathbf{r}_M$ is the vector from the Earth to the Moon, and $\mathbf{\Delta}_M=\mathbf{r}-\mathbf{r}_M$. The other forces are:

\begin{itemize}
    \item $\mathbf{F}_{3rd}$ is the third-body gravitational perturbation from the Sun and the planets. In this study, the orbits of these celestial bodies are based on JPL DE 441 ephemeris \autocite{Park_2021}. The expression of $\mathbf{F}_{3rd}$ is
\begin{equation}
\mathbf{F}_{3rd}=-\sum\mu_i\left(\frac{\mathbf{\Delta}_{i}}{{\Delta}_{i}^3}+\frac{\mathbf{r}_{i}}{{r}_{i}^3}\right),
\end{equation}
where $\mu_i$ is the gravitational constant of the perturbing body, $\mathbf{r}_i$ is the vector from the Earth to the perturbing body, and $\mathbf{\Delta}_i=\mathbf{r}-\mathbf{r}_i$.

    \item $\mathbf{F}_{ns}$ is the non-spherical gravity of the Earth and the Moon. In our work, these two non-spherical gravities are truncated at degree and order 5. The Earth gravity field used is the EGM 2008 \autocite{Pavlis_2012} and the Moon gravity field used is the GRAIL product GGGRX1200a (or called GRGM1200A) \autocite{KLOKOCNIK_2022}. Taking the Earth's non-spherical gravity as an example, it can be expressed as:
\begin{equation}
\begin{aligned}\left\{\begin{aligned}
\mathbf{F}_{ns}&=-\mathbf{C}_E\left[\frac{\partial(\Delta{V})}{\partial \mathbf{R}}\right],\\
\Delta{V}&=\frac{\mu_E}{R}\sum_{l=2}^{\infty}\sum_{m=0}^{l}\left(\frac{a_e}{R}\right)^{l}P_{lm}(\sin\phi)[C_{lm}\cos{m\lambda}+S_{lm}\sin{m\lambda}],\\
\end{aligned}\right.\end{aligned}
\end{equation}
\\ where $\mathbf{R}$ is the vector of the field point in the Earth's body-fixed frame, $a_e=R_E$ is the reference radius of the EGM 2008 model, $\mu_E$ is the Earth gravitational constant, $C_{lm}$ and $S_{lm}$ are spherical harmonic coefficients, $P_{lm}$ is the associated Legendre polynomial, $\mathbf{C}_E$ is the rotation matrix from the Earth's body-fixed frame to the E-BCRS reference frame, $\phi$ is the geographic latitude, and $\lambda$ is the geographic longitude. Since the coordinate origin is centered at the Earth, we remark that the indirect contribution from the Moon's non-spherical gravity through accelerating the Earth should also be considered.
    
    \item $\mathbf{F}_{PN}$ is the post-Newtonian effect (relativistic correction). Since we use the TDB time and the E-BCRS frame, this correction is given by the EIH equation. For the satellite, it takes the form \autocite{Deep_Space_ch4}:
\begin{equation}
\begin{aligned}
\mathbf{F}_{PN}^{s}=&-\sum_{i\in S}\mu_i\frac{\mathbf{r}-\mathbf{r}_i}{\Delta_i^3}\left\{-\frac{2(\beta+\gamma)}{c^2}\sum_{k\in S}\frac{\mu_k}{\Delta_k}-\frac{2\beta-1}{c^2}\sum_{k\neq{i}}\frac{\mu_k}{\Delta_{ki}}+\frac{\gamma{v}^2}{c^2}\right.\\
&\left.-\frac{2(1+\gamma)}{c^2}(\dot{\mathbf{r}}\cdot\ddot{\mathbf{r}}_i)-\frac{3}{2c^2}\left[\frac{(\mathbf{r}-\mathbf{r}_i)\cdot\dot{\mathbf{r}}_i}{\Delta_i}\right]^2-\frac{1}{2c^2}(\mathbf{r}-\mathbf{r}_i)\cdot\ddot{\mathbf{r}}_i\right\}\\
&+\sum_{i\in{S}}\frac{\mu_i}{c^2\Delta_i}\left\{\frac{3+4\gamma}{2}\ddot{\mathbf{r}}_i+\frac{[(\mathbf{r}-\mathbf{r}_i)\cdot((2+2\gamma))\dot{\mathbf{r}}-(1+2\gamma)\dot{\mathbf{r}}_i](\dot{\mathbf{r}}-\dot{\mathbf{r}}_i)}{\Delta_i^2}\right\},\nonumber
\end{aligned}
\end{equation} 
$\beta$ and $\gamma$ are parameters which equal 1 in this work, $v$ is the speed of the satellite, and $S=[1,2,\ldots,11]$ represents the set of massive celestial bodies in the solar system. As for the Earth ($j=3$), its post-Newtonian perturbation takes the following form:
\begin{equation}
\begin{aligned}
\mathbf{F}_{PN}^{E}=&-\sum_{i\neq j}\mu_i\frac{\mathbf{r}_j-\mathbf{r}_i}{\Delta_i^3}\left\{-\frac{2(\beta+\gamma)}{c^2}\sum_{k\neq j}\frac{\mu_k}{\Delta_{kj}}-\frac{2\beta-1}{c^2}\sum_{k\neq{i}}\frac{\mu_k}{\Delta_{ki}}+\frac{\gamma{v}_j^2}{c^2}+\frac{(1+\gamma){v}_i^2}{c^2}\right.\\
&\left.-\frac{2(1+\gamma)}{c^2}(\dot{\mathbf{r}_j}\cdot\ddot{\mathbf{r}}_i)-\frac{3}{2c^2}\left[\frac{(\mathbf{r}-\mathbf{r}_i)\cdot\dot{\mathbf{r}}_i}{\Delta_i}\right]^2-\frac{1}{2c^2}(\mathbf{r}_j-\mathbf{r}_i)\cdot\ddot{\mathbf{r}}_i\right\}\\
&+\sum_{i\in{S}}\frac{\mu_i}{c^2\Delta_{ij}}\left\{\frac{3+4\gamma}{2}\ddot{\mathbf{r}}_{i}+\frac{[(\mathbf{r}_j-\mathbf{r}_i)\cdot((2+2\gamma))\dot{\mathbf{r}}_j-(1+2\gamma)\dot{\mathbf{r}}_i](\dot{\mathbf{r}}_j-\dot{\mathbf{r}}_i)}{\Delta_{ij}^2}\right\},\nonumber
\end{aligned}
\end{equation}  
As a result, the relativistic correction of the satellite with respect to the Earth ($\mathbf{F}_{PN}$) in E-BCRS should be expressed as: 
\begin{equation}
\mathbf{F}_{PN} = \mathbf{F}_{PN}^{s} - \mathbf{F}_{PN}^{E}.
\end{equation}

    \item $\mathbf{F}_{tide}$ is the Earth's tidal force. According to the IERS convention 2010, the correction to the spherical harmonic coefficients can be expressed as \autocite{petit_2010}:
\begin{equation}
\Delta\overline{C}_{lm}-i\Delta\overline{S}_{lm}=\frac{k_{lm}}{2l+1}\sum_{j=10,11}\frac{\mu_j}{\mu_E}\left(\frac{a_e}{r_j}\right)^{l+1}\overline{P}_{lm}\left(\sin{\Phi_j}\right)e^{-im\lambda_j}.
\end{equation}
In this study, only the gravitational effects of the Moon and the Sun are considered, and the subscript $j$ is the number of the celestial body: when $j=10$, it means the effect of the Moon; and when $j=11$, it means the effect of the Sun. The other symbols in the equation are defined as follows: $i$ is the imaginary unit, $r_j$ is the distance from the center of the celestial body to the geocenter, $\Phi_{j}$ is the geocentric latitude of the celestial body in the Earth body-fixed frame, $\lambda_j$ is the longitude (from Greenwich) of the celestial body in the Earth body-fixed frame, $\overline{P}_{lm}$ is the normalized associated Legendre function, and $k_{lm}$ is the Love number for degree $l$ and order $m$. In this work, only the second order solid Earth tide is considered, and the love numbers $k_{lm}$ are taken for an anelastic Earth model according to the IERS convention 2010 \autocite{petit_2010}:
\begin{equation}
k_{20}=0.30190, \quad k_{21}=0.29830-0.00144i, \quad k_{22}=0.30102-0.00130i.
\nonumber
\end{equation}

\item $\mathbf{F}_{QE}$ is the coupling term between the Earth's non-spherical gravity and the major celestial bodies.  
\begin{equation}
\mathbf{F}_{QE}=\mathbf{C}_E\sum_{i=10,11}\frac{\mu_i}{\mu_E}\left(\frac{\partial\Delta{V}}{\partial{\mathbf{R}}}\mathbf{r}_i^E\right)
\end{equation}
where $\mathbf{r}_i^E$ is the vector of the $i^{th}$ perturbing celestial body in Earth's body-fixed frame. Due to the long distance between the Earth and the major celestial bodies, we only consider the Earth's non-spherical gravity at the second order in this work.
    
    \item $\mathbf{F}_{srp}$ is the solar radiation pressure (SRP). Several SRP models can be considered for cislunar objects, such as the cannonball model, the ECOM model \autocite{fliegel_1996} and the ENCM model , which is an empirical model specifically developed by us for cislunar satellites \autocite{Li_2026_SRP}. To remove the SRP's modeling error, the simple cannonball model is employed throughout this work in both the observation simulation and the AOD process, i.e., we assume perfect knowledge on the SRP model. The cannonball model is expressed as follows:
\begin{equation}
\mathbf{F}_{srp}=-P(1+\rho)\frac{S}{mc}\frac{\mathbf{r}_s}{{r}_s}.
\end{equation}
Here, $\mathbf{r}_s$ is the vector from the Sun to the satellite, $P=P_0\cdot{D^2}/{r_s^2}$ is the solar radiation pressure at distance $r_s$, $D$ is the average Sun-Earth distance (1 AU), $\rho$ is the specular reflectivity, $S$ is the equivalent area of the satellite, $m$ is the mass of the satellite and $c$ is the vacuum speed of light. 

\end{itemize} 

\subsection{Observation data}
\par The observable in this work is the inter-satellite range, denoted by $\rho$. Accounting for the time delay caused by the finite speed of light, we define $t_1$ and $t_2$ as the signal transmission time and reception time in the TDB timescale, respectively. The range $\rho$ is then given by the following equation:
\begin{equation}
\rho=r_{12}^{BCRS}+\Delta\rho_{\mathrm{GB}}+\epsilon,
\label{rho}
\end{equation}
where $r_{12}^{BCRS}$ is the geometrical range between the transmitting satellite and the receiving satellite; $\Delta\rho_{\mathrm{GB}}$ is the range correction due to the gravitational delay and $\epsilon$ is the observation error. In this work, the observation error $\epsilon$ is modeled as white Gaussian noise. Let $\mathbf{r}_1(t_1)$ and $\mathbf{r}_2(t_2)$ denote the BCRS position vectors of the transmitting and receiving satellites at times $t_1$ and $t_2$, respectively. Then, $r_{12}^{\mathrm{BCRS}}$ is given by \autocite{Deep_Space_ch5}:
\begin{equation}
r_{12}^{BCRS}=|\mathbf{r}_2^{BCRS}(t_2)-\mathbf{r}_{1}^{BCRS}(t_1)|.\nonumber
\end{equation}

\par For near-Earth satellites, $\Delta\rho_{\mathrm{GB}}$ is typically negligible. However, for cislunar objects this term should be accounted for, as its magnitude is on the order of 10 meters. The expression for $\Delta\rho_{\mathrm{GB}}$, considering the contributions from these bodies, is derived as follows \autocite{Deep_Space_ch4}:
\begin{equation}
\Delta\rho_{GB}=\frac{(1+\gamma)\mu_s}{c^2}\ln\left[\frac{r_1^{s}+r_2^{s}+r_{12}^{s}+\frac{(1+\gamma)\mu_s}{c^2}}{r_1^{s}+r_2^{s}-r_{12}^{s}+\frac{(1+\gamma)\mu_s}{c^2}}\right]+\frac{(1+\gamma)\mu_m}{c^2}\ln\left[\frac{r_1^{m}+r_2^{m}+r_{12}^{m}}{r_1^{m}+r_2^{m}-r_{12}^{m}}\right],\nonumber
\label{gravitational_delay}
\end{equation}
where $\mu_s$ and $\mu_m$ are the gravitational parameters of the Sun and the Moon, respectively; $c$ is the vacuum speed of light; $r$ is the distance from the satellite to the massive celestial body. The superscript represents the massive celestial body (the Sun or the Moon) and the subscript represents the satellite. In addition, $r_{12}=|\mathbf{r_2}-\mathbf{r_1}|$.
\par These equations reveal a fundamental reciprocity: given precise orbits, one can fix the reception time $t_2$ and solve iteratively for the transmission time $t_1$, or vice versa. In this work, only the signal propagation delays are considered. The influence of clock errors is ignored. Signal block due to occultation by the Earth or the Moon is also not considered for simplicity.

\subsection{Batch Filter and Covariance Matrix}
\par In this study, the batch filter is employed for AOD. We define $\mathbf{X}_0$ as the initial state vector at the reference epoch, $\mathbf{C}$ as the vector of dynamical parameters, $\mathbf{Y}_{\mathrm{o}}$ as the set of all observations, and $\mathbf{Y}_{\mathrm{e}}$ as the set of corresponding estimated observation values. For the $i^{th}$ epoch $t^i$, the observation and the estimation are denoted as $Y^i_\mathrm{o} \in \mathbf{Y}_\mathrm{o}$ and $Y^i_\mathrm{e} \in \mathbf{Y}_\mathrm{e}$, respectively. The corresponding residual $y^i$ can be defined as:
\begin{equation}
y^i=Y^{i}_\mathrm{e}-Y^{i}_\mathrm{o}=\mathbf{H}_\mathbf{x}^i\hat{\mathbf{x}}^{0}+\mathbf{H}_\mathbf{c}^i\hat{\mathbf{c}}+\bm{\varepsilon}.\nonumber
\end{equation}
In the following, $\hat{\mathbf{x}}_{0}$ is the correction to the estimated orbit, $\hat{\mathbf{c}}$ is the correction to the dynamic parameters, and $\bm{\varepsilon}$ represents the observation error at the epoch $t^i$. The matrices in the equation are:
\begin{equation}
\begin{aligned}
\mathbf{H}_\mathbf{x}^i=\frac{\partial Y^i_\mathrm{e}}{\partial \mathbf{X}^i}\frac{\partial \mathbf{X}^i}{\partial \mathbf{X}^0}=\mathbf{\widetilde{H}}_\mathbf{x}^i\bm{\Phi}^i;& \quad \bm{\Phi}^i=\frac{\partial \mathbf{X}^i}{\partial \mathbf{X}^0};\quad \mathbf{\widetilde{H}}_\mathbf{x}^i=\frac{\partial Y^{i}_\mathrm{e}}{\partial \mathbf{X}^i},\\
\mathbf{H}_\mathbf{c}^i=\frac{\partial Y^{i}_\mathrm{e}}{\partial \mathbf{X}^i}\frac{\partial \mathbf{X}^i}{\partial \mathbf{C}}=\mathbf{\widetilde{H}}_\mathbf{x}^i\bm{\Phi}_\mathbf{c}^i;& \quad \bm{\Phi}_\mathbf{c}^i=\frac{\partial \mathbf{X}^i}{\partial \mathbf{C}}.
\nonumber
\end{aligned}
\end{equation}
Here, $\mathbf{X^i}$ is the state vector at the time $t^i$ for the $i^{th}$ observation, $\bm{\Phi}^i=\bm{\Phi}(\mathbf{X}^i,t^i)$ is the state transition matrix (STM), and $\bm{\Phi}_\mathbf{c}^i=\bm{\Phi}_\mathbf{c}(\mathbf{X}^i,t^i)$ is the parameter sensitivity matrix (PSM). These matrices are computed by integrating the following variational equations along with the orbit:
\begin{equation}
\begin{aligned}\left\{\begin{aligned}
& \bm{\Dot{\Phi}}(\mathbf{X},t)=\frac{\partial \Dot{\mathbf{X}}}{\partial \mathbf{X}}\bm{\Phi}(\mathbf{X},t),\\
& \bm{\Dot{\Phi}}_\mathbf{c}(\mathbf{X},t)=\frac{\partial \Dot{\mathbf{X}}}{\partial \mathbf{X}}\bm{\Phi}_\mathbf{c}(\mathbf{X},t)+\frac{\partial{ \Dot{\mathbf{X}}}}{\partial \mathbf{C}}.
\end{aligned}\right.\end{aligned}
\end{equation}
Defining $\mathbf{H}^i = [\mathbf{H}_\mathrm{x}^i,\mathbf{H}_\mathrm{c}^i]^T$ and assuming all observation data are equally weighted, $\hat{\mathbf{z}}=[\hat{\mathbf{x}}_{0} \quad \hat{\mathbf{c}}]^T$ can be estimated using the following formula:
\begin{equation}
\hat{\mathbf{z}}=\left[\hat{\mathbf{x_0}}\quad\hat{\mathbf{c}} \right]^T=\left[\sum^{N}_{i=1}(\mathbf{H}^{iT}\mathbf{H}^i)\right]^{-1}\sum^{N}_{i=1}\mathbf{H}^{iT}y.
\label{OD}
\end{equation}
In our study, when $\hat{\mathbf{z}}$ is smaller than a prescribed threshold, or when the number of iterations reaches the prescribed upper limit, the OD process stops \autocite{schutz_2004}. The covariance matrix $\mathbf{P_z}$ of $\hat{\mathbf{z}}$ is defined as:
\begin{equation}
\mathbf{P_z}=\left[\sum^{N}_{i=1}(\mathbf{H}^{iT}\mathbf{H}^i)\right]^{-1}.
\label{Cov_matrix}
\end{equation}
The smaller the covariance matrix, the higher the OD accuracy is \autocite{Gao2022}. 

\par Theoretically, the result of OD does not depend on the choice of coordinate system. Define the coordinate transformation between two coordinate systems as
\begin{equation}
    \mathbf{X'}(t)=\mathbf{T}(t)\mathbf{X}(t), \nonumber
\end{equation}
where $\mathbf{X}(t)$ and $\mathbf{X'}(t)$ are the state vectors of the original coordinate system and the new coordinate system at time $t$, respectively, $\mathbf{T}(t)$ is the coordinate transform matrix. The transformation of the STM can be expressed as
\begin{equation}
\bm{\Phi}^{syn}_{t,t_0}=\mathbf{T}^{-1}(t)\bm{\Phi}_{t,t_0}\mathbf{T}(t_0),
\end{equation}
where $\bm{\Phi}_{t,t_0}$ is the STM from time $t_0$ to $t$ in the original coordinate system and $\bm{\Phi}^{syn}_{t,t_0}$ is the STM in the new coordinate system.

\subsection{AOD performance for different constellations}
\par As mentioned above, different constellation configurations exhibit different AOD performance. Four different constellation configurations are provided here, denoted constellations A to D. Eight different orbits (two L3 Lissajous orbits, two L4 TLPOs, two L5 TLPOs, one DRO and one NRHO) are used in these constellations, whose initial states are shown in Appendix A. The suffix '-P' means 'planar', indicating that this orbit almost lies in the lunar orbit plane. The suffix '-S' means 'space', indicating that the orbit has a significant out-of-plane motion component.
\par
\begin{figure}[!h]
\centering   
\subfigure[Constellation A] 
{
\begin{minipage}[b]{.45\linewidth} 
\centering
\includegraphics[scale=0.3]{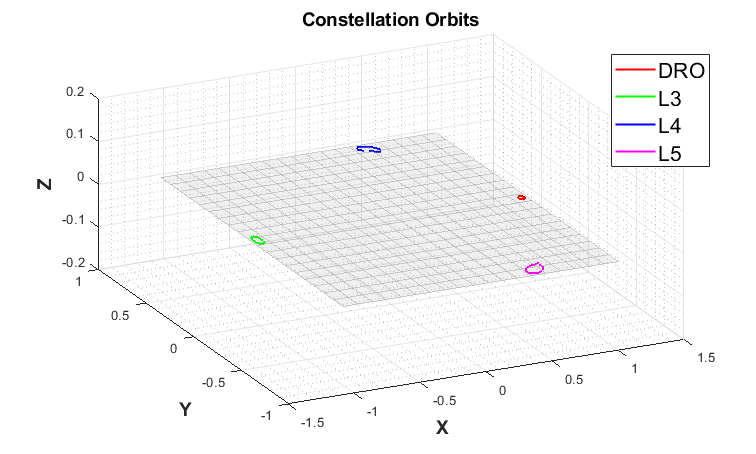}
\end{minipage}
}
\subfigure[Constellation B]
{
\begin{minipage}[b]{.45\linewidth}
\centering
\includegraphics[scale=0.3]{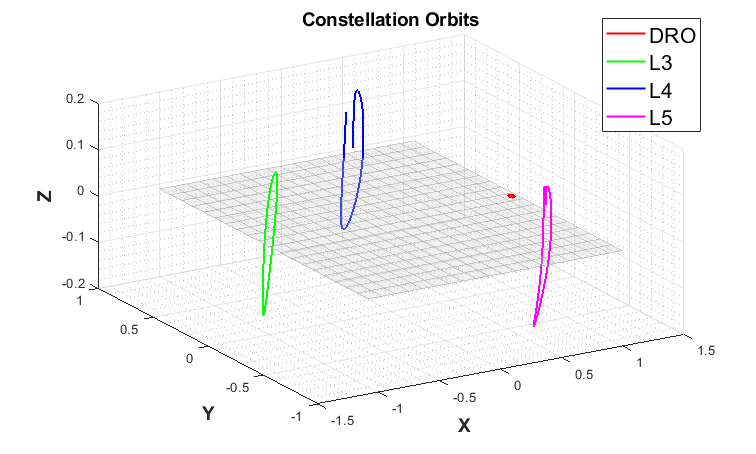}
\end{minipage}
}
\subfigure[Constellation C]
{
\begin{minipage}[b]{.45\linewidth}
\centering
\includegraphics[scale=0.3]{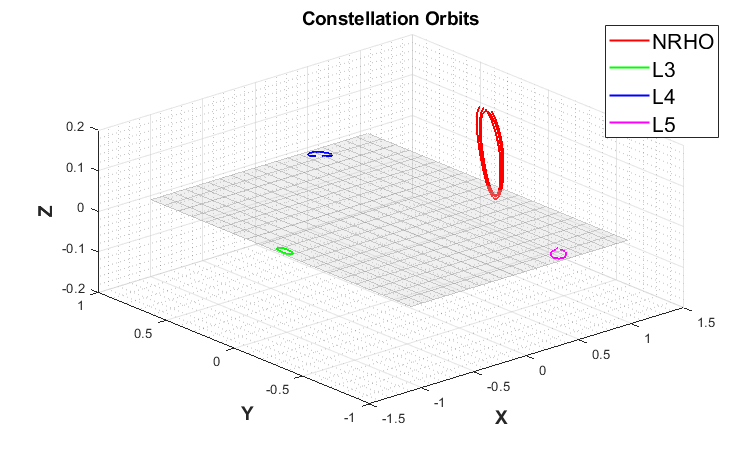}
\end{minipage}
}
\subfigure[Constellation D]
{
\begin{minipage}[b]{.45\linewidth}
\centering
\includegraphics[scale=0.3]{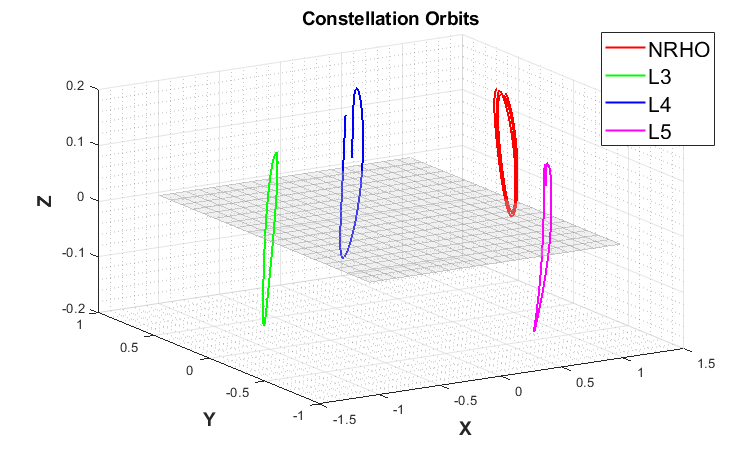}
\end{minipage}
}
\caption{3D configuration of Constellations A to D in 30 days}
\label{Fig_Constellations}
\end{figure}

 Fig.~\ref{Fig_Constellations} shows the 30-day orbits of the four constellations in the synodic frame. constellation A consists of L3-P, L4-P, L5-P and DRO; constellation B consists of L3-S, L4-S, L5-S and DRO; constellation C consists of L3-P, L4-P, L5-P and NRHO; and constellation D consists of L3-S, L4-S, L5-S and NRHO.

\par Numerical simulations are carried out to compare the AOD performance of these constellations. Each AOD simulation starts from an initial orbit with an error of 50,000 m in position and 1 m/s in velocity. The measurements are sampled at a 5-minute interval. Two different observation arcs are used: 3 days and 7 days. For a constellation with $M$ satellites, the overall AOD error of the constellation is represented by the root mean square (RMS) of the position errors of all the satellites, denoted as $\overline{\text{RMS}}_r$, which is expressed as:
\begin{equation}
    \overline{\text{RMS}}_r = \sqrt{\frac{1}{M}\sum_{i=1}^M\frac{1}{n}\sum_{k=1}^n\Delta{r}_{i,k}^{2}}.
    \nonumber
\end{equation}
Here $n$ is the total number of observations for each satellite. $\Delta{r}_{i,k}$ is the position error of the $i^{th}$ satellite at the $k^{th}$ observation. 

\begin{table}[htp]
    \renewcommand\arraystretch{1.2}
    \caption{\textbf{Comparison of the AOD results of four Constellations}}
    \centering
    \begin{tabular}{|c|c|c|}
    \hline
    \textbf{Constellation ID} & {3-day arc $\overline{\text{RMS}}_r$/m} & {7-day arc $\overline{\text{RMS}}_r$/m} \\ \hline
    {Constellation A} & 369.230 & 88.046 \\ \hline
    {Constellation B} & 53.822 & 25.180 \\ \hline
    {Constellation C} & 171.580 & 36.109 \\ \hline
    {Constellation D} & 168.921 & 26.972 \\ \hline
    \end{tabular}
    \label{OD_constellation}
\end{table}
\par Table~\ref{OD_constellation} shows the AOD results for different constellations. For the 3-day observation arc, difference in the AOD accuracy significantly varies with the constellation configuration, which means the constellation configuration has a non-negligible influence on the AOD accuracy for short AOD arcs. For the 7-day observation arc, the difference still exists, but becomes smaller. In the subsequent sections, the reason for this difference will be analyzed step by step.

\section{Orbit Types of the constellation}
\par As mentioned above, the four-satellite configuration consists of three satellites around the L3, L4, and L5 points, and one satellite close to the Moon, for which both DROs and NRHOs are popular candidates. 

\subsection{Lissajous Orbits around L3}
\par The Lissajous orbit is a viable option for a navigation satellite around the L3 point. Denote the deviation of the satellite from the L3 point as $\bm{\rho}=(\xi,\eta,\zeta)=\mathbf{r}-\mathbf{r_L}$, where $\mathbf{r_L}=(x_L,y_L,z_L)$ and $\mathbf{r}=(x,y,z)$ are coordinates of the libration point and the satellite in the synodic frame, respectively. In the circular restricted three-body problem (CRTBP), it can be analytically expressed as \autocite{JORBA1999}:
\begin{equation}
\begin{aligned}\left\{\begin{aligned}
{\xi}(t) &= \sum_{i+j=1}^{\infty}\left[\sum_{|k|\leq i,|m|\leq j} \xi_{ijkm}\cos{(k\theta_1+m\theta_2)}\right]\alpha^i\beta^j, \\
{\eta}(t) &= \sum_{i+j=1}^{\infty}\left[\sum_{|k|\leq i,|m|\leq j} \eta_{ijkm}\cos{(k\theta_1+m\theta_2)}\right]\alpha^i\beta^j, \\
{\zeta}(t) &= \sum_{i+j=1}^{\infty}\left[\sum_{|k|\leq i,|m|\leq j} \zeta_{ijkm}\cos{(k\theta_1+m\theta_2)}\right]\alpha^i\beta^j, \\
\end{aligned}\right.\end{aligned}
\label{Lissajous_orbit}
\end{equation}
where the value of $i+j$ represents the order of the analytical solution; $\alpha$ and $\beta$ are the in-plane and out-of-plane amplitude parameters of the Lissajous orbit; $\xi_{ijkm}$, $\eta_{ijkm}$, and $\zeta_{ijkm}$ are the coefficients of the corresponding angle combinations; $\theta_1$ and $\theta_2$ are in-plane and out-of-plane libration angles, which are defined as: 
\begin{equation}
\theta_1=\omega t+\phi_1;\quad \theta_2=\upsilon t+\phi_2,
\nonumber
\end{equation}
where $\phi_1$ and $\phi_2$ are phase angles, and $\omega$ and $\upsilon$ are the in-plane and out-of-plane frequencies which can be expanded as
\begin{equation}
\omega=\sum_{i,j=0}^{\infty}\omega_{i,j}\alpha^i\beta^j;\quad
\upsilon=\sum_{i,j=0}^{\infty}\upsilon_{i,j}\alpha^i\beta^j.
\nonumber
\end{equation}
The $0^{th}$ order components are the most significant frequencies in $\omega$ and $\upsilon$. For L3, the normalized value of the $0^{th}$ order coefficients is given by: 
\begin{equation}
\omega_{0,0}^\mathrm{L3} \approx 1.0031;\quad
\upsilon_{0,0}^\mathrm{L3} \approx 1.0093.
\label{eq:omega-L3}
\end{equation}
The unit of the frequencies is the normalized frequency, for which value 1 means the period of 30 days.
\par
\begin{figure}[htp]
    \centering
    \includegraphics[width=7.5cm]{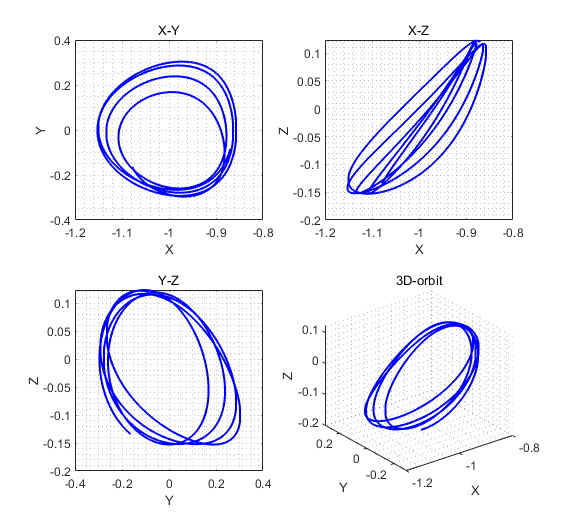}
    \caption{An example of an L3 Lissajous orbit lasting 180 days in the high-fidelity model of the Earth-Moon system}
    \label{Fig_L3}
\end{figure}

 In the real Earth-Moon system, due to perturbations from the Moon's orbit eccentricity and the Sun's gravity, dynamics around the L3 point are different from those of Eq.~\eqref{Lissajous_orbit}. Nevertheless, the deviation for the collinear libration points is not significant \autocite{hou_2011}. Lissajous orbits in the real Earth-Moon system can be obtained by numerical shooting methods \autocite{Howell1987,Gomez1998}, starting from an initial guess given by Eq.~\eqref{Lissajous_orbit}, Fig.~\ref{Fig_L3} shows a 3D L3 Lissajous orbit lasting 180 days in the synodic frame, which is continued from the CRTBP to a high-fidelity model of the real Earth-Moon system.

\subsection{Orbits around the L4 and L5 Points}
\par Different from the collinear libration points, dynamics around the triangular libration points in the real Earth-Moon system exhibit qualitative differences from those of the CRTBP. The triangular points are no longer equilibrium points in the synodic frame. Around them there are some special quasi-periodic orbits called dynamical substitutes (DS). Denote the deviation of the substitute from the triangular libration point as $\bm{\rho}=(\xi,\eta,\zeta)^T$. Motions around the triangular libration points in the real Earth-Moon system can be separated as forced dynamical substitutes $\bm{\bar{\rho}}=(\bar{\xi},\bar{\eta},\bar{\zeta})$ and free deviations $\Delta\bm{\rho}=(\Delta\xi,\Delta\eta,\Delta\zeta)$. As a result, the full orbits can be expressed as:
\begin{equation}
\begin{aligned}\left\{\begin{aligned}
{\xi} &= \bar{\xi}+\Delta\xi, \\
{\eta} &= \bar{\eta}+\Delta\eta,\\
{\zeta} &= \bar{\zeta}+\Delta\zeta.
\end{aligned}\right.\end{aligned}\nonumber
\end{equation}
According to \autocite{hou_2010}, in the ephemeris Earth-Moon  system, the DS can be expressed as
\begin{equation}
\begin{aligned}\left\{\begin{aligned}
\bar{\xi} = \sum_{ijkl} \bar{C}_{ijkl} \cos\bar{\theta}_{ijkl} + \bar{S}_{ijkl} \sin\bar{\theta}_{ijkl}, \\
\bar{\eta} = \sum_{ijkl} \bar{C}'_{ijkl} \cos\bar{\theta}_{ijkl} + \bar{S}'_{ijkl} \sin\bar{\theta}_{ijkl}, \\
\bar{\zeta} = \sum_{ijkl} \bar{C}''_{ijkl} \cos\bar{\theta}_{ijkl} + \bar{S}''_{ijkl} \sin\bar{\theta}_{ijkl},
\end{aligned}\right.\end{aligned}
\end{equation}
where $\bar{C}_{ijkl}$ and $\bar{S}_{ijkl}$ are coefficients associated with the angle $\theta_{ijkl}$, which is expressed as:
\begin{equation}
\bar{\theta}_{ijkl}=\bar{\omega}_{ijkl}\cdot t=(i\omega_1+j\omega_2+k\omega_3+l\omega_4)t.
\nonumber
\end{equation}
$\omega_1$, $\omega_2$, $\omega_3$ and $\omega_4$ are four basic frequencies of the lunar orbital motion \autocite{hou_2010}, given by:
\begin{equation}
\begin{aligned}\left\{\begin{aligned}
\omega_1&=0.99154828857,\\
\omega_2&=0.07480066375,\\
\omega_3&=0.92519871658,\\
\omega_4&=1.00402177967.\\
\end{aligned}\right.\end{aligned}\nonumber
\end{equation}
The unit of the frequencies is the same as that in Eq.~\eqref{eq:omega-L3}.

\begin{figure}[!h]
\centering   
\subfigure[L4 TLPO] 
{
\begin{minipage}[b]{.45\linewidth} 
\centering
\includegraphics[scale=0.45]{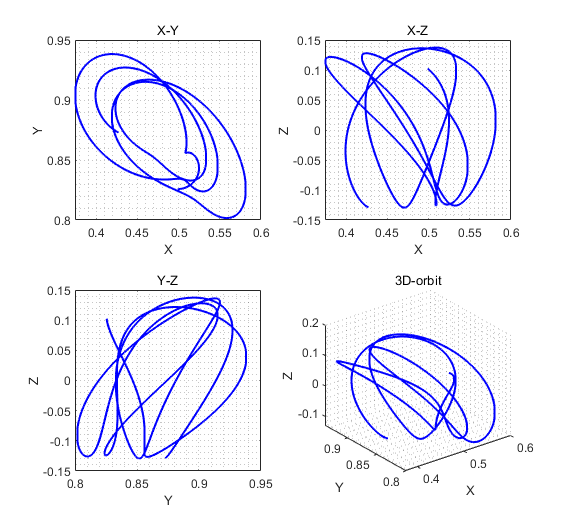}
\end{minipage}
}
\subfigure[L5 TLPO]
{
\begin{minipage}[b]{.45\linewidth}
\centering
\includegraphics[scale=0.45]{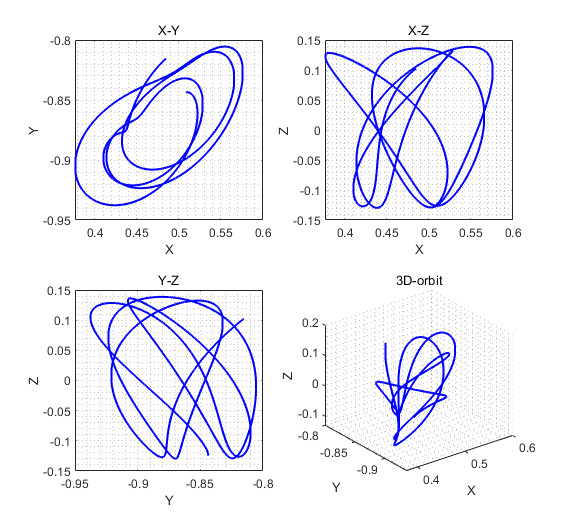}
\end{minipage}
}
\caption{Examples of L4 (a) and an example of L5 (b) TLPOs lasting 180 days in the high-fidelity model of the Earth-Moon system}
\label{Fig_L4L5}
\end{figure}

Truncated at the first order, the free motion can be expressed as \autocite{hou_2010}:
\begin{equation}
\begin{aligned}\left\{\begin{aligned}
\Delta \xi =& \alpha \sum_{ijkl} \left[ A_{ijkl}^{c,1} \cos\left( \bar{\theta}_{ijkl} + \theta_1 \right) + A_{ijkl}^{s,1} \sin\left( \bar{\theta}_{ijkl} + \theta_1 \right) \right] \\ &+ \beta \sum_{ijkl} \left[ A_{ijkl}^{c,2} \cos\left( \bar{\theta}_{ijkl} + \theta_2 \right) + A_{ijkl}^{s,2} \sin\left( \bar{\theta}_{ijkl} + \theta_2 \right) \right], \\
\Delta \eta =& \alpha \sum_{ijkl} \left[ B_{ijkl}^{c,1} \cos\left( \bar{\theta}_{ijkl} + \theta_1 \right) + B_{ijkl}^{s,1} \sin\left( \bar{\theta}_{ijkl} + \theta_1 \right) \right] \\ &+ \beta \sum_{ijkl} \left[ B_{ijkl}^{c,2} \cos\left( \bar{\theta}_{ijkl} + \theta_2 \right) + B_{ijkl}^{s,2} \sin\left( \bar{\theta}_{ijkl} + \theta_2 \right) \right], \\
\Delta \zeta =& \alpha \sum_{ijkl} \left[ C_{ijkl}^{c,1} \cos\left( \bar{\theta}_{ijkl} + \theta_1 \right) + C_{ijkl}^{s,1} \sin\left( \bar{\theta}_{ijkl} + \theta_1 \right) \right] \\ &+ \beta \sum_{ijkl} \left[ C_{ijkl}^{c,2} \cos\left( \bar{\theta}_{ijkl} + \theta_2 \right) + C_{ijkl}^{s,2} \sin\left( \bar{\theta}_{ijkl} + \theta_2 \right) \right].
\end{aligned}\right.\end{aligned}
\end{equation}
Here, $A_{ijkl}^{c,1}$, $A_{ijkl}^{s,1}$, $A_{ijkl}^{c,2}$, $A_{ijkl}^{s,2}$, $B_{ijkl}^{c,1}$, $B_{ijkl}^{s,1}$, $B_{ijkl}^{c,2}$, $B_{ijkl}^{s,2}$, $C_{ijkl}^{c,1}$, $C_{ijkl}^{s,1}$, $C_{ijkl}^{c,2}$, and $C_{ijkl}^{s,2}$ are the coefficients; $\alpha$ and $\beta$ are the amplitudes of the free motion; $\theta_1$ and $\theta_2$ are the angles of the free motion and can be written as:
\begin{equation}
\theta_1=\upsilon_1 t+\phi_1;\quad \theta_2=\upsilon_2 t+\phi_2,
\nonumber
\end{equation}
where $\phi_1$ and $\phi_2$ are the initial phases of in-plane motion and out-of-plane motion respectively, which can be arbitrarily chosen; $\upsilon_1$ and $\upsilon_2$ are free frequencies, which are approximately 0.30256 and 1.00406. As a result, to design a TLPO around the dynamical substitute, we need four parameters: $\alpha$, $\beta$, $\phi_1$ and $\phi_2$. Fig.~\ref{Fig_L4L5} shows example orbits of an example of L4 TLPO and an example of L5 TLPO in the synodic frame. 

\subsection{Near-Moon Orbits}
\subsubsection{DRO}
\par
\begin{figure}[h]
\centering   
\subfigure[DRO family in the CRTBP model] 
{
\begin{minipage}[b]{.45\linewidth} 
\centering
\includegraphics[scale=1]{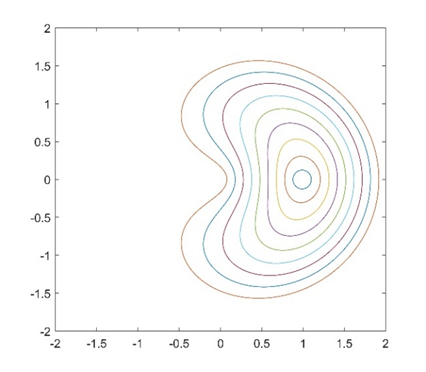}
\end{minipage}
}
\subfigure[An example of a DRO lasting 180 days in the high-fidelity model of the Earth-Moon system] 
{
\begin{minipage}[b]{.45\linewidth}
\centering
\includegraphics[scale=0.45]{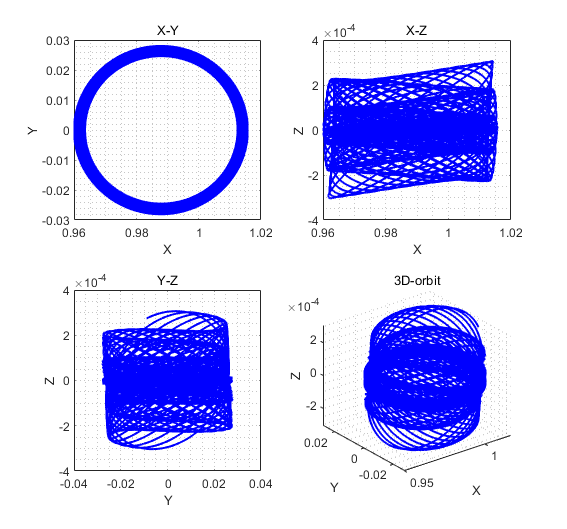}
\end{minipage}
}
\caption{DRO family in the CRTBP (a) and an example of a DRO in the high-fidelity model (b)}
\label{Fig_DRO}
\end{figure}

 In the CRTBP model, Distant Retrograde Orbits (DROs) refer to the family of periodic orbits that are generated from infinitesimal retrograde circumlunar orbits in the cislunar synodic frame, as shown in Fig.~\ref{Fig_DRO}(a). The period of a DRO changes with its amplitude. In the high-fidelity model, the DRO still exists but becomes quasi-periodic due to perturbations of different frequencies. An example of a DRO in the high-fidelity model is shown in Fig.~\ref{Fig_DRO}(b). 

\par To design a DRO, we need two parameters: the planar orbit amplitude $A_p$ and the initial planar phase angle $\phi_p$.

\subsubsection{NRHOs}
\par
\begin{figure}[h]
\centering   
\subfigure[Four halo families in the CRTBP] 
{
\begin{minipage}[b]{.45\linewidth} 
\centering
\includegraphics[scale=0.9]{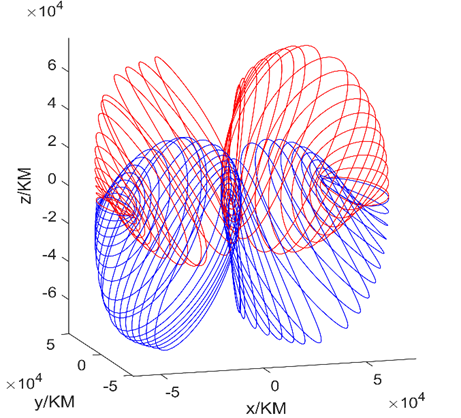}
\end{minipage}
}
\subfigure[An example of a L1-north NRHO lasting 60 days in the high-fidelity model of the Earth-Moon system]
{
\begin{minipage}[b]{.45\linewidth}
\centering
\includegraphics[scale=0.45]{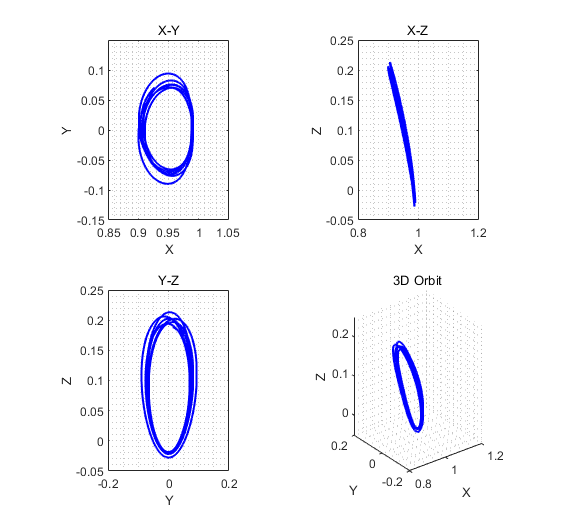}
\end{minipage}
}
\caption{Four families of halo orbits in CRTBP model (a) and an example of a L1-north NRHO in the high-fidelity model}
\label{Fig_NRHO}
\end{figure}

 The nearly-rectilinear halo orbits (NRHOs) belong to the halo orbit family \autocite{zimovan_2020}. Since there are two collinear libration points L1 and L2 close to the Moon, and each point has two halo families which are symmetric with respect to the $x-y$ plane, there are a total of four halo families. The four halo families around L1 and L2 in the CRTBP are shown in Fig.~\ref{Fig_NRHO}(a). The red orbits are the north group, and the blue ones are the south group. NRHOs are defined as halo orbits that are stable in the CRTBP model \autocite{Spreen2017}. As a result, there are four groups of NRHOs in the CRTBP model. In the high-fidelity model, the NRHOs still exist, but become quasi-periodic due to perturbations of different frequencies. Moreover, they are no longer stable due to these perturbations. Fig.~\ref{Fig_NRHO}(b) shows an example of a L1-north NRHO orbit lasting 60 days. 

\par To design an NRHO, we first need to choose one group out of the four NRHO groups: L1-North, L1-South, L2-North and L2-South. Then we need two parameters: one is the orbital amplitude (we use the perilune height as the parameter), and the other is the initial phase. Once we have chosen an NRHO in the CRTBP model, we can numerically continue it from the CRTBP to the high-fidelity model.

\subsubsection{Comparison between DRO and NRHO}
\par The four-satellite constellation is composed of one L3 Lissajous satellite, one L4 TLPO satellite, one L5 TLPO satellite, and one near-Moon satellite. Several aspects should be considered when choosing the orbit type for the near-Moon satellite. One factor to be considered in this choice is the AOD accuracy. As shown in Table~\ref{OD_constellation}, when the L3/L4/L5 satellites move on planar orbits (constellations A and C), the NRHO-based constellation C achieves better 3-day AOD accuracy than the DRO-based constellation A. However, when the L3/L4/L5 satellites move on orbits with significant out-of-plane amplitudes (constellations B and D), the DRO-based constellation B achieves a notably better accuracy than the NRHO-based constellation D. More importantly, the best AOD performance among all four constellations is attained by the DRO-based spatial constellation B.
\par
\begin{figure}[h]
    \centering
    \includegraphics[width=8cm]{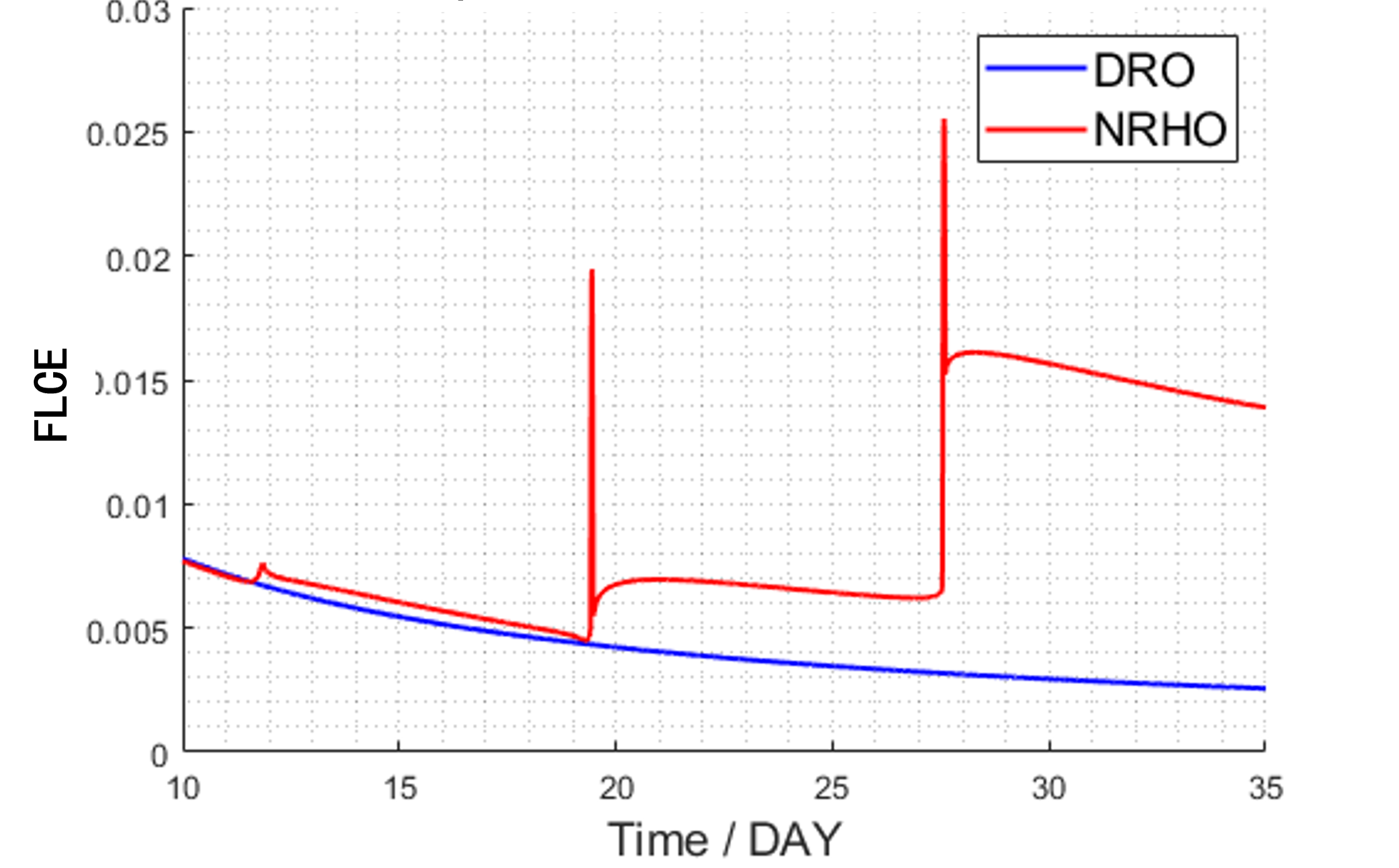}
    \caption{Evolution of the FLCE of the example DRO and NRHO orbits}
    \label{FLCE}
\end{figure}

 Orbital stability is another important factor. The fast Lyapunov characteristic exponent (FLCE) $\lambda(t)$ is a useful indicator for measuring the stability of an orbit within a finite time \autocite{Wang2022}. The FLCE can be expressed as:

\begin{equation}
\lambda(t)=\frac{1}{t}\ln{\frac{||\delta{X(t)}_{max}||}{||\delta{X_0}||}}=\frac{1}{t}\ln
\sqrt{{|\Lambda(t)|_{max}}}.\nonumber
\end{equation}
Here, $\delta{X_0}$ is the state error at the initial time $t_0$, $\delta{X(t)}$ is the state error at the time $t$, $\Lambda(t)$ is the eigenvalue of a positive definite matrix $\Phi^T\Phi$, where $\Phi=\Phi(t)$ is STM at the time $t$ and the subscript $max$ means the maximum value. The larger the FLCE value is, the stronger the orbital instability is. The FLCEs of the example DRO (see Fig.~\ref{Fig_DRO}b) and the example NRHO (see Fig.~\ref{Fig_NRHO}b) are shown in Fig.~\ref{FLCE}. Clearly, the instability of DRO is much weaker than that of the NRHO. Actually, long-time integration of the example DRO shows that it is practically stable in the high-fidelity model.

\par Based on the above two factors, we are apt to choose the DRO as the orbit type for the near-Moon satellite. The final four-satellite constellation is composed of one L3-Lissajous satellite, one L4-TLPO satellite, one L5-TLPO satellite and one DRO satellite. This constellation has also been used in some previous studies \autocite{XU2024}. In the following sections, we proceed to analyze the AOD performance of the constellation based on this constellation architecture.

\section{AOD Accuracy Analysis}

\subsection{Orbit Parameter Analysis of a Constellation}
\par
\begin{figure}[htp]
    \centering
    \includegraphics[width=16cm]{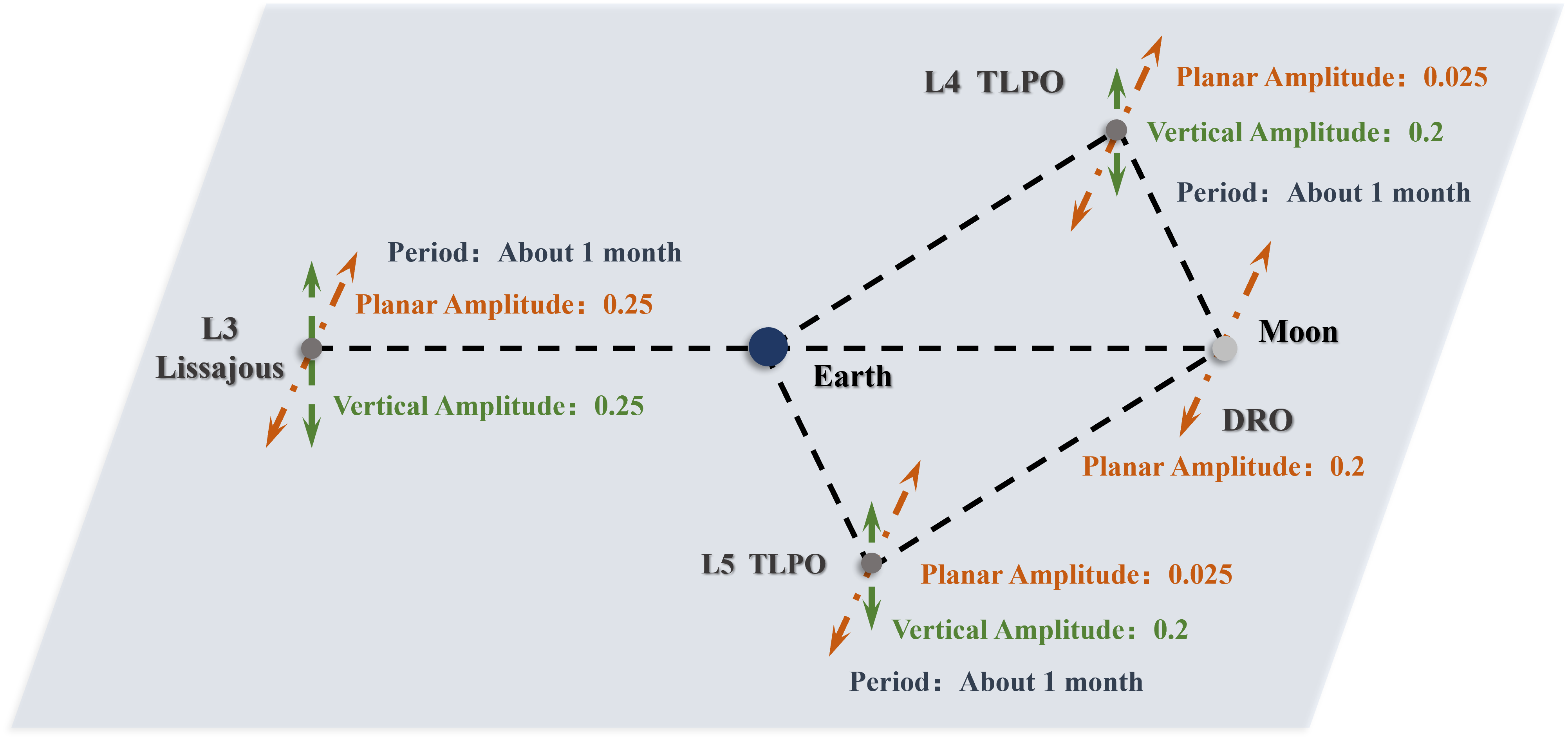}
    \caption{Relative geometry of the L3+L4+L5+DRO constellation and the upper limits of the amplitude parameters used in our analysis}
    \label{fig_constellation}
\end{figure}

 For the L3+L4+L5+DRO constellation, a key question is how to select these orbital parameters to optimize the AOD performance of the constellation. A schematic diagram is displayed in Fig.~\ref{fig_constellation} to show the orbit parameters to be considered. For each of the L3, L4, and L5 satellites, there are two amplitude parameters (the in-plane and the out-of-plane) and two associated initial phase angles. For the DRO satellite, only the in-plane amplitude and the initial phase angle are considered. Consequently, the constellation optimization involves a total of 14 free parameters. Before we start the AOD analysis, we need to put some constraints on these free parameters to make the analysis tractable. To simplify the expression, the normalized unit of distance is taken as the Earth-Moon distance.
\begin{itemize} 
\item[(1)] For the L3 Lissajous orbit, an upper limit of about 0.25 Earth-Moon distance (about 100,000km) for both the in-plane and the out-of-plane amplitudes is set in our analysis. It should be emphasized that the in-plane amplitude corresponds to the motion amplitude in the $x$-direction.
\item[(2)] For the L4 and L5 TLPOs, previous studies show that the dynamical substitutes are actually unstable \autocite{hou_2010} and the instability mainly appears in the in-plane motion. The size of the dynamical substitute itself is already 0.1 Earth-Moon distance. Both in-plane and out-of-plane free motions are allowed to exist around the dynamical substitute. Recent studies indicate that the out-of-plane amplitude $\beta$ of quasi-periodic orbits surrounding the dynamical substitute is limited by some resonances \autocite{liu_2024_stability}. For the interval $\beta\in[0.2, 0.4]$, it is difficult to construct long lasting quasi-periodic orbits surrounding the unstable dynamical substitute due to a strong resonance \autocite{liu_2024_frequency}. As a result, we put an upper limit for the out-of-plane amplitude as $\beta_{max}=0.2$ (about 76,000km). For the in-plane motion, we simply require the in-plane amplitude  $\alpha\in[0, 0.025]$ (about 10,000km).
\item[(3)] For the DROs, they are almost planar orbits. Previous studies show that these orbits become unstable in the high-fidelity force model when the in-plane amplitudes reach about 0.2 Earth-Moon distance (about 76,000km) \autocite{Bezrouk2017}. As a result, an upper limit of 0.2 Earth-Moon distance is set in our analysis.
\end{itemize} 
\par The in-plane amplitude of the orbits is much smaller than the in-plane size of the whole constellation. Our numerical experiments indicate that variations in the in-plane amplitudes have negligible influence on the AOD accuracy of the constellation. Therefore, the design parameter space is reduced to three pairs of out-of-plane parameters: the out-of-plane amplitudes (denoted $A_3$, $A_4$, $A_5$) and the initial out-of-plane phase angles (denoted $\phi_3$, $\phi_4$, $\phi_5$) for the L3 Lissajous, L4 TLPO, and L5 TLPO orbits, respectively.
\par Furthermore, considering the fact that the L3, L4 and L5 orbits have very close frequencies, the phase-angle differences between them drift slowly. Consequently, we further modify the parameter set to the following five parameters:
\begin{itemize} 
\item $\mathbf{A_3}$: the out-of-plane amplitude of the L3 Lissajous satellite.
\item $\mathbf{A_4}$: the out-of-plane amplitude of the L4 TLPO satellite.
\item $\mathbf{A_5}$: the out-of-plane amplitude of the L5 TLPO satellite.
\item $\mathbf{\Delta\phi_4}=\phi_4-\phi_3$: the difference between the initial out-of-plane phase angles of the L3 satellite and the L4 TLPO satellite.
\item $\mathbf{\Delta\phi_5}=\phi_5-\phi_3$: the difference between the initial out-of-plane phase angles of the L3 satellite and the L5 TLPO satellite.
\end{itemize}

\subsection{The $DAGDOP$ factor}
\par Based on the covariance matrix $\mathbf{P_z}$, Gao $\&$ Hou proposed the $DAGDOP$ factor to describe the OD accuracy under different conditions \autocite{Gao2022}. Denote the covariance matrix $P_z$ as:
\begin{equation}
\mathbf{P_z}=\begin{bmatrix}
    \sigma_{11}^2 & Cov(\hat{z}_1,\hat{z}_2) & \cdots & Cov(\hat{z}_1,\hat{z}_n) \\ Cov(\hat{z}_2,\hat{z}_1) & \sigma_{22}^2 & \cdots & Cov(\hat{z}_2,\hat{z}_n) \\ \vdots & \vdots & \ddots & \vdots \\ Cov(\hat{z}_n,\hat{z}_1) & Cov(\hat{z}_n,\hat{z}_2) & \cdots & \sigma_{nn}^2
\end{bmatrix}.\nonumber
\end{equation}
Here, $\hat{z}_i$ is the $i^{th}$ element of $\mathbf{\hat{z}}$ in Eq.~\eqref{OD} and $\sigma_{ii}^2=Cov(\hat{z}_i,\hat{z}_i)$ is the variance of $\hat{z}_i$. In the case of a single satellite and only the state vector being determining, $n=6$. In such a case, $DAGDOP$ can be expressed as:
\begin{equation}
\begin{aligned}\left\{\begin{aligned}
DAGDOP_{p} &= \sqrt{\sigma_{11}^2+\sigma_{22}^2+\sigma_{33}^2}, \\
DAGDOP_{v} &= \sqrt{\sigma_{44}^2+\sigma_{55}^2+\sigma_{66}^2}.
\end{aligned}\right.\end{aligned}\nonumber
\end{equation}
Obviously, $DAGDOP_p$ represents the standard deviation of the estimated position and $DAGDOP_v$ represents the standard deviation of the estimated velocity. It is clear that the smaller the $DAGDOP_p$ is, the better the OD accuracy is. Therefore, in order to improve the AOD accuracy, we need to choose orbital parameters to make the $DAGDOP_p$ as small as possible. For an $M$-satellite constellation, dimension of the covariance matrix is $n=6M$, and the $DAGDOP_p$ can be defined as:
\begin{equation}
DAGDOP_{p} = \sqrt{\sum_{i=1}^M\frac{\left(\sigma_{xi}^2+\sigma_{yi}^2+\sigma_{zi}^2\right)}{M}},
\label{DAGDOPp}
\end{equation}
where $\sigma_{xi}^2$, $\sigma_{yi}^2$, and $\sigma_{zi}^2$ represent the estimated position variances along the $x$-, $y$-, and $z$-axes for the $i^{th}$ satellite, respectively. In the following sections, unless otherwise specified, the term $DAGDOP$ refers specifically to $DAGDOP_p$ defined in Eq.~\eqref{DAGDOPp}.
\par According to Eq.~\eqref{Cov_matrix}, calculation of $\mathbf{P_z}$ requires inverting the matrix (denoted as matrix $\mathbf{P_z}'$): 
\begin{equation}
\mathbf{P_z}'=\sum^{N}_{i=1}(\mathbf{H}^{iT}\mathbf{H}^i). \nonumber
\end{equation}
The matrix $\mathbf{P_z}'$ is an $n{\times}n$ symmetric positive semi-definite matrix. The determinant of the matrix $\mathbf{P_z}'$ satisfies:
\begin{equation}
|\mathbf{P_z}'| = \prod_{k=1}^{n}\lambda_k', 
\label{Cov_martix_inv_Det}
\end{equation}
where $\lambda_k'$ is an eigenvalue of $\mathbf{P_z}'$, and satisfies:
\begin{equation}
\lambda_k' \geq 0 \quad (k=1,2,\ldots,n).
\label{Limitation-1}
\end{equation}
Therefore, the relationship between each element of $\mathbf{P_z}$ ($\sigma_{ij}^2$) and its inverse matrix $\mathbf{P_z}'$ can be expressed as:
\begin{equation}
\sigma_{ij}=\frac{A_{ji}}{|\mathbf{P_z}'|}=A_{ji}{/}\prod_{i=1}^{n}\lambda_i',  
\label{Minimum_inv_matrix}
\end{equation}
where ${A}_{ji}$ is the cofactor of $\sigma_{ij}$. Therefore, in order to minimize the $DAGDOP_p$, the value of $\sigma_{ii}$ should be minimized and $|\mathbf{P_z}'|$ should be maximized. From Eq.~\eqref{Cov_martix_inv_Det}, it can be observed that the determinant $|\mathbf{P_z}'|$ can be expressed as the product of the eigenvalues $\lambda_i'$. Therefore, maximizing the smallest eigenvalue $\lambda_{i,min}'$ leads to the greatest improvement in the value of $|\mathbf{P_z}'|$.

\subsection{Two Basic Theorems}
\par The following theorems pertain to the eigenvalues. Leveraging these theorems, we can derive a mathematical approximation for the smallest eigenvalue $\lambda_{i,min}'$ of the matrix $\mathbf{P_z}'$.

\begin{itemize} 
\item \textbf{Gershgorin Circle Theorem}
\end{itemize} 

\par For a $n{\times}n$ matrix $\mathbf{A}$, all the eigenvalues $\lambda_i$ of $\mathbf{A}$ should be located in at least one Gershgorin disc in the complex plane, which can be expressed as \autocite{1994_Brualdi}: 
\begin{equation}
\begin{aligned}\left\{\begin{aligned}
\lambda &\in \cup_{i=1}^{n} \left[z \in \mathbb{R},|z-a_{i,i}|\leq\sum_{j\neq i}|a_{i,j}|\right] ,\\
\lambda &\in \cup_{i=1}^{n} \left[z \in \mathbb{R},|z-a_{i,i}|\leq\sum_{j\neq i}|a_{j,i}|\right] .
\end{aligned}\right.\end{aligned}
\label{Limitation-2}
\end{equation}
Here, the first expression is known as the row disc and the second as the column disc.

\begin{itemize} 
\item \textbf{Rayleigh Quotient and Rayleigh-Ritz Theorem}
\end{itemize} 
\par For a real symmetric matrix $\mathbf{A}$ and any non-zero vector $x$, the Rayleigh quotient $R(\mathbf{A},x)$ is defined as \autocite{SUNAR_2001} :
\begin{equation}
R(\mathbf{A},x)=\frac{x^{T}\mathbf{A}x}{x^{T}x}.
\nonumber
\end{equation}
The Rayleigh-Ritz theorem can be written as
\begin{equation}
\lambda_{min} \leq \frac{x^{T}\mathbf{A}x}{x^{T}x} \leq \lambda_{max},
\nonumber
\end{equation}
where $\lambda_{min}$ is the minimum eigenvalue of matrix $\mathbf{A}$ and $\lambda_{max}$ is the maximum eigenvalue. When we choose a vector $x$ as the unit vector along the $i^{th}$ row, we can obtain the following expression: 
\begin{equation}
\lambda_{min} \leq min(a_{i,i}).
\label{Limitation-3}
\end{equation}

\subsection{The range of the minimum eigenvalue $\lambda_{min}$}
\par Denoting the element of the matrix $\mathbf{P_z}'$ as $p_{ij}$, we can derive the range of the minimum eigenvalue $\lambda_{i,\min}'$ of the matrix $\mathbf{P_z}'$ from Eqs.~\eqref{Limitation-1}, \eqref{Limitation-2} and \eqref{Limitation-3}.
\begin{equation}
\max\left[0,\min(p_{i,i})-\sum_{i\neq j}|p_{i,j}|\right]\leq \lambda_{i,min}' \leq \min(p_{i,i}).
\label{Limitation}
\end{equation}
To determine the bounds of $\lambda_{i,min}'$, we now focus on analyzing the left-hand side of Eq.~\eqref{Limitation}. Following Eq.~\eqref{Cov_matrix}, each element $p_{i,j}$ is computed as:
\begin{equation}
p_{i,j}=\sum_{k=1}^{N} \left[ \sum_{l=1}^{6n} \left(\frac{\partial Y^k}{\partial X_l^k} \frac{\partial X_l^k}{\partial X_i^0}\right) \times \sum_{l=1}^{6n} \left(\frac{\partial Y^k}{\partial X_l^k} \frac{\partial X_l^k}{\partial X_j^0}\right) \right].
\label{pij}
\end{equation}
Here $N$ is the number of observations, $n$ is the number of satellites, and $\mathbf{X}=[\mathbf{X}_1,\mathbf{X}_2,\ldots,\mathbf{X}_n]$ is the state vector of all the satellites. The superscript $k$ indicates the observation epoch, while $0$ refers to the initial epoch. The indices $i$, $j$, and $l$ denote positions within the state vector. For example, $X_l^k$ means the $l^{th}$ element in the state vector at the time of the $k^{th}$ observation.
\par Define $q_{k,*} = \sum_{\ell=1}^{6n} \left( \frac{\partial Y}{\partial X_{\ell}^k} \cdot \frac{\partial X_{\ell}^k}{\partial X_{*}^0} \right)$, $p_{i,j}^{k} = q_{k,i} {\times} q_{k,j}$, and $\mathbf{P}^{k}{'} = \mathbf{H}_k^T\mathbf{H}_k$. For the $k^{th}$ observation, the smallest absolute value among all $q_{k,*}$ is denoted by $q_{k,i_m}$, where $i_m$ is the corresponding state vector index. Since $p_{i,i}^{k} = (q_{k,i})^2$, it follows that $p_{i_m, i_m}^{k}$ is the minimal diagonal element of the matrix $\mathbf{P}^{k}{'}$ for this observation. In the local $6\times6$ submatrix corresponding to each satellite (taking the $p^{th}$ satellite as an example), there exists a partial minimal diagonal element $p_{i_{mp},i_{mp}}^k$. Here $i_{mp}$ represents a global index and ranges from 1 to $6n$. Clearly, $p_{i_m ,i_m}^{k}$ satisfies:
\begin{equation}
p_{i_m ,i_m}^{k}=\min\left[p_{i_{m1},i_{m1}}^k \quad p_{i_{m2},i_{m2}}^k \quad \ldots \quad p_{i_{mp},i_{mp}}^k\right].\nonumber
\end{equation}
For each submatrix, denote $j_m$ as one of the indexs other than $i_m$, then:
\begin{equation}
p_{i_{mp},i_{mp}}^k \leq |p_{i_{mp},j_{m}}^k| \leq \sum_{j \neq i_{mp}} |p_{i_{mp},j}^k|.\nonumber
\end{equation}
Owing to the relatively stable geometric configuration of the satellite constellation, the ordinal relationship of the $q_{k,*}$ values remains within each submatrix (proved in Section 4.5). In other words, within each submatrix, the relative index of the minimal diagonal element remains unchanged. After superimposing multiple observations, the minimal diagonal element $\min{(p_{i,i})}$ is still among the set of fixed indices:
\begin{equation}
\min{(p_{i,i})}=p_{i_{m},i_{m}}=\min\left[p_{i_{m1},i_{m1}} \quad p_{i_{m2},i_{m2}} \quad \ldots \quad p_{i_{mp},i_{mp}}\right].\nonumber
\end{equation}
Here $i_{m}$ is the global index of the global minimal diagonal element $\min{(p_{i,i})}$. Similarly, $\min{(p_{i,i})}$ satisfies:
\begin{equation}
\min(p_{i,i})=p_{i_{m},i_{m}} \leq \sum_{j \neq i_{m}}|p_{i_m,j}|.\nonumber
\end{equation}
Therefore, Eq.~\eqref{Limitation} can be rewritten as:
\begin{equation}
0 \leq \lambda_{i,\min}' \leq \min(p_{i,i}).
\label{eq:minimum_eigenvalue_range}
\end{equation}
Consequently, leveraging the inequality $0 \leq \lambda_{i,\min}' \leq \min(p_{i,i})$, we can obtain a larger $\lambda_{i,\min}'$ by maximizing its upper bound $\min(p_{i,i})$. Thus, maximizing $\min(p_{i,i})$ is adopted as the proxy objective, which helps minimize the covariance and the $DAGDOP$. We remark that this conclusion is only a qualitative approximation. Nevertheless, it suffices as a tool for qualitative analysis.

\subsection{The minimum diagonal element of $\mathbf{P_z}'$}
\par The next step is to calculate the diagonal elements of $\mathbf{P_z}'$ and locate the minimum one. We begin by deriving an explicit expression for a diagonal element $p_{i,i}$. Setting $i = j$ in Eq.~\eqref{pij} yields the expression for the diagonal elements $p_{i,i}$. In this formula, $\bm{\Phi}(t_k)={\partial \mathbf{X}^{k}}/{\partial \mathbf{X}^{0}}$ is the STM and ${\partial Y^k}/{\partial \mathbf{X}^{k}}$ is the observation matrix. For the inter-satellite range data between satellite $p$ and satellite $q$, the observation matrix with respect to the state vector $\mathbf{X_p}$ can be expressed as:
\begin{equation}
\frac{\partial Y(\mathbf{X}_p,\mathbf{X}_q)}{\partial \mathbf{X}_p}=\frac{\mathbf{r}_p-\mathbf{r}_q}{\sqrt{(x_p-x_q)^2+(y_p-y_q)^2+(z_p-z_q)^2}}=\left[ \frac{x_p-x_q}{Y}\;\,\frac{y_p-y_q}{Y}\;\,\frac{z_p-z_q}{Y}\;\, 0 \;\, 0 \;\, 0\right].
\nonumber
\end{equation}
Evidently, the first three elements of $\partial{Y}/\partial{X_p}$ are precisely the unit vector pointing from satellite $q$ to satellite $p$. Thus, $p_{i,i}$ can be approximately expressed as a function of the out-of-plane parameters. For the $i_m^{th}$ diagonal element of the $m^{th}$ spacecraft, which, in the global $6n\times6n$ matrix, corresponds to the index $i = 6(m-1) + i_m$, the value $p_{i,i}$ can be expressed using Eq.~\eqref{pij} as:
\begin{equation}
p_{i,i} = \sum_{k=1}^{\mathrm{N}} \left[ \sum_{l=1}^{6n} \left(\frac{\partial Y^k}{\partial X_l^k} \frac{\partial X_l^k}{\partial X_i^0}\right)\right]^2 = \sum_{k=1}^N \left( \frac{\partial Y^k}{\partial x_m^k} \frac{\partial x_m^k}{\partial X_{m,i_m}^0} + \frac{\partial Y^k}{\partial y_m^k} \frac{\partial y_m^k}{\partial X_{m,i_m}^0} + \frac{\partial Y^k}{\partial z_m^k} \frac{\partial z_m^k}{\partial X_{m,i_m}^0}  \right)^2.
\label{pii}
\end{equation}
Here $X_i^0=X_{m,i_m}^0$ is the $i_m^{th}$ initial state component of the $m^{th}$ satellite, $\mathbf{r}^k_m=[x^k_m,y^k_m,z^k_m]$ is the position vector of the $m^{th}$ satellite at time $t^k$. Consequently, the value of $p_{i,i}$ is  determined only by $\partial Y/ \partial \mathbf{r}^k_m$ and the upper $3 \times 6$ block of the $m^{th}$ satellite's STM (denoted as $\bm{\Phi}_m^{upper}$), which is defined as $\bm{\Phi}_m^{upper}=[{\partial\mathbf{r}}/{\partial\mathbf{r_0}},{\partial\mathbf{r}}/{\partial\mathbf{\dot{r}_0}}]$.
\par To analyze these unit vectors, we introduce a simplified geometry model. In the Earth-Moon synodic coordinate system, with distances normalized to the Earth-Moon distance $L_{\text{EM}}$, we can define the vertical coordinates of L3, L4, and L5 satellites at time $t$ as $z_3$, $z_4$ and $z_5$ respectively. As mentioned in Section~4.1, $|z_3|\leq0.25$, $|z_4|\leq0.2$, and $|z_5|\leq0.2$. Therefore, the out-of-plane motions are all small parameters compared with the constellation size. Consequently, the simplified estimation of the coordinates of the four satellites in the synodic frame can be approximated as: L3 satellite $(-1,0,z_3)$, L4 satellite $(1/2,\sqrt{3}/2,z_4)$, L5 satellite $(1/2,-\sqrt{3}/2,z_5)$, and DRO satellite $(1,0,0)$. With these approximations, we can derive the relative vector between any pair of the two satellites. For example, the relative vectors from the L3 satellite to the L4 satellite, from the L3 satellite to the DRO satellite and from the L4 satellite to the L5 satellite are denoted as $\mathbf{\hat{r}}_{L3\xrightarrow{}L4}$, $\mathbf{\hat{r}}_{L3\xrightarrow{}DRO}$, and $\mathbf{\hat{r}}_{L4\xrightarrow{}L5}$ respectively, which can be simplified to:
\begin{equation}
\begin{aligned}\left\{\begin{aligned}
&\mathbf{\hat{r}}_{L3\xrightarrow{}L4}=\left[\frac{\sqrt{3}}{2}-\frac{(z_4-z_3)^2}{4\sqrt{3}},\frac{1}{2}-\frac{(z_4-z_3)^2}{12},\frac{z_4-z_3}{\sqrt{3}}-\frac{(z_4-z_3)^3}{6\sqrt{3}}\right];\\
&\mathbf{\hat{r}}_{L3\xrightarrow{}DRO}=\left[1-\frac{(z_3)^2}{8},0,-\frac{z_3}{2}+\frac{z_3^3}{16}\right];\\
&\mathbf{\hat{r}}_{L4\xrightarrow{}L5}=\left[0,1-\frac{(z_5-z_4)^2}{6},\frac{z_5-z_4}{\sqrt{3}}-\frac{(z_5-z_4)^3}{6\sqrt{3}}\right].
\end{aligned}\right.\end{aligned}
\label{unit_vector}
\end{equation}
In these expressions, the $z$-components $\partial{Y}/\partial{z_i}$ are of order $\mathcal{O}(z_i - z_j)$, while the non-zero $x$-components $\partial{Y}/\partial{x_i}$ and $y$-components $\partial{Y}/\partial{y_i}$ are of order $\mathcal{O}(1)$. Given that $|z_i - z_j| < 1$, the $z$-components are therefore higher-order small quantities compared to the leading-order in-plane components. In fact, this conclusion applies to all the relative position vectors between each of two satellites in the aforementioned constellation.

\par The value of the STM also affects the diagonal elements $p_{i,i}$. Truncating at the first order, we can approximate the out-of-plane motions around the L3, L4, and L5 satellites as:
\begin{equation}
\begin{aligned}\left\{\begin{aligned}
z_3(t) &= A_3\cos{(\omega t+\phi_3)}, \\
z_4(t) &= A_4\cos{(\omega t+\phi_3+\Delta\phi_4)}, \\
z_5(t) &= A_5\cos{(\omega t+\phi_3+\Delta\phi_5)}. \\
\end{aligned}\right.\end{aligned} 
\label{out-of-plane_motion}
\end{equation}
Here, $\omega$ is the frequency of the $z$-direction motion. For the L3, L4 and L5 satellites, their out-of-plane frequencies are almost the same. If one month is normalized to $2\pi$, then $\omega \approx 1$. Truncating at the first order, the planar motion of the $m^{th}$ satellite can be approximated as:
\begin{equation}
\begin{aligned}\left\{\begin{aligned}
x_m(t) &= A_m^p\cos{(\omega_m^p t+\phi_m^p)}+x_m^c, \\
y_m(t) &= A_m^p\sin{(\omega_m^p t+\phi_m^p)}+y_m^c.
\end{aligned}\right.\end{aligned}
\label{in-plane_motion}
\end{equation}
The superscript $p$ indicates that the parameter pertains to planar motion. $x_m^c$ and $y_m^c$ are the planar coordinates of the reference point of the $m^{th}$ satellite. For the DRO satellite, the reference point is the Moon; for the Lagrange point satellites, the reference points are the corresponding Lagrange point.
\par Truncated at the first order, Eqs.~\eqref{out-of-plane_motion} and \eqref{in-plane_motion} show that the initial in-plane position and velocity $[x^0,y^0,\dot{x}^0,\dot{y}^0]$ predominantly govern the in-plane motion $[x,y,\dot{x},\dot{y}]$. For these four terms of the $m^{th}$ satellite, taking $x$ as an example, Eq.~\eqref{pij} can be rewritten as:
\begin{equation}
p_{i_x,i_x} = \sum_{k=1}^N \left( \frac{\partial Y^k}{\partial x_m^k} \frac{\partial x_m^k}{\partial x_m^0} + \frac{\partial Y^k}{\partial y_m^k} \frac{\partial y_m^k}{\partial x_m^0} \right)^2,
\label{pxx-pyy}
\end{equation}
where $i_x=6m-5$ is the index corresponding to $x_m$. Moreover, the initial out-of-plane position and velocity $[z^0,\dot{z}^0]$ predominantly govern the out-of-plane motion $[z,\dot{z}]$. Similarly, taking $z$ as an example, the expression is:
\begin{equation}
p_{i_z,i_z} = \sum_{k=1}^N \left( \frac{\partial Y^k}{\partial z_m^k} \frac{\partial z_m^k}{\partial z_m^0} \right)^2,
\label{pzz}
\end{equation}
where the $i_z=6m-3$ is the index corresponding to  $z_m$. From the Eqs.~\eqref{out-of-plane_motion} and \eqref{in-plane_motion}, under a first-order approximation, $\partial{x_m}/\partial{x^0_m}$, $\partial{x_m}/\partial{y^0_m}$, $\partial{y_m}/\partial{x^0_m}$, $\partial{y_m}/\partial{y^0_m}$ and $\partial{z_m}/\partial{z^0_m}$ are roughly of the same order of magnitude. However, $\partial{Y}/\partial{z_m}$ is one order of magnitude smaller than $\partial{Y}/\partial{x_m}$ and $\partial{Y}/\partial{y_m}$. The same holds for the velocity terms. For this reason, the diagonal elements $p_{i_{z},i_{z}}$ corresponding to $z$ and $p_{i_{\dot{z}},i_{\dot{z}}}$ corresponding to $\dot{z}$ are smaller than the diagonal elements corresponding to $[x,y]$ and $[\dot{x},\dot{y}]$. In other words, for each satellite, the smaller of $p_{i_{z},i_{z}}$ and $p_{i_{\dot{z}},i_{\dot{z}}}$ is the minimum diagonal element.
\par Using Eqs.~\eqref{out-of-plane_motion}$\&$\eqref{in-plane_motion}, ${\partial \mathbf{r}_{m}(t)}/{\partial z_{m}^0}$ and ${\partial \mathbf{r}_{m}(t)}/{\partial \dot{z}_{m}^0}$ can be expressed as:
\begin{equation}
\begin{aligned}
&\frac{\partial \mathbf{r}_m(t)}{\partial z_m^0} = \left[\frac{\partial x_m(t)}{\partial z_m^0} \quad \frac{\partial y_m(t)}{\partial z_m^0} \quad \frac{\partial z_m(t)}{\partial z_m^0}\right] = \left[0 \quad 0 \quad \cos{\omega{t}}\right],
\\ &\frac{\partial \mathbf{r}_m(t)}{\partial \dot{z}_m^0} = \left[\frac{\partial x_m(t)}{\partial \dot{z}_m^0} \quad \frac{\partial y_m(t)}{\partial \dot{z}_m^0} \quad \frac{\partial z_m(t)}{\partial \dot{z}_m^0}\right] = \left[0 \quad 0 \quad \frac{\sin{\omega{t}}}{\omega}\right].
\end{aligned}
\label{STM_simple}
\end{equation}
Here, ${\partial \mathbf{r}_m(t)}/{\partial z_m^0}$ is dimensionless, while ${\partial \mathbf{r}_m(t)}/{\partial \dot{z}_m^0}$ has the unit of time. It should be noted that the numerical values of the STM elements presented below are obtained under a different time normalization. In our software, the time unit is chosen to be $\sqrt{R_E^3/\mu_E}\approx{806.81}$ s rather than the lunar synodic period , where $R_E$ is the radius of the Earth. Under this normalization, for the cislunar satellites, the value of $\omega$ is small enough that ${\partial z_m(t)}/{\partial \dot{z}_m^0}$ is larger than ${\partial z_m(t)}/{\partial z_m^0}$ during a short arc. As a result, Eq.~\eqref{STM_simple} approximately holds only in a short arc. Here we give two examples: a 1-day STM of the L3 Lissajous satellite $\bm{\Phi}_{L_3}^{upper}(1d)$, given by:
\begin{equation}
\bm{\Phi}_{L_3}^{upper}(1d)=
\begin{bmatrix}
1.18 & -5.24 & 0.00 & 5.09 & -0.32 & 0.00 \\
4.06 & 1.24 & 0.00 & -0.42 & 4.26 & 0.00 \\
0.00 & 0.00 & 0.76 & 0.00 & 0.00 & 4.09
\end{bmatrix},
\nonumber
\end{equation}
and a 3-day STM of the L3 Lissajous satellite $\bm{\Phi}_{L_3}^{upper}(3d)$, as:
\begin{equation}
\bm{\Phi}_{L_3}^{upper}(3d)=
\begin{bmatrix}
226.82 & -322.93 & -0.12 & 324.10 & 224.61 & 0.03 \\
233.23 & 172.23 & -0.13 & -171.22 & 234.06 & 0.04 \\
-0.07 & 0.13 & 1.22 & 0.03 & -0.07 & 285.94
\end{bmatrix}.
\nonumber
\end{equation}
The third and sixth columns of these matrices approximately follow the form of Eq.~\eqref{STM_simple}. As time increases, Eq.~\eqref{in-plane_motion} no longer holds. Hence, the conclusion that $p_{i_z,i_z}$ is the minimum diagonal element for each satellite, and that is mainly determined by the $z$-component motion is only valid over a short arc.

\subsection{Relative Planarity Factor $P_\text{r}$}
In the previous subsection, the minimum diagonal element of $\mathbf{P_z}'$ in the synodic coordinate system was determined to be the $p_{i_z,i_z}$ term. It is not difficult to find that for any positive semi-definite matrix similar to $\mathbf{P_z}'$ (denoted as matrix $\mathbf{B}$), Eq.~\eqref{eq:minimum_eigenvalue_range} still holds. Therefore, finding the minimum diagonal element of any positive semi-definite matrix $\mathbf{B}$ similar to $\mathbf{P_z}'$ can provide a more accurate approximation for the minimum eigenvalue $\lambda_{i,\min}'$. The relationship between the matrix $\mathbf{B}$ and the matrix $\mathbf{P_z}'$ is given by:
\begin{equation}
\mathbf{B} = \mathbf{V}^{-1}\mathbf{P_z}'\mathbf{V}. 
\label{eq:Similar_Matrix_Pz'}
\end{equation}
To ensure that matrix $\mathbf{B}$ is positive semi-definite, $\mathbf{V}$ must be an orthogonal matrix, i.e.:
\begin{equation}
\mathbf{V}^{-1}=\mathbf{V}^{\text{T}}. 
\label{eq:fm}
\end{equation}
For the $m^{th}$ spacecraft, the value of $p_{i_z,i_z}$ is much smaller than its remaining five diagonal elements. Therefore, if there exists a smaller diagonal element $f_m$ in the corresponding position of matrix $\mathbf{B}$, it should also be located at the position corresponding to the $p_{i_z,i_z}$ term. It is assumed that there is only a slight deviation between $f_m$ and $p_{i_z,i_z}$. From Eqs.~\eqref{eq:Similar_Matrix_Pz'} and \eqref{pij}, $f_m$ can be expressed as:
\begin{equation}
f_m=\nu_{i_z}^2p_{i_z,i_z}+\sum_{\substack{j_1,j_2 = 6m-5 \\ j_1 \neq i_z \ \text{or} \ j_2 \neq i_z}}^{6m}\nu_{j_1}\nu_{j_2}p_{i_z,j_1}p_{j_2,i_z},
\nonumber
\end{equation}
where the coefficients $\nu_i$ satisfy
\begin{equation}
\sum\nu_i^2 = 1, 
\nonumber
\end{equation}
and due to the small deviation,
\begin{equation}
\nu_{i_z} \gg \nu_j \quad (j \neq i_z). 
\nonumber
\end{equation}
Using the approximation of Eq.~\eqref{STM_simple}, the corresponding velocity terms and the contributions of the STM in Eq.~\eqref{pxx-pyy} can be further neglected. Therefore, the minimum diagonal element $f_m$ depends only on the relative position vectors between spacecraft. Let $\mathbf{X}_m^k$ be the position of spacecraft $m$ and $\mathbf{X}_q^k$ be the position of spacecraft $q$ at the $k^{th}$ observation epoch. For the case with $N$ observation and $M$ satellites, $f_m$ can be expressed as:
\begin{equation}
f_m =  \sum_{k=1}^{N}\sum_{q\neq{m}}^M\left[\frac{\mathbf{v}^{\text{T}}\cdot(\mathbf{X}_m^k-\mathbf{X}_q^k)}{|\mathbf{X}_m^k-\mathbf{X}_q^k|}\right]^2 = \mathbf{v}^{\text{T}} \sum_{k=1}^{N}\sum_{q\neq{m}}^M\frac{\left(\mathbf{X}_m^k-\mathbf{X}_q^k\right)\left(\mathbf{X}_m^k-\mathbf{X}_q^k\right)^{\text{T}}}{|\mathbf{X}_m^k-\mathbf{X}_q^k|^2} \mathbf{v},
\label{eq:Minimum_diagonal_element_planar_single}
\end{equation}
where $\mathbf{v}=[\nu_1,\nu_2,\nu_3]^{\text{T}}$ is a unit vector satisfying
\begin{equation}
\mathbf{v}^{\text{T}}\mathbf{v} = 1. 
\nonumber
\end{equation}
Here, $\nu_1$, $\nu_2$, $\nu_3$ are the projections of the unit vectors along the $x$, $y$, and $z$ axes of the synodic system onto the direction of $\mathbf{v}$, geometrically corresponding to a rotation of the coordinate system. Since all spacecraft share the same coordinate system and $\mathbf{v}$ geometrically represents the direction of this global rotation, $\mathbf{v}$ should take the same value for different spacecraft.
\par To optimize AOD performance, it is necessary to minimize ${DAGDOP}$, which corresponds to maximizing the minimum diagonal element of matrix $\mathbf{B}$. The minimum diagonal element among the six diagonal elements corresponding to the $m^{th}$ spacecraft is $f_m$. Therefore, for a constellation of $M$ spacecraft ($M=4$ here), it is desirable to maximize each of $f_m$ $(m=1,2,\ldots,M)$. Given the symmetric configuration and similar observation conditions among spacecraft, maximizing the sum of each $f_m$ is approximately equivalent to maximizing their sum $f$, which satisfies:
\begin{equation}
\begin{aligned}
f(\mathbf{v}) = \sum_{m=1}^{M}f_m &= \sum_{k=1}^{N}\sum_{m=1}^{M}\sum_{q\neq{m}}^M\left[\frac{\mathbf{v}^{\text{T}}\cdot(\mathbf{X}_m^k-\mathbf{X}_q^k)}{|\mathbf{X}_m^k-\mathbf{X}_q^k|}\right]^2 \\
&= \mathbf{v}^{\text{T}} \sum_{k=1}^{N}\sum_{m=1}^{M}\sum_{q\neq{m}}^M\frac{\left(\mathbf{X}_m^k-\mathbf{X}_q^k\right)\left(\mathbf{X}_m^k-\mathbf{X}_q^k\right)^{\text{T}}}{|\mathbf{X}_m^k-\mathbf{X}_q^k|^2} \mathbf{v}=\mathbf{v}^{\text{T}}\mathbf{S}\mathbf{v} .
\label{eq:Minimum_diagonal_element_planar}
\end{aligned}
\end{equation}
To simplify the expression, define the matrix $\mathbf{S}$ as:
\begin{equation}
\mathbf{S} = \sum_{k=1}^{N}\sum_{m=1}^{M}\sum_{q\neq{m}}^M\frac{\left(\mathbf{X}_m^k-\mathbf{X}_q^k\right)\left(\mathbf{X}_m^k-\mathbf{X}_q^k\right)^{\text{T}}}{|\mathbf{X}_m^k-\mathbf{X}_q^k|^2}. 
\label{eq:Matrix_S_RPF}
\end{equation}
A suitable vector $\mathbf{v}$ needs to be chosen to minimize $f(\mathbf{v})$. The Lagrange multiplier method is employed here. Introducing the Lagrange multiplier $\lambda$, the problem reduces to finding the minimum of $F(\mathbf{v},\lambda)$, where
\begin{equation}
F(\mathbf{v},\lambda) = f(\mathbf{v})-{\lambda}(\mathbf{v}^{\text{T}}\mathbf{v}-1). 
\nonumber
\end{equation}
Setting $\partial{F(\mathbf{v},\lambda)}/\partial{\mathbf{v}}=0$ yields:
\begin{equation}
\frac{\partial{F(\mathbf{v},\lambda)}}{\partial{\mathbf{v}}} = 2\left(\mathbf{S}\mathbf{v}-\lambda\mathbf{v}\right)=0. 
\label{eq:lagrange_stationary}
\end{equation}
Clearly, when the unit vector $\mathbf{v}$ is an eigenvector of $\mathbf{S}$, $F(\mathbf{v},\lambda)$ attains an extreme value. In this case, the Lagrange multiplier $\lambda$ is the corresponding eigenvalue of $\mathbf{S}$. Consequently, it can be shown that the minimum of $F(\mathbf{v},\lambda)$ is exactly the minimum eigenvalue $\lambda_{\min}^\mathbf{S}$ of matrix $\mathbf{S}$. Its geometric interpretation is the minimum component of the real-time relative unit vectors between satellites in the constellation along a spatial direction. We define this quantity as the Relative Planarity Factor (RPF) $P_\text{r}$, i.e.:
\begin{equation}
P_\text{r} = \lambda_{\min}^\mathbf{S}. 
\label{eq:RPF}
\end{equation}
\par Using Eq.~\eqref{out-of-plane_motion}, $P_\text{r}$ can be simplified to a function of the out-of-plane parameters $A_3$, $A_4$, $A_5$, $\Delta\phi_4$, $\Delta\phi_5$, $\phi_3$ and observation time $t$. Accordingly, Eq.~\eqref{eq:minimum_eigenvalue_range} can be rewritten as:
\begin{equation}
0 \leq \lambda_{i,\min}' \leq P_\text{r}(A_3,A_4,A_5,\Delta\phi_4,\Delta\phi_5;\phi_3,t).
\label{eq:minimum_eigenvalue_range_RPF}
\end{equation}
When $P_\text{r}$ is small, all relative position vectors are nearly perpendicular to a certain fixed direction, implying that the spacecraft are approximately coplanar. Let $\theta$ be the angle between the eigenvector corresponding to $P_\text{r}$ and the $+z$ axis. Calculation shows that for the constellation composed of L3+L4+L5+DRO, the maximum value of $\theta$ is approximately 9 degrees. In this case, $\nu_{i_z}=\cos\theta\approx0.989$. Hence, the deviation of the minimum eigen-direction from the $z$-axis is extremely small, and the simplified calculation described above can be fully utilized to analyze the minimum diagonal element.
\par Through the above analysis, under short-arc observations, the variation of $P_\text{r}$ for different constellation configurations should theoretically be negatively correlated with the variation of ${DAGDOP}$. By fixing the OD arc length and observation interval and varying the constellation design parameters, the variation of $P_\text{r}$ can be obtained analytically, thereby enabling the analysis of the influence of different factors on the AOD performance of the constellation. 
\par From Eqs.~\eqref{eq:Minimum_diagonal_element_planar} and \eqref{eq:Matrix_S_RPF}, another fact is obvious: In order to get high accuracy in the short arc AOD, at least four satellites are required. This is because the coplanar condition can always be satisfied if there are only three satellites in the constellation. In a short arc, three satellites cannot provide a large value of $P_\text{r}$. Hence, it is recommended to use a constellation of at least four satellites to avoid the coplanarity in a short arc.

\section{Numerical Simulations}
\par This section presents numerical simulations to validate the feasibility as well as the limitations of the aforementioned theoretical analysis. We devide the analysis into two different scenarios: the short-arc AOD scenario and the long-arc AOD scenario, and then analyze the effects of the five relative invariant out-of-plane parameters and the variable parameter $\phi_3$. In the numerical simulations, an arc of three days or less is considered short; otherwise, it is considered long. We only consider the random errors in the inter-satellite range data, with a standard deviation of 1 meter. The observation intervals $\Delta t$ are the same, which is set as 5 minutes. For the subsequent simulations, a baseline parameter configuration is established. Unless otherwise stated, default simulation parameters are provided in Table~\ref{Fundamental Set 0}. In the short-arc AOD scenario, the arc length $t_{arc}$ is set as 3 days.
\begin{table}[htp]
    \caption{\textbf{The Fundamental Set of the simulations}}
    \renewcommand\arraystretch{1.2}
    \centering
    \begin{tabular}{|c|c|c|c|c|} \hline
    \multirow{2}{*}{\textbf{Orbit Types}} & \multicolumn{2}{c}{\textbf{Out-of-plane Parameters}} & \multicolumn{2}{|c|}{In-plane Parameters} \\ \cline{2-5}
    & Amplitude & initial phase & Amplitude & initial phase \\ \hline
    L3 Lissajous & $A_3=50000\text{km}$ & ${\phi_3=0}$ & $A_3^p=5000\text{km}$ & $\phi_3^p=0$  \\ \hline
    L4 TLPO & ${A_4=50000\text{km}}$ & ${\Delta\phi_4=\pi}$ & $A_4^p=5000\text{km}$ & $\phi_4^{p}=0$ \\ \hline
    L5 TLPO & ${A_5=50000\text{km}}$ & ${\Delta\phi_5=\pi}$ & $A_5^p=5000\text{km}$ & $\phi_5^p=0$ \\ \hline
    DRO & \multicolumn{2}{c|}{\diagbox{}} & $A_{DRO}^p=10000\text{km}$ & $\phi_{DRO}^p=\pi$ \\ \hline
    \end{tabular}
    \label{Fundamental Set 0}
\end{table}
\par The correlation coefficient is a useful metric for quantifying the negative correlation between $P_\text{r}$ and $DAGDOP$. For two parameters $X$ and $Y$, the correlation coefficient between them $r(X,Y)$ is defined as:
\begin{equation}
r(X,Y) = \frac{Cov(X,Y)}{\sqrt{\text{Var}(X)\cdot\text{Var}(Y)}}.
\label{Correlation coefficient}
\end{equation}
$Cov(X,Y)$ is the covariance between $X$ and $Y$. $\text{Var}(X)$ and $\text{Var}(Y)$ are the variances of $X$ and $Y$ respectively. The value of $r(X,Y)$ lies in $[-1,1]$. When $r(X,Y)<0$, it means that $X$ and $Y$ are negatively correlated; when $r(X,Y)=0$, $X$ and $Y$ are uncorrelated; and when $r(X,Y)>0$, $X$ and $Y$ are positively correlated.

\subsection{Short-arc AOD Scenario}
\subsubsection{Influence of the out-of-plane amplitudes: $A_3$, $A_4$ and $A_5$}
\par
\begin{figure}[!h]
\centering   
\subfigure[Variation of $P_\text{r}$ with respect to $A_3$] 
{
\begin{minipage}[b]{.45\linewidth} 
\centering
\includegraphics[scale=0.55]{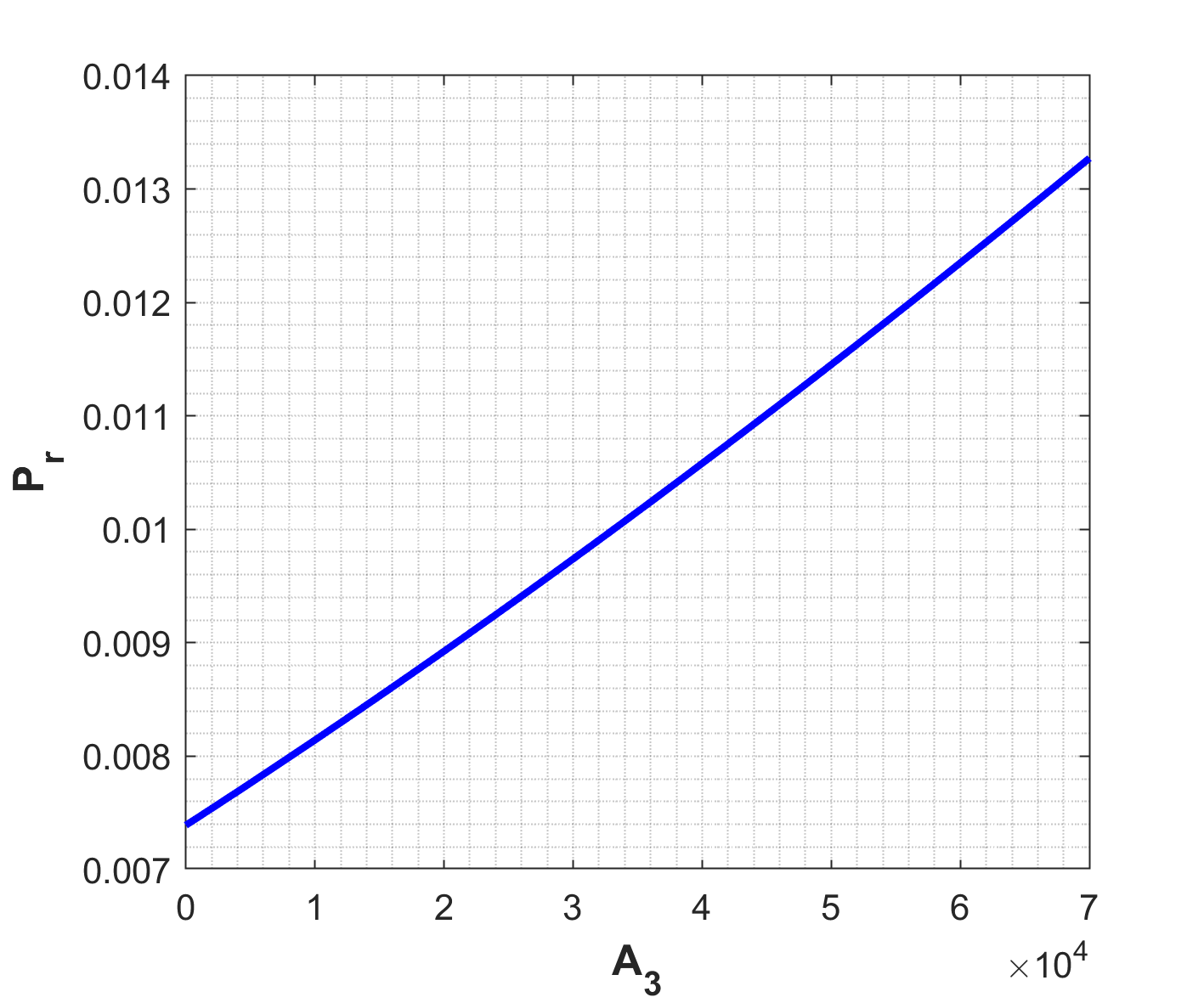}
\end{minipage}
}
\subfigure[Variation of $DAGDOP$ with respect to $A_3$]
{
\begin{minipage}[b]{.45\linewidth}
\centering
\includegraphics[scale=0.55]{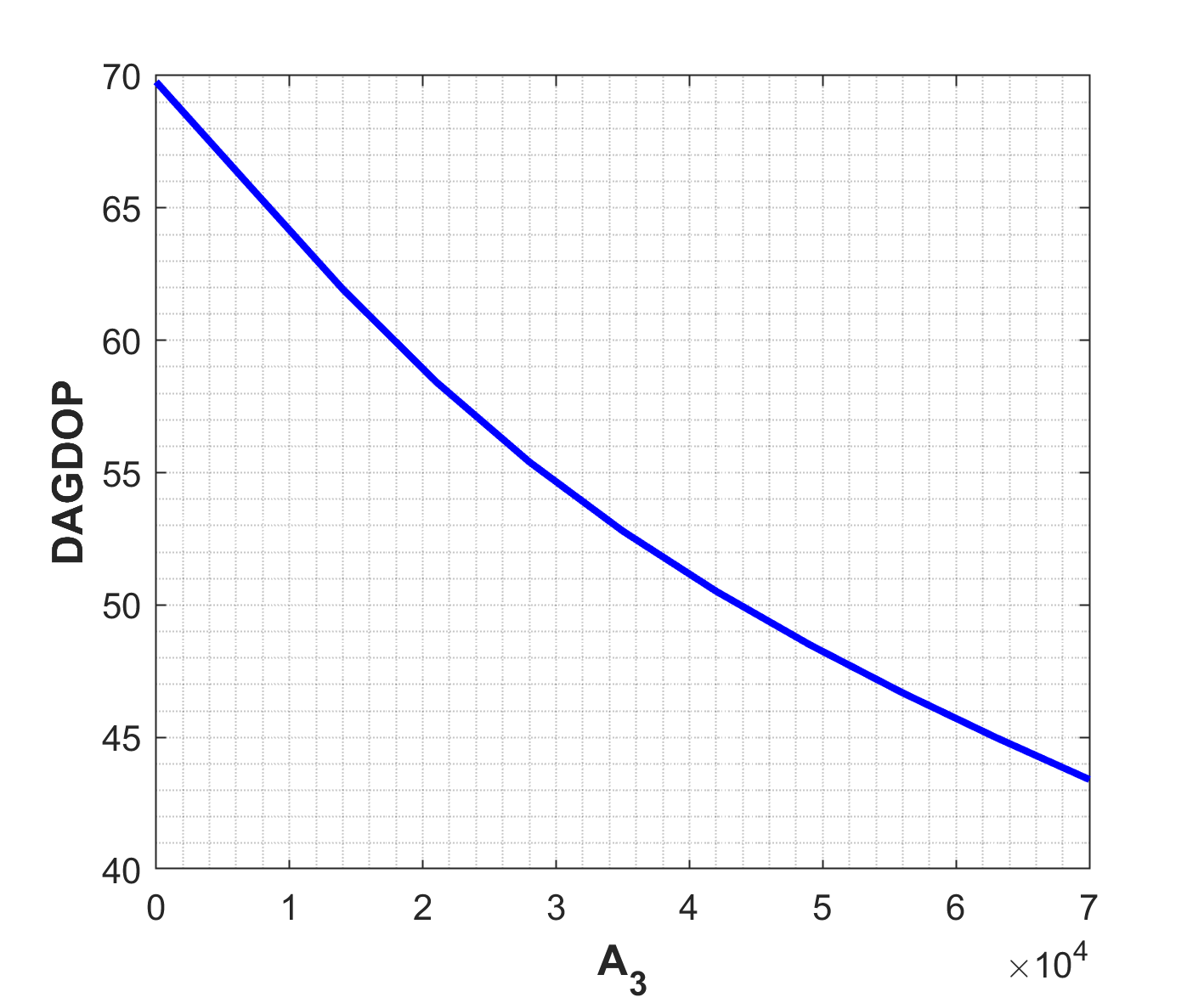}
\end{minipage}
}
\centering
\caption{Variation of $P_\text{r}$ and $DAGDOP$ with respect to $A_3$ in simulation 1 \\ $r(P_\text{r},DAGDOP)=-0.9806$}
\label{Result_A3}
\end{figure}

 We first analyze the influence of $A_3$ (corresponding to simulation 1), and then extend the analysis to the influence of $A_4$ and $A_5$ (corresponding to simulation 2). For simulation 1, the parameter $A_3$ varies in the range:
\begin{equation}
A_3\in[0,70000]~\text{km}.
\nonumber
\end{equation}
In Fig.~\ref{Result_A3}, subplot (a) shows the variation of $P_\text{r}$ with respect to $A_3$, and subplot (b) depicts the variation of the $DAGDOP$ factor. A strong negative correlation exists between $P_\text{r}$ and the $DAGDOP$ factor, which agrees with the above analysis. As $A_3$ increases, $P_\text{r}$ increases monotonically, whereas the $DAGDOP$ factor decreases. 

\par
\begin{figure}[h!]
\centering   
\subfigure[Variation of $P_\text{r}$ with respect to $A_4$ and $A_5$] 
{
\begin{minipage}[b]{.45\linewidth} 
\centering
\includegraphics[scale=0.6]{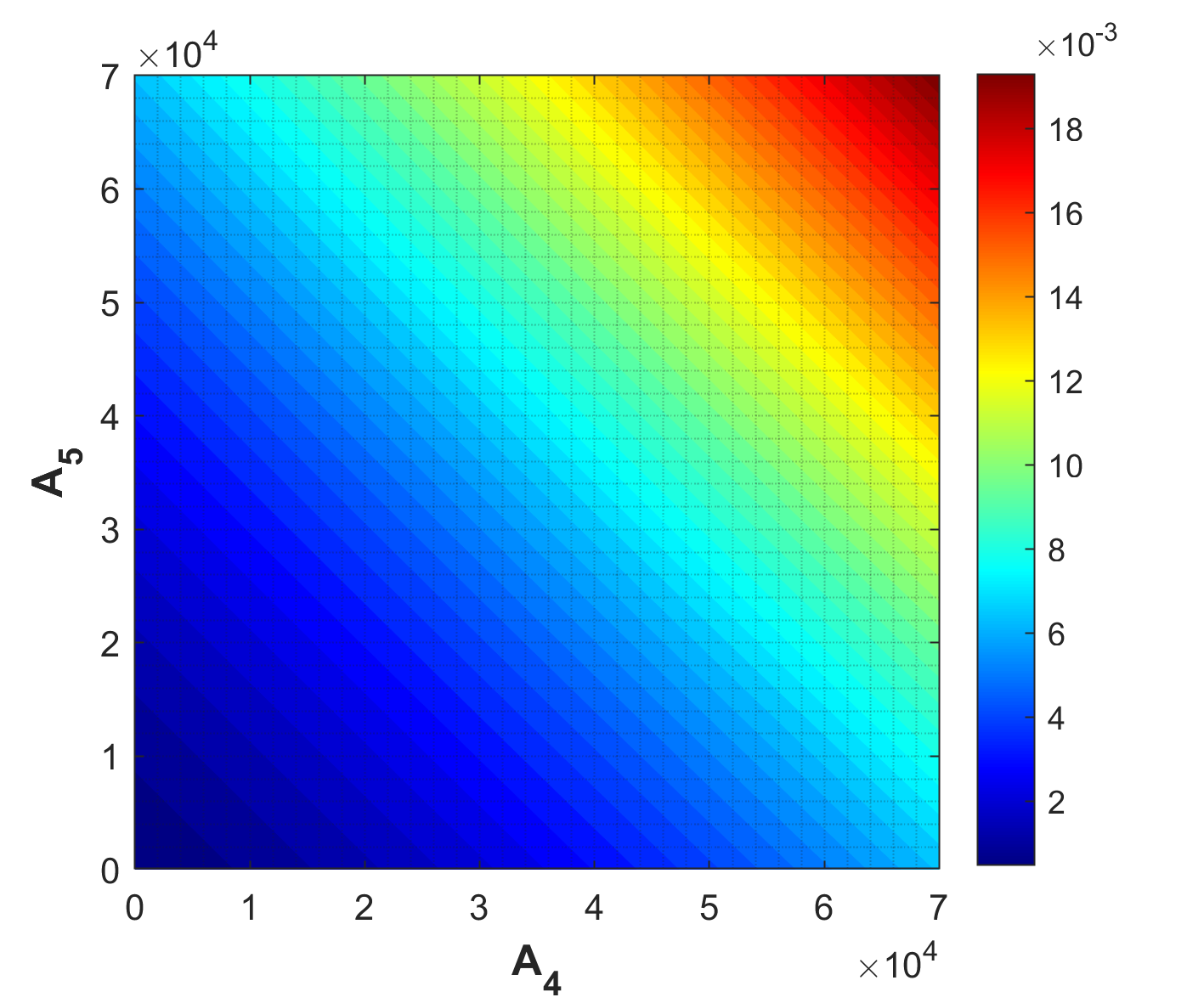}
\end{minipage}
}
\subfigure[Variation of $DAGDOP$ with respect to $A_4$ and $A_5$]
{
\begin{minipage}[b]{.45\linewidth}
\centering
\includegraphics[scale=0.6]{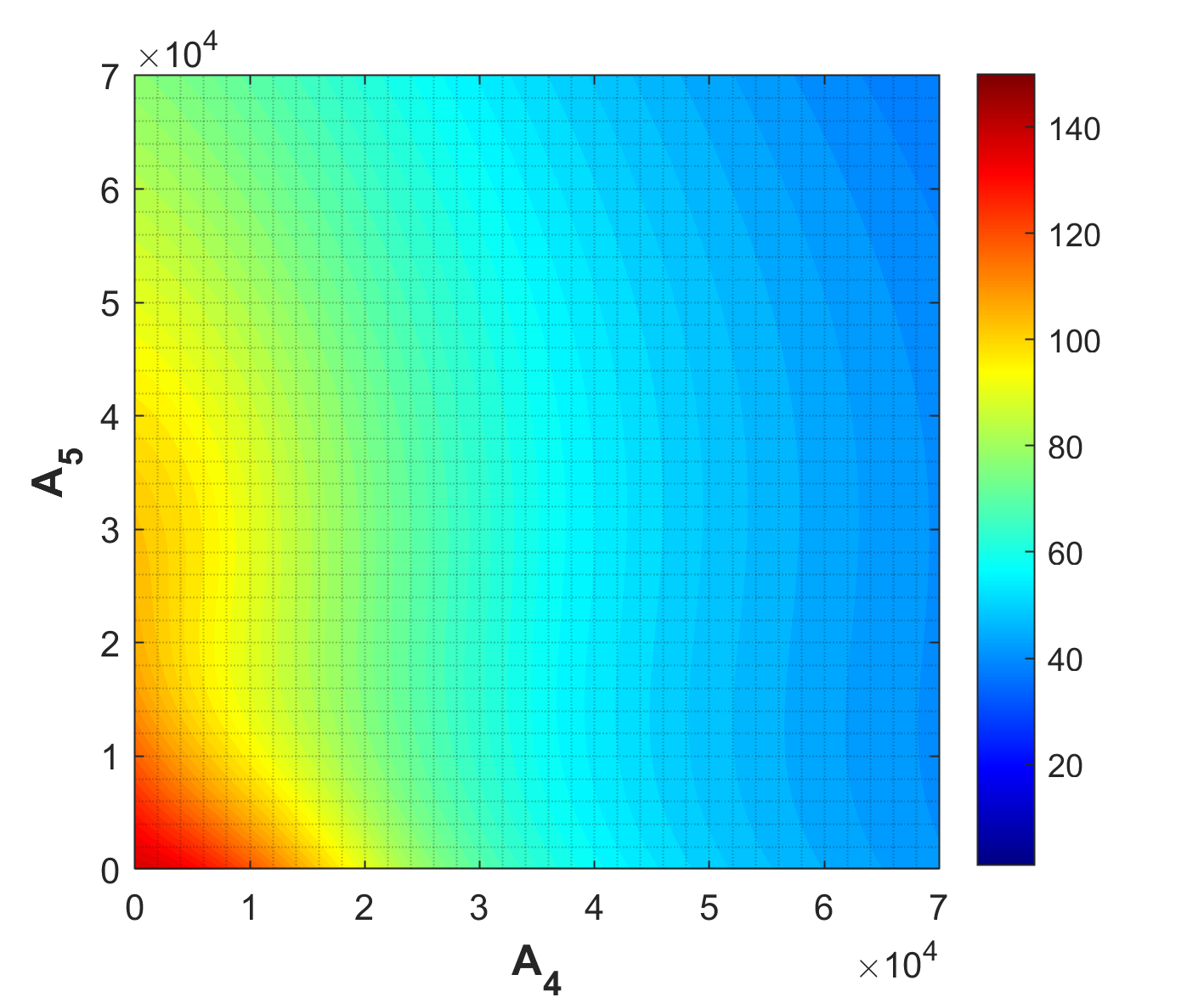}
\end{minipage}
}
\caption{Contour maps of $P_\text{r}$ and $DAGDOP$ with respect to $A_4$ and $A_5$ in simulation 2  \\ $r(P_\text{r},DAGDOP)=-0.7520$}
\label{Result_A4A5}
\end{figure}

 Simulation 2 shows the influence of $A_4$ and $A_5$. The parameters $A_4$ and $A_5$ vary in the range:
\begin{equation}
A_4\in[0,70000]~\text{km},\quad A_5\in[0,70000]~\text{km}.
\nonumber
\end{equation}
Fig.~\ref{Result_A4A5}(a) and \ref{Result_A4A5}(b) show the trends of $P_\text{r}$ and the $DAGDOP$ factor respectively. Same as simulation 1, the negative correlation between $P_\text{r}$ and the $DAGDOP$ factor still exists, which also agrees with the above analysis. According to simulations 1 and 2, with the increase of $A_3$, $A_4$ and $A_5$, $P_\text{r}$ increases and the $DAGDOP$ factor decreases. In other words, increasing out-of-plane amplitude improves the short-arc AOD performance.

\subsubsection{Influence of the out-of-plane initial relative phase angles: $\Delta\phi_4$ and $\Delta\phi_5$}
\par
\begin{figure}[h!]
\centering   
\subfigure[Variation of $P_\text{r}$ with respect to $\Delta\phi_4$ and $\Delta\phi_5$] 
{
\begin{minipage}[b]{.45\linewidth} 
\centering
\includegraphics[scale=0.6]{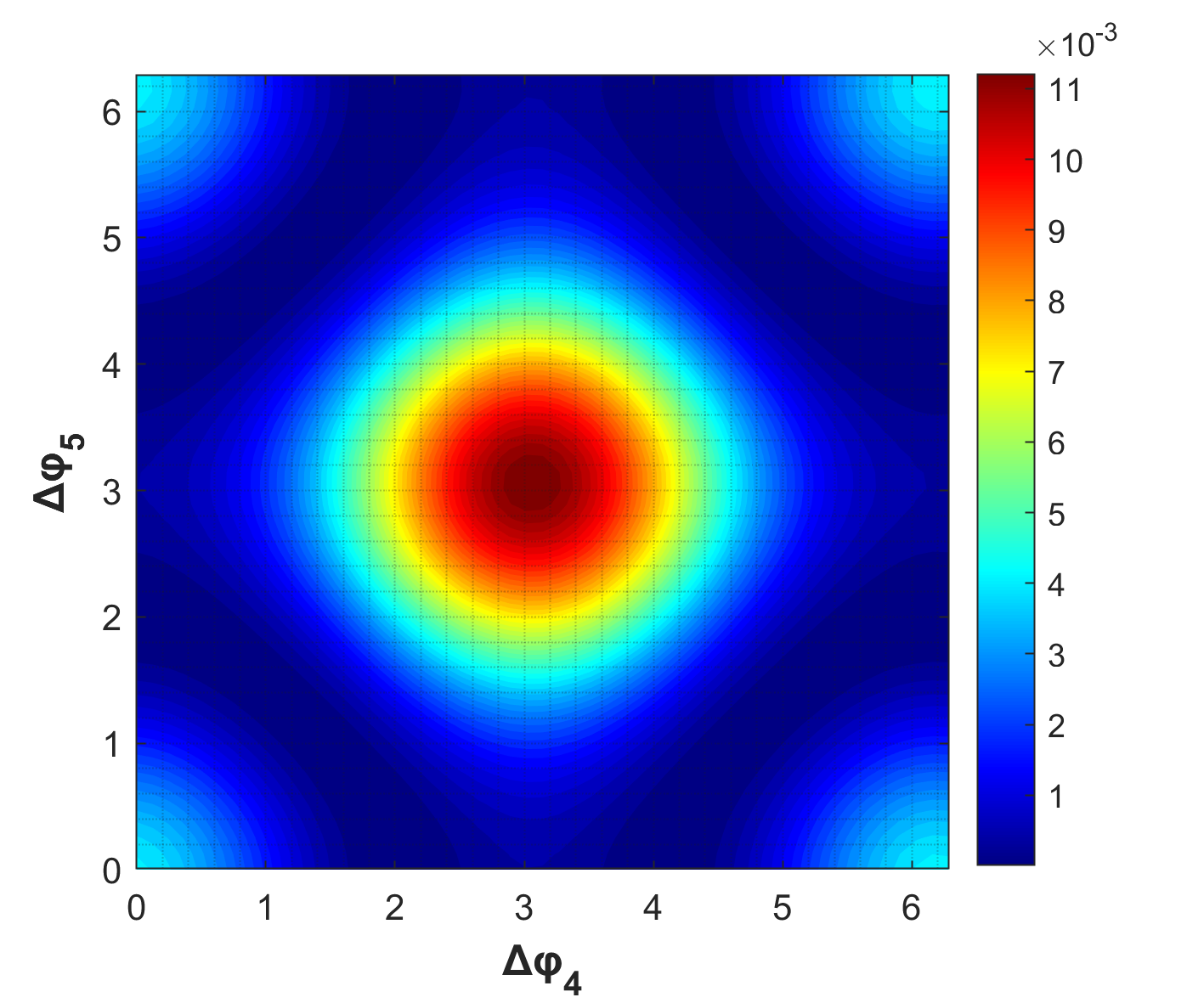}
\end{minipage}
}
\subfigure[Variation of $DAGDOP$ with respect to  $\Delta\phi_4$ and $\Delta\phi_5$]
{
\begin{minipage}[b]{.45\linewidth}
\centering
\includegraphics[scale=0.5]{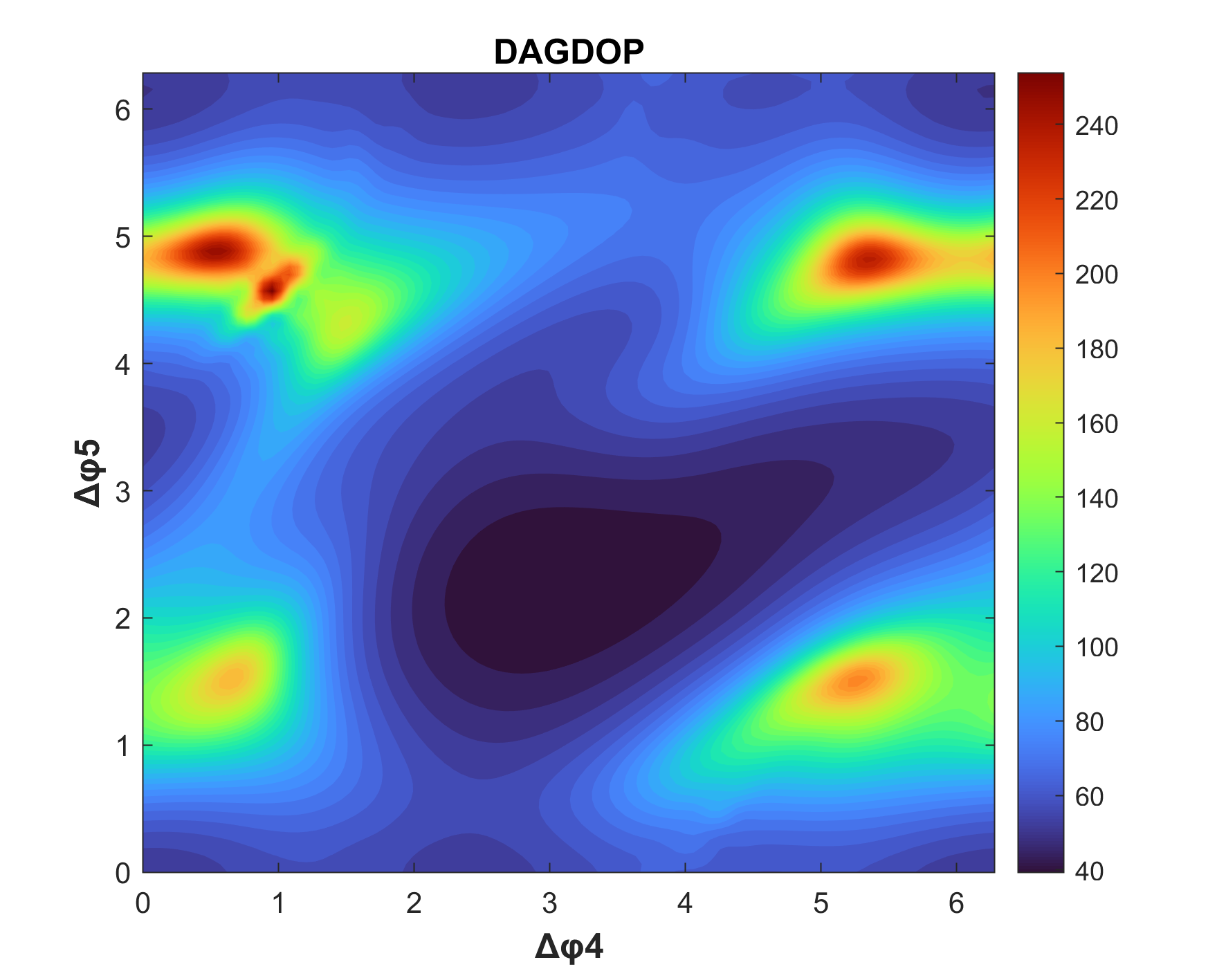}
\end{minipage}
}
\caption{Contour maps of $P_\text{r}$ and $DAGDOP$ in simulation 3 ($\phi_3=0$)  \\ $r(P_\text{r},DAGDOP)=-0.4436$}
\label{Result_P4P5_1}
\end{figure}

\begin{figure}[h!]
\centering   
\subfigure[Variation of $P_\text{r}$ with respect to $\Delta\phi_4$ and $\Delta\phi_5$] 
{
\begin{minipage}[b]{.45\linewidth} 
\centering
\includegraphics[scale=0.6]{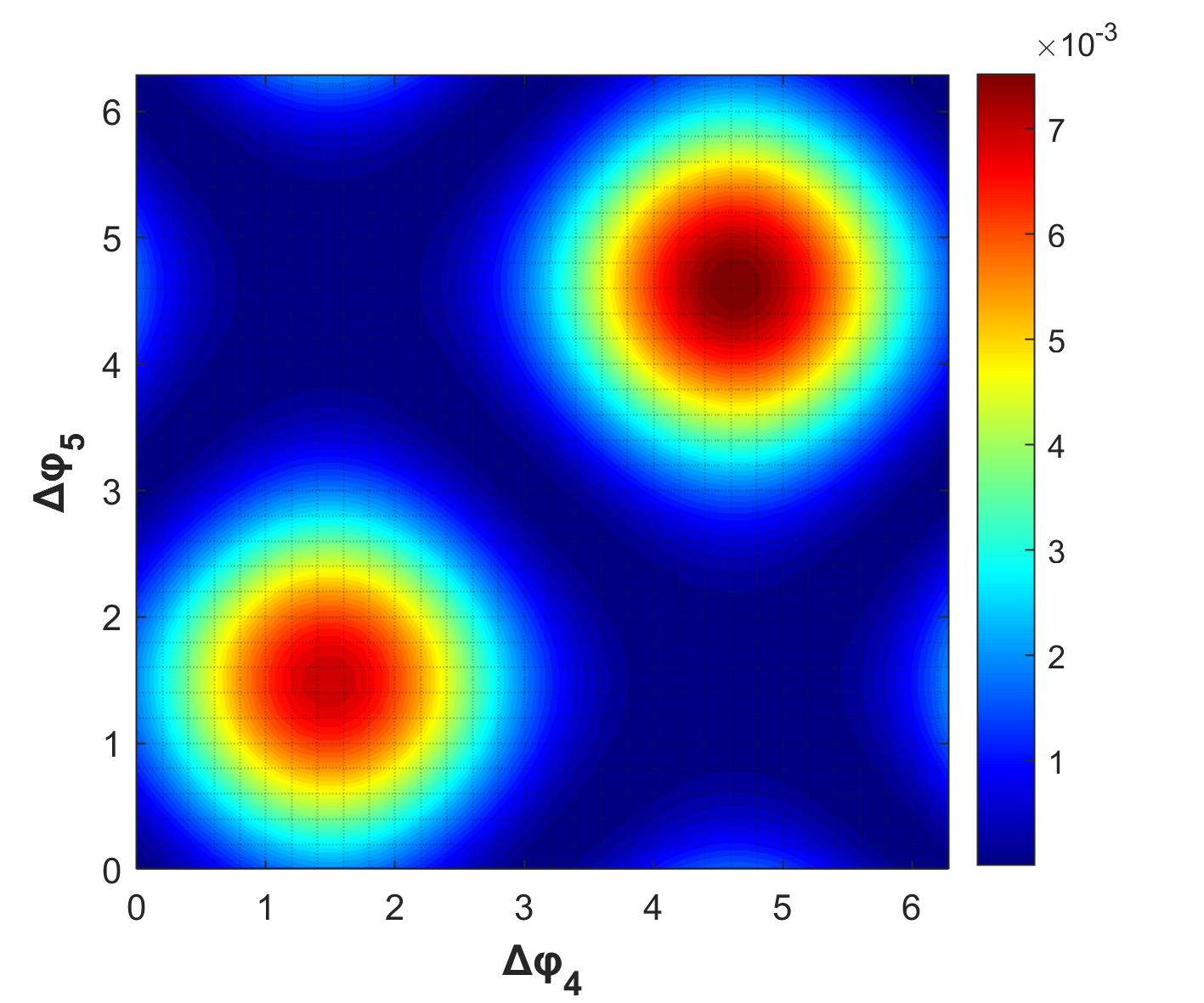}
\end{minipage}
}
\subfigure[Variation of $DAGDOP$ with respect to $\Delta\phi_4$ and $\Delta\phi_5$]
{
\begin{minipage}[b]{.45\linewidth}
\centering
\includegraphics[scale=0.5]{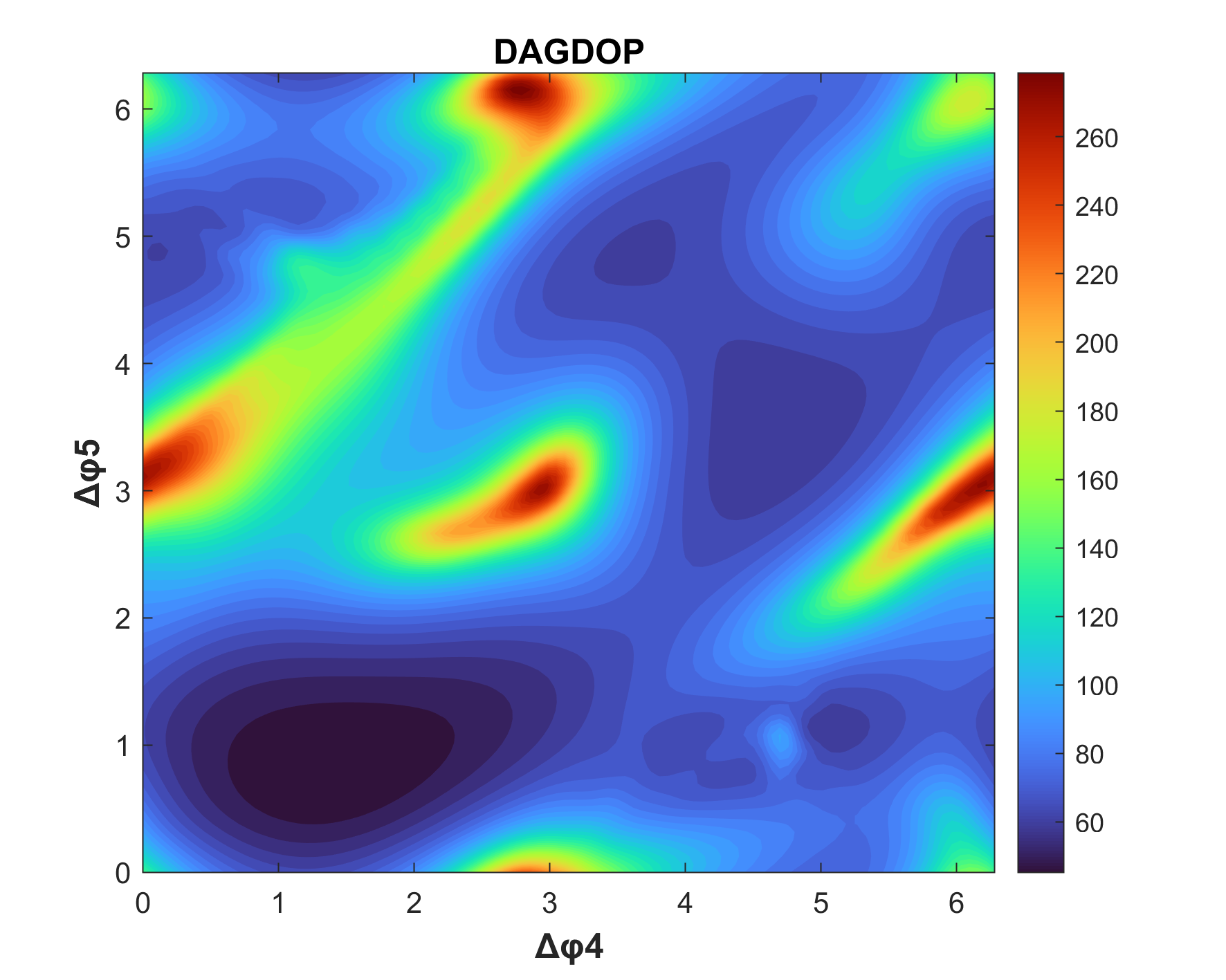}
\end{minipage}
}
\caption{Contour maps of $P_\text{r}$ and $DAGDOP$ in simulation 4 ($\phi_3=\pi/2$)  \\ $r(P_\text{r},DAGDOP)=-0.4442$}
\label{Result_P4P5_2}
\end{figure}

 Next, we study the effects on the $DAGDOP$ factor of the initial relative phase angles $\Delta\phi_4$ and $\Delta\phi_5$. In this subsection, we fix $\phi_3$ and change $\Delta\phi_4$ and $\Delta\phi_5$ to analyze their influence. The difference between simulation 3 and simulation 4 lies in $\phi_3$: in situation 3, $\phi_3=0$ and in situation 4, $\phi_3=\pi/2$. The parameters $\Delta\phi_4$ and $\Delta\phi_5$ vary over the ranges:
\begin{equation}
\Delta\phi_4\in[0,2\pi],\quad \Delta\phi_5\in[0,2\pi].
\nonumber
\end{equation}
Values of other parameters can be found in Table~\ref{Fundamental Set 0}. Results of simulations 3 and 4 are shown in Figs.~\ref{Result_P4P5_1} and \ref{Result_P4P5_2}.
\par As shown in Fig.~\ref{Result_P4P5_1}, a local minimum value of $P_\text{r}$ approximately corresponds to a local maximum of $DAGDOP$. an approximate inverse correlation exists between $P_\text{r}$ and $DAGDOP$. More specifically, (1) The maximum of $P_\text{r}$ occurs at $\Delta\phi_4 = \Delta\phi_5 = \pi$, where $DAGDOP$ approximately attains its minimum. (2) Conversely, the minimum of $P_\text{r}$ occurs for the phase combinations $[\Delta\phi_4,\Delta\phi_5]=[\pi/2+k_1\pi,\pi/2+k_2\pi]$,  $(k_1,k_2=0,1)$. These combinations also fall within the parameter region where $DAGDOP$ reaches its maximum. These phenomena confirm the negative correlation between $P_\text{r}$ and $DAGDOP$.

\par Similar conclusions can still be drawn from Fig.~\ref{Result_P4P5_2}. Although the negative correlation between $P_\text{r}$ and $DAGDOP$ is slightly weaker, the approximate negative correlation still holds. The partial maxima of $DAGDOP$ are located near $[\Delta\phi_4,\Delta\phi_5]=[k_1\pi,k_2\pi]$, $(k_1,k_2=0,1,2)$, while $P_\text{r}$ attains its minimum value. In these configurations, L3, L4, L5 and DRO satellites are nearly coplanar in the $x-y$ plane, resulting in a negligible out-of-plane component. Conversely, the partial minima of $DAGDOP$ also locate approximately at the positions of the partial maxima of $P_\text{r}$, which correspond to $[\Delta\phi_4,\Delta\phi_5]=[\pi/2+k_1\pi,\pi/2+k_2\pi]$, $(k_1,k_2=0,1)$. Along the lines where $\Delta{\phi_5}=\Delta{\phi_4}+\pi$ and $\Delta{\phi_5}=\Delta{\phi_4}-\pi$, $P_\text{r}$ exhibits a local minimum across each entire line, which is also manifested in $DAGDOP$ as a distinct local maximum variation. In addition, on the line with $\Delta{\phi_5}+\Delta{\phi_4}=\pi$, $P_\text{r}$ also exhibits a local minimum, while the $DAGDOP$ factor shows a less pronounced local maximum variation.

\subsubsection{Influence of the out-of-plane initial reference phase angles: $\phi_3$}
\par In section 5.1.1 and 5.1.2, the influence of the five main parameters mentioned at the end of section 4.1 has been analyzed. Comparing Fig.~\ref{Result_P4P5_1}(b) and \ref{Result_P4P5_2}(b), the locations of the local minima of $DAGDOP$ shifts as $\phi_3$ varies, indicating that $\phi_3$ also influences the $DAGDOP$ factor. Since the L3, L4, and L5 satellites have approximately the same orbital period, the variation of $\phi_3$ from $0$ to $2\pi$ corresponds to the evolution of the constellation in one month. Thus, by analyzing the variation of $DAGDOP$ with respect to $\phi_3$, the performance variation of a given constellation over time can be characterized.
\par
\begin{figure}[!h]
\centering   

\subfigure[Variation of $P_\text{r}$ (Test 1)]{%
\includegraphics[scale=0.32]{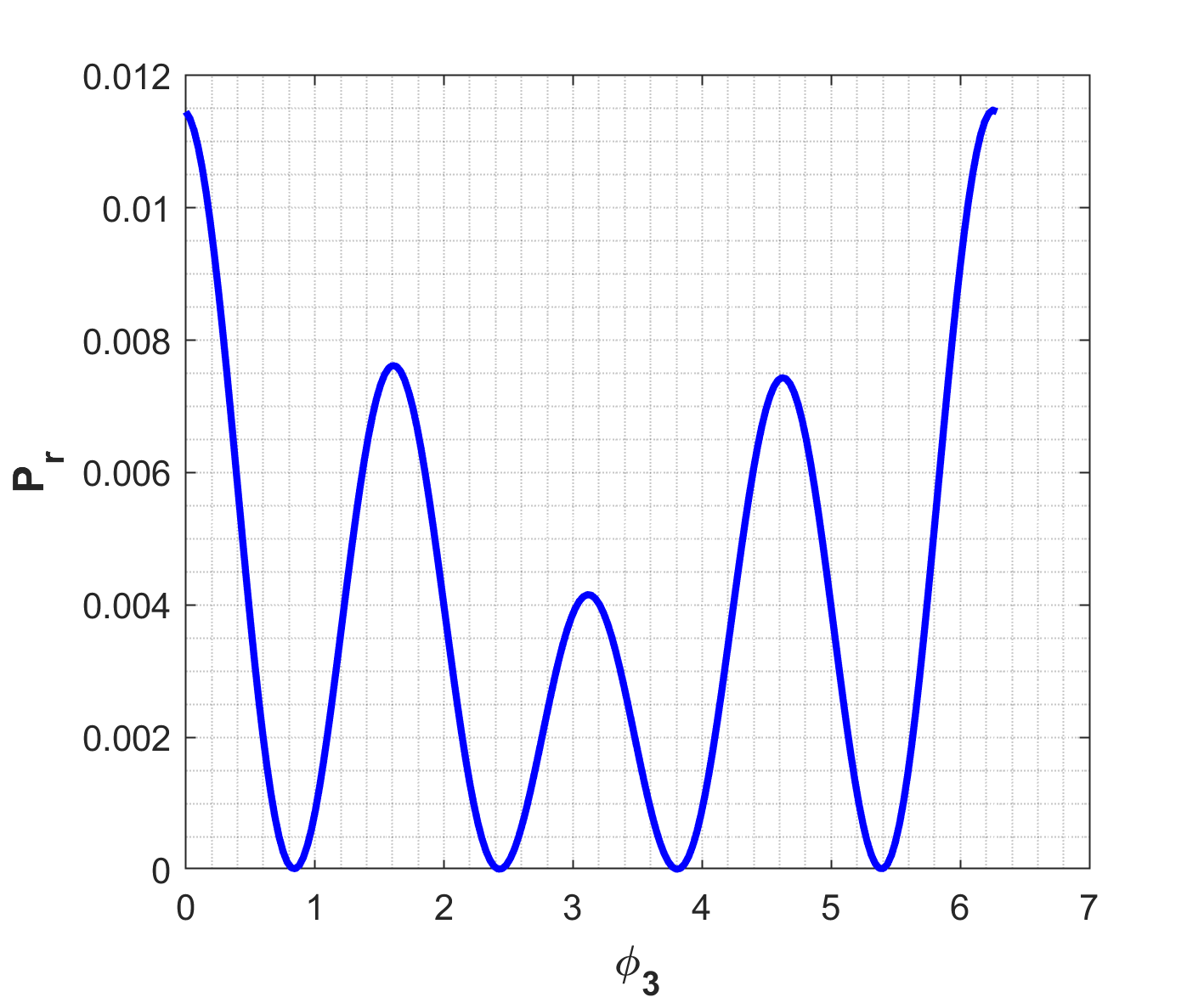}%
}%
\hfill
\subfigure[Variation of $P_\text{r}$ (Test 2)]{%
\includegraphics[scale=0.32]{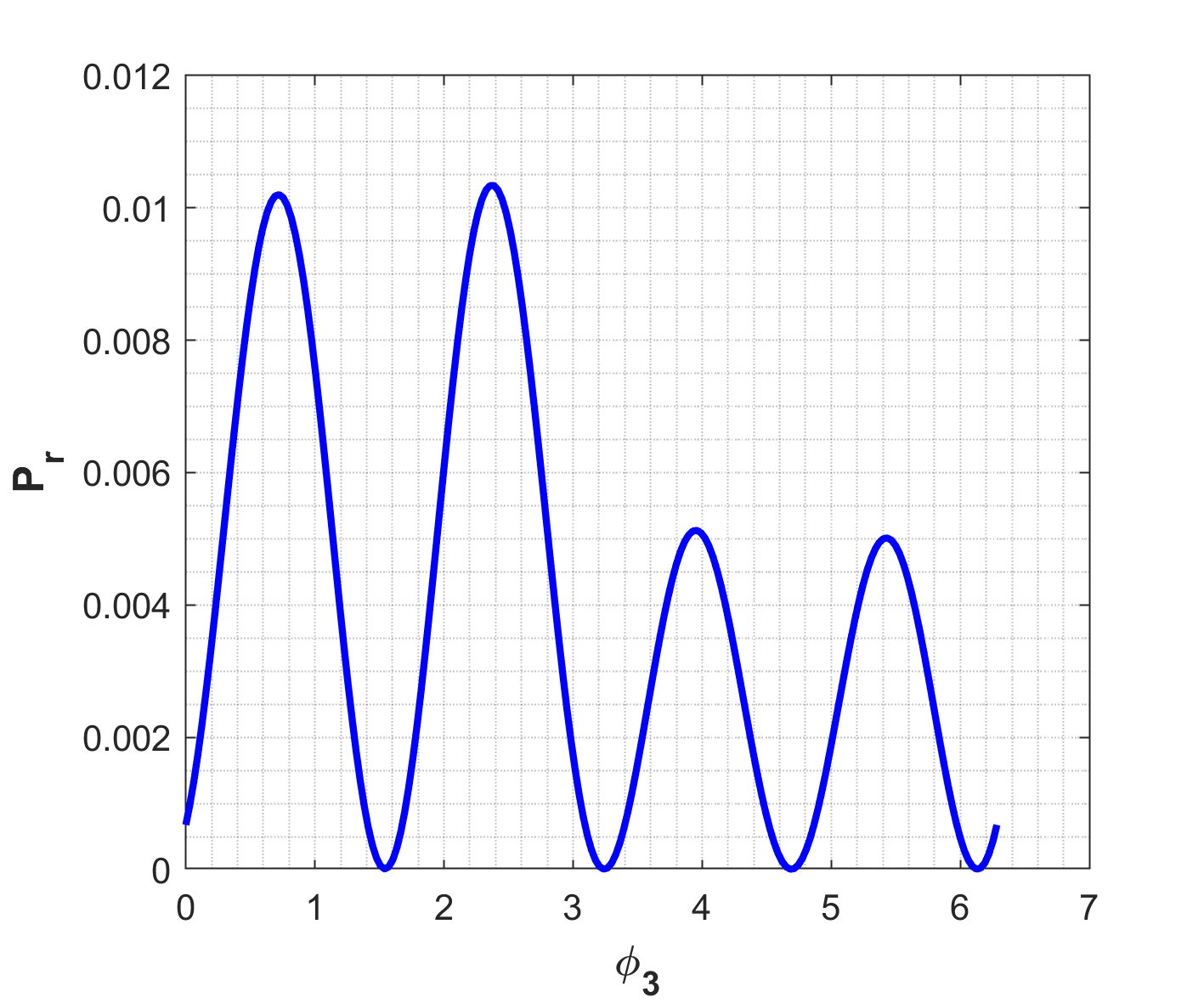}%
}%
\hfill
\subfigure[Variation of $P_\text{r}$ (Test 3)]{%
\includegraphics[scale=0.32]{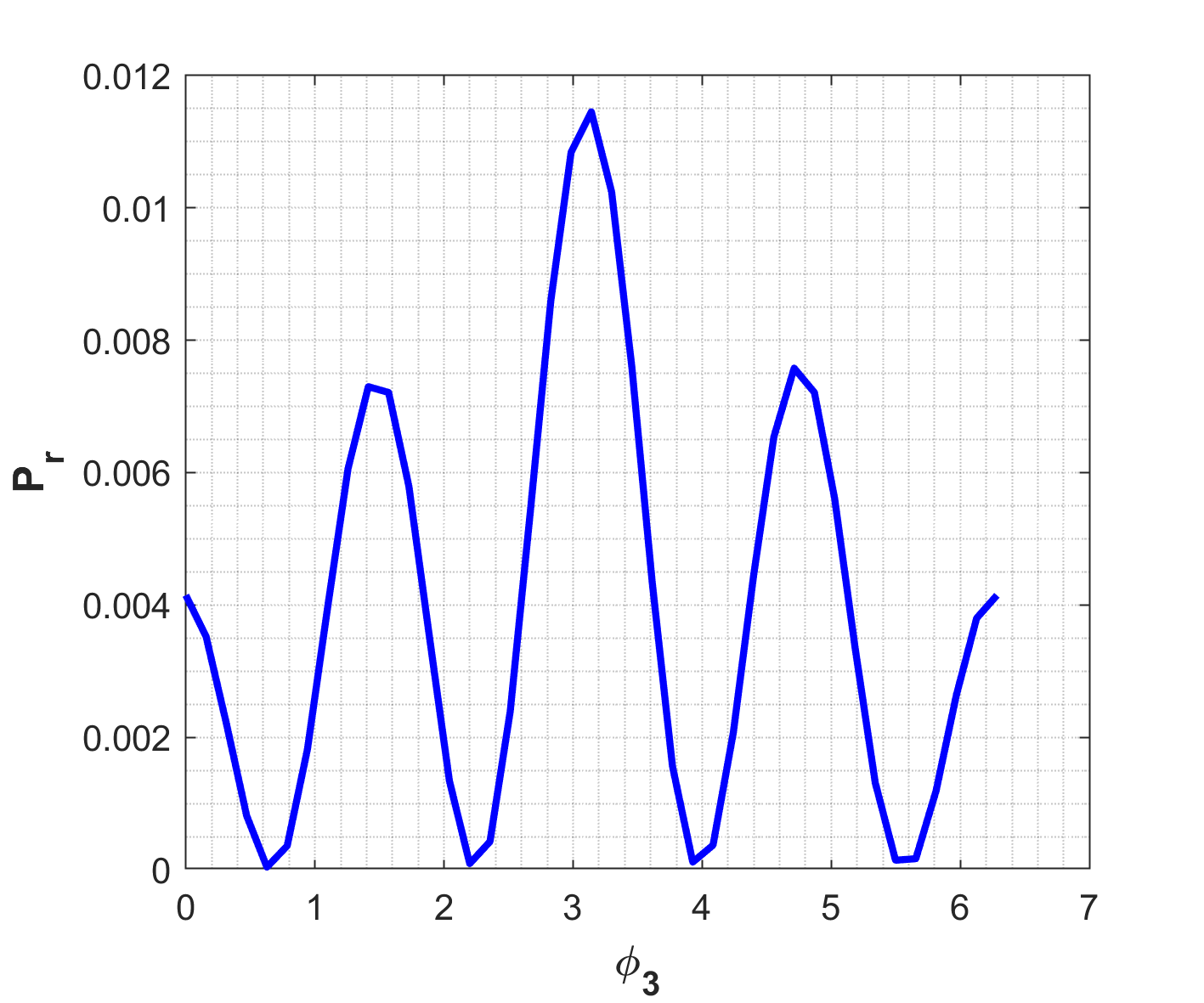}%
}%
\hfill
\subfigure[Variation of $P_\text{r}$ (Test 4)]{%
\includegraphics[scale=0.32]{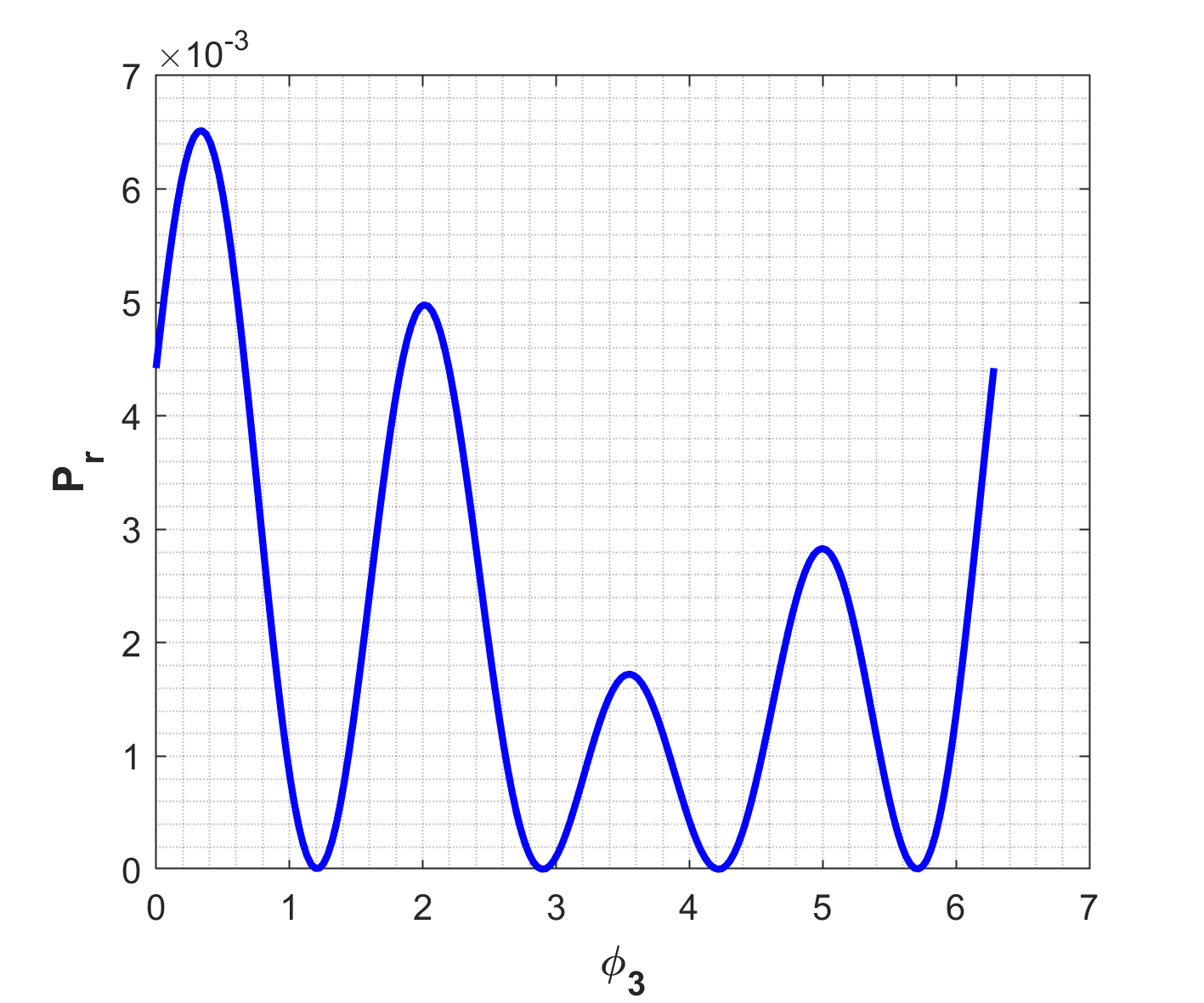}%
}

\vspace{0.3cm}

\subfigure[Variation of $DAGDOP$ (Test 1)]{%
\includegraphics[scale=0.32]{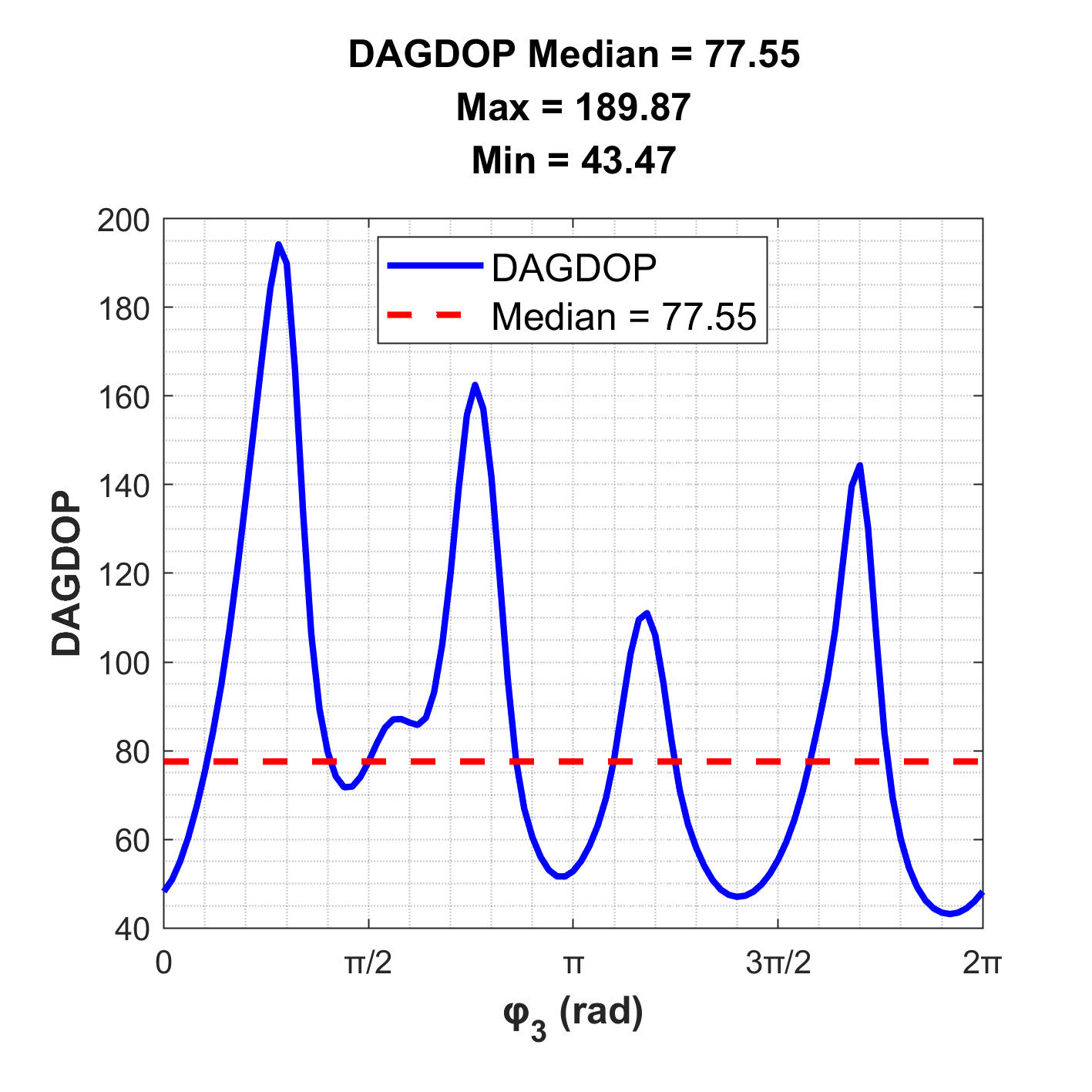}%
}%
\hfill
\subfigure[Variation of $DAGDOP$ (Test 2)]{%
\includegraphics[scale=0.32]{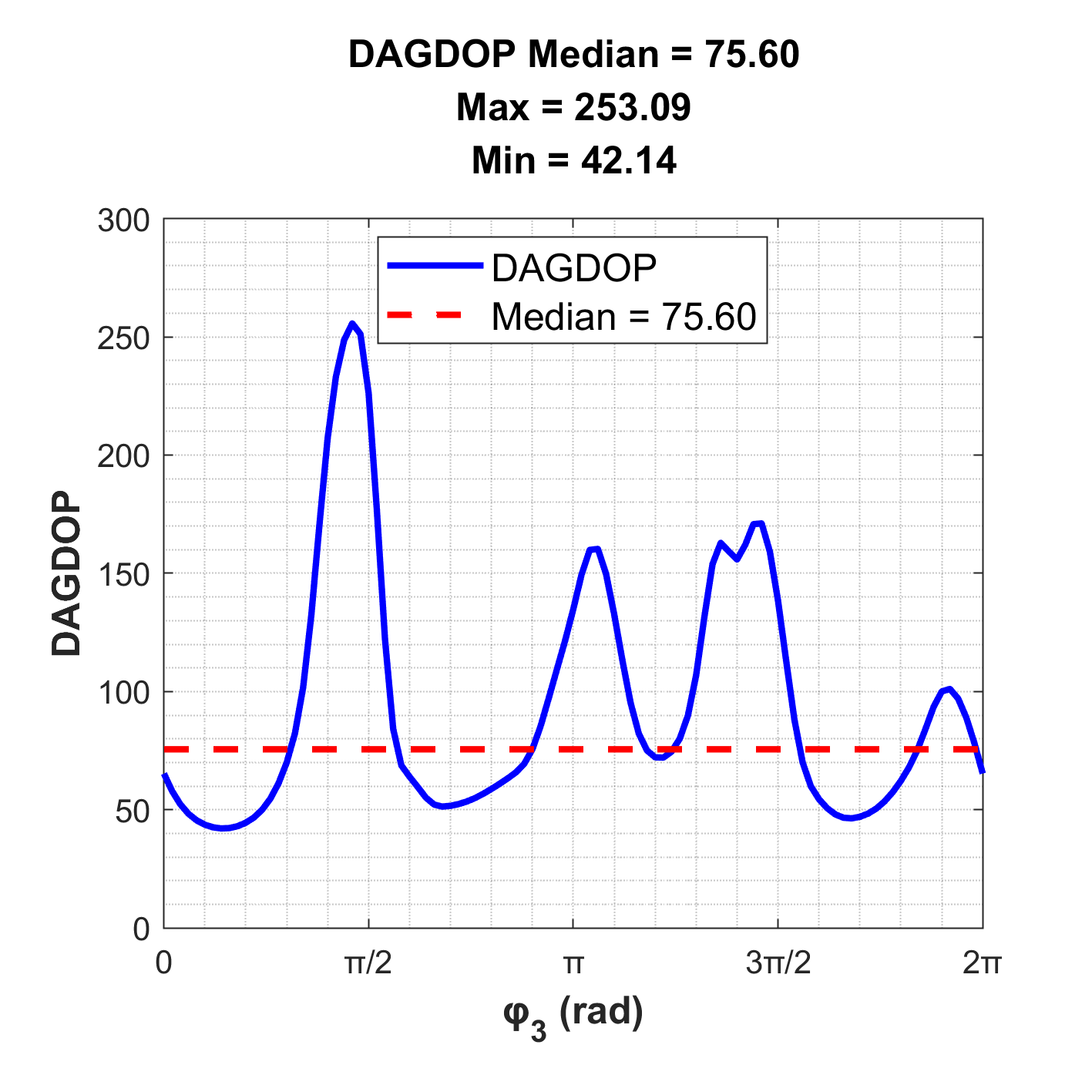}%
}%
\hfill
\subfigure[Variation of $DAGDOP$ (Test 3)]{%
\includegraphics[scale=0.32]{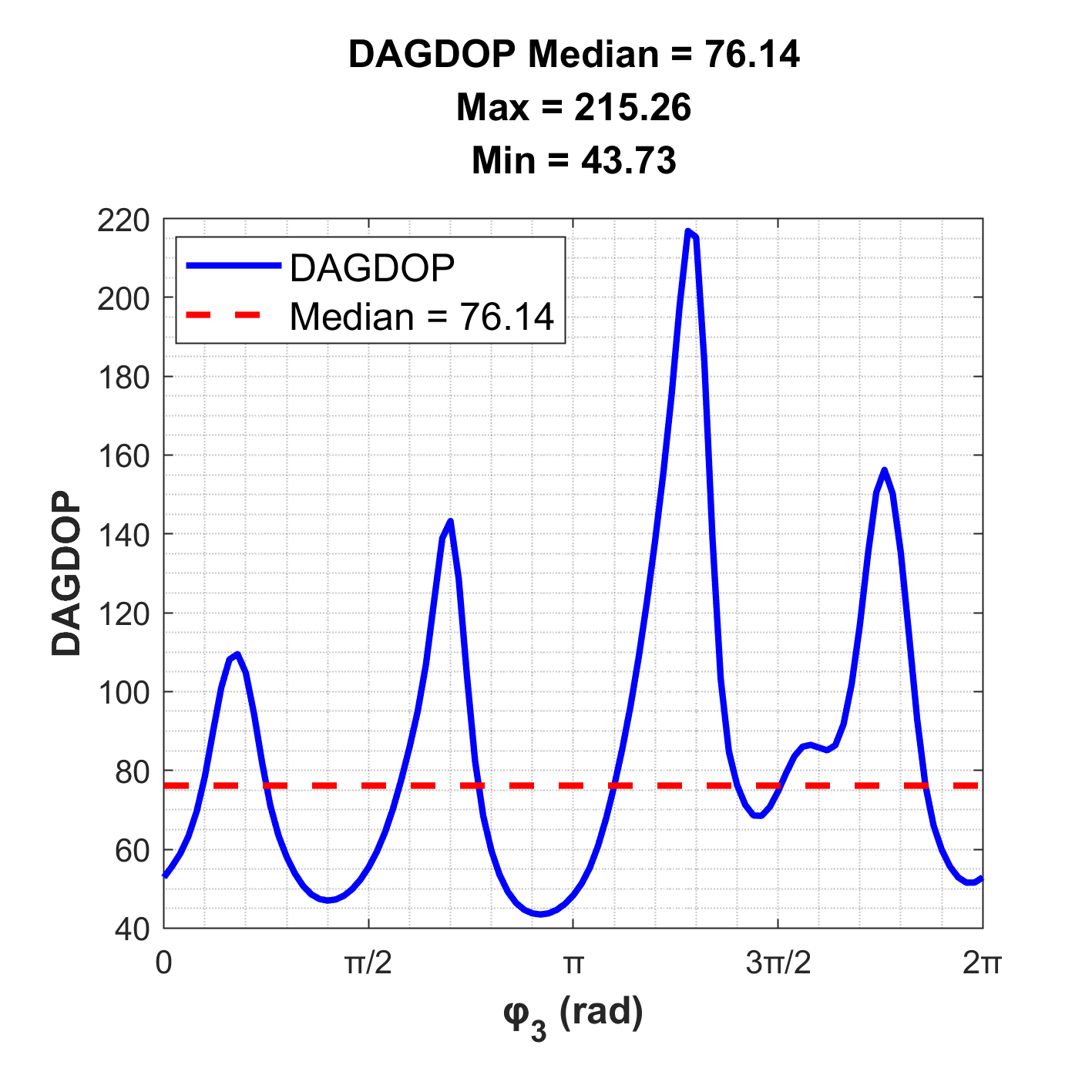}%
}%
\hfill
\subfigure[Variation of $DAGDOP$ (Test 4)]{%
\includegraphics[scale=0.32]{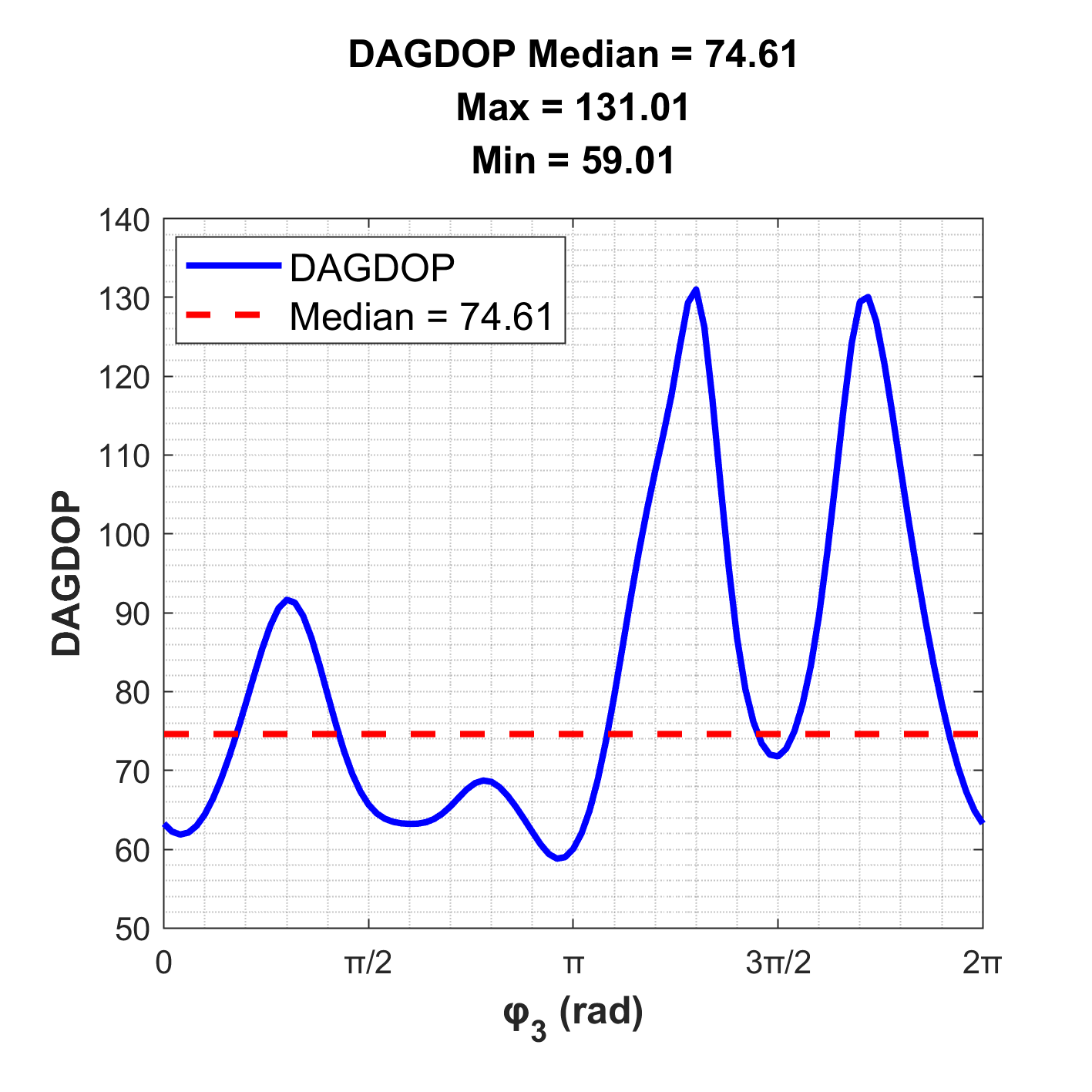}%
}

\caption{Variation of $P_\text{r}$ and $DAGDOP$ with respect to $\phi_3$ in simulation 5}
\label{Result_P3}
\end{figure}

 Table~\ref{Simulation 5} and Fig.~\ref{Result_P3} show the results of three different tests. Here, $\Delta\phi_4$ and $\Delta\phi_5$ are approximately fixed. For different $\Delta\phi_4$ and $\Delta\phi_5$ combinations, $DAGDOP$ oscillates with the variation of $\phi_3$, albeit in a non-uniform manner. To characterize this variation, we select the median, maximum, and minimum values of $DAGDOP$ to represent its typical, worst-case, and best-case outcomes, respectively. The approximate inverse correlation between $DAGDOP$ and $P_\text{r}$ still exists. The locations of the maxima of $DAGDOP$ correspond the the minima of $P_\text{r}$. 
\begin{table}[!h]
    \caption{\textbf{The Set and Results of simulation 5}}
    \renewcommand\arraystretch{1.2}
    \centering
    \begin{tabular}{|c|c|c|c|c|c|c|} \hline
    \multirow{2}{*}{\textbf{Test ID}} & \multicolumn{2}{c}{\textbf{Sets}} & \multicolumn{3}{|c|}{$DAGDOP$ Results/m} & \multirow{2}{*}{\textbf{$r(P_\text{r},DAGDOP)$}} \\ \cline{2-6}
    & $\Delta\phi_4$ & $\Delta\phi_5$ & Median & Maximum & Minimum & \\ \hline
    1 & $\pi$ & $\pi$ & \textbf{77.55} & \textbf{189.87} & \textbf{43.47} & \textbf{-0.6771} \\ \hline
    2 & $\pi/2$ & $\pi/2$ & \textbf{70.03} & \textbf{226.18} & \textbf{43.71} & \textbf{-0.5673} \\ \hline
    3 & $0$ & $0$ & \textbf{76.14} & \textbf{215.26} & \textbf{43.73} & \textbf{-0.5091} \\ \hline
    4 & $\pi/2$ & $\pi$ & \textbf{78.48} & \textbf{131.01} & \textbf{60.05} & \textbf{-0.4433} \\ \hline
    \end{tabular}
    \label{Simulation 5}
\end{table}

\par A comparison of Tests 1, 2, 3, and 4 shows that the variation range of $DAGDOP$ in Tests 1, 2, and 3 is significant: their optimal values are similar and much better to those of Test 4; whereas their worst values are much larger than those of Test 4. This is because, under these configurations, four satellites are nearly coplanar at all times. In contrast, in Test 4, since $\Delta\phi_4 \neq \Delta\phi_5$, at least one satellite always deviates from the lunar orbital plane, resulting in a smoother variation of $DAGDOP$, causing its optimal value and median to be smaller than those in Tests 1, 2, and 3. Correspondingly, the maximum $P_\text{r}$ values in Tests 1, 2, and 3 are much larger than the maximum $P_\text{r}$ value in Test 4, while their minimum $DAGDOP$ values are much smaller than the corresponding minimum in Test 4. However, the median values of all these tests don't have significant differences. This indicates that different combinations of $\Delta{\phi}_4$ and $\Delta{\phi}_5$ affect significantly only at the local maxima and minima. 
\par By comparing the variation curves of $DAGDOP$ and $P_\text{r}$ across the tests, a strong negative correlation between the two can be observed: where $DAGDOP$ reaches a local minimum, $P_\text{r}$ attains a local maximum, and vice versa. The values of correlation coefficients between them also prove this result. However, this does not imply that the global maximum of $P_\text{r}$ strictly corresponds to the global minimum of $DAGDOP$. Consistent with the conclusion in section 4, this relationship merely reflects the trend of variation.
\par Simulation 5 further reveals the complex influence of the variation of $\phi_3$ on the AOD performance of the constellation under the phase combinations of $\Delta\phi_4$ and $\Delta\phi_5$. Based on all simulation results, the following conclusions can be drawn:
\begin{itemize} 
\item[(1)] When $\phi_3 = 0$ or $\phi_3 = \pi$, the optimal short-arc AOD accuracy is achieved in the neighborhood of $\Delta\phi_4 = \Delta\phi_5 = \pi$. However, with the variation of $\phi_3$, the worst-case accuracy also occurs under the same geometric condition. This indicates that the AOD accuracy varies significantly with $\phi_3$ when $\Delta\phi_4 = \Delta\phi_5$.
\item[(2)] When $\Delta\phi_4 \neq \Delta\phi_5$, the variation of $DAGDOP$ with respect to $\phi_3$ is much smoother. This configuration ensures that the AOD accuracy remains at an acceptable level throughout all phases of the constellation's operation.
\end{itemize}

\subsection{Summary of short-arc AOD}
\par According to the simulations in subsection 5.1.1 $-$ 5.1.3, the following conclusion can be drawn: The negative correlation between $P_\text{r}$ and $DAGDOP$ holds. Hence, the local minima of $DAGDOP$ should correspond to the local maxima of $P_\text{r}$, and vice versa. Therefore, by searching for the maxima of $P_\text{r}$, the interval in which the local minimum of $DAGDOP$ lies can be quickly determined. This method can be used to provide useful guidance when designing the constellation. However, it should be noted that this does not mean the global optimal values of the two are in one-to-one correspondence. The reason is that the negative correlation between $P_\text{r}$ and $DAGDOP$ is based on Eq.~\eqref{Minimum_inv_matrix}. This approximate inverse relationship cannot accurately reflect the linear variation.

\subsection{Long-arc AOD Scenario}
\par As already mentioned in Section 4.5 and 4.6, when the AOD arc is long, $P_\text{r}$ no longer reflects the variation of $DAGDOP$. This is because the simplification of the STM in Eq.~\eqref{STM_simple} becomes invalid. Under this condition, the magnitude of $\partial x_m/\partial z^0_m$, $\partial y_m/\partial z^0_m$ and $\partial z_m/\partial z^0_m$ can be regarded as the same order of magnitude. 
\par
\begin{figure}[h]
\centering   
\subfigure[$DAGDOP$ in a 7-day arc with $\phi_3=0$] 
{
\begin{minipage}[b]{.45\linewidth} 
\centering
\includegraphics[scale=0.6]{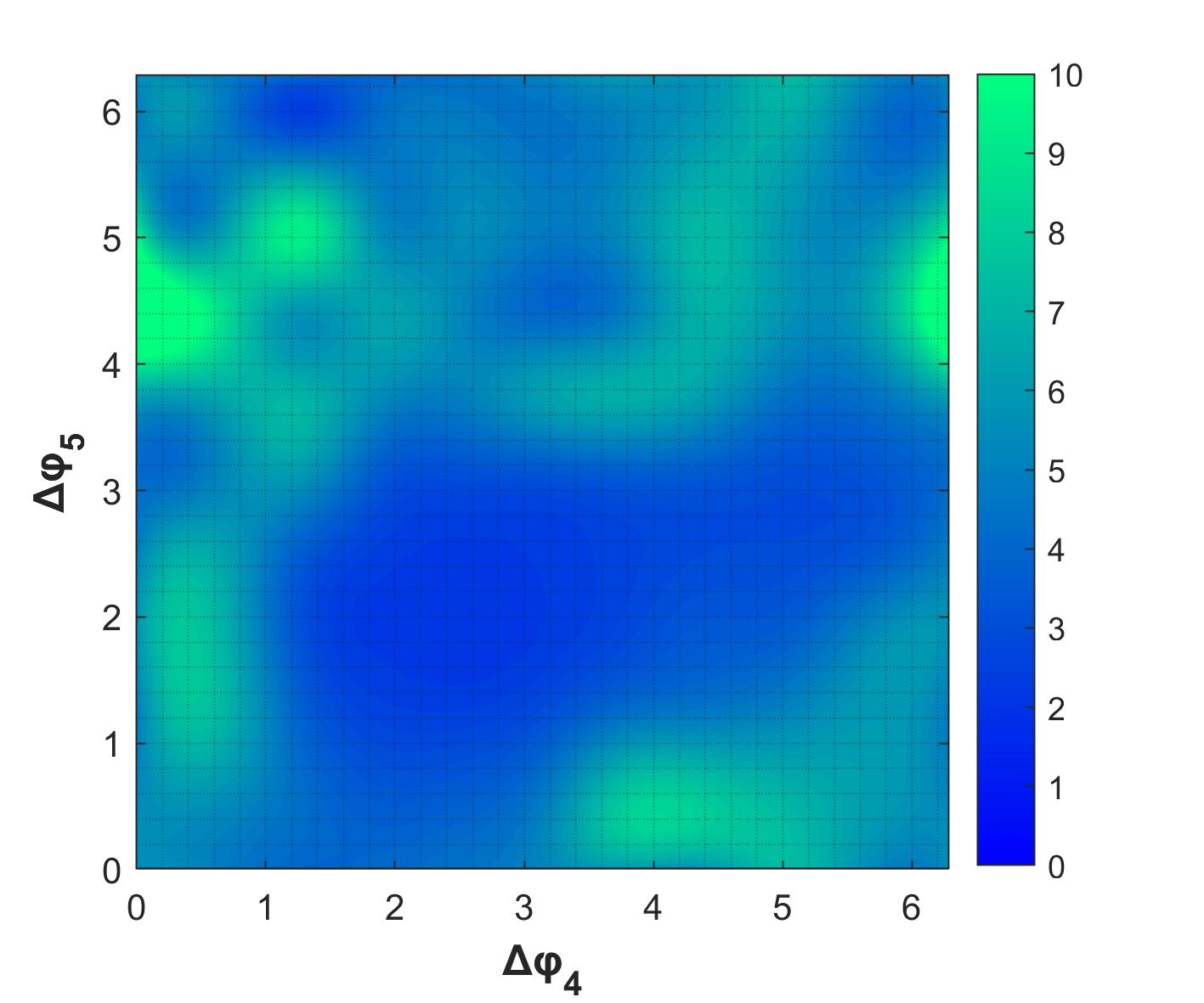}
\end{minipage}
}
\subfigure[$DAGDOP$ in a 7-day arc with $\phi_3=\pi/2$]
{
\begin{minipage}[b]{.45\linewidth}
\centering
\includegraphics[scale=0.6]{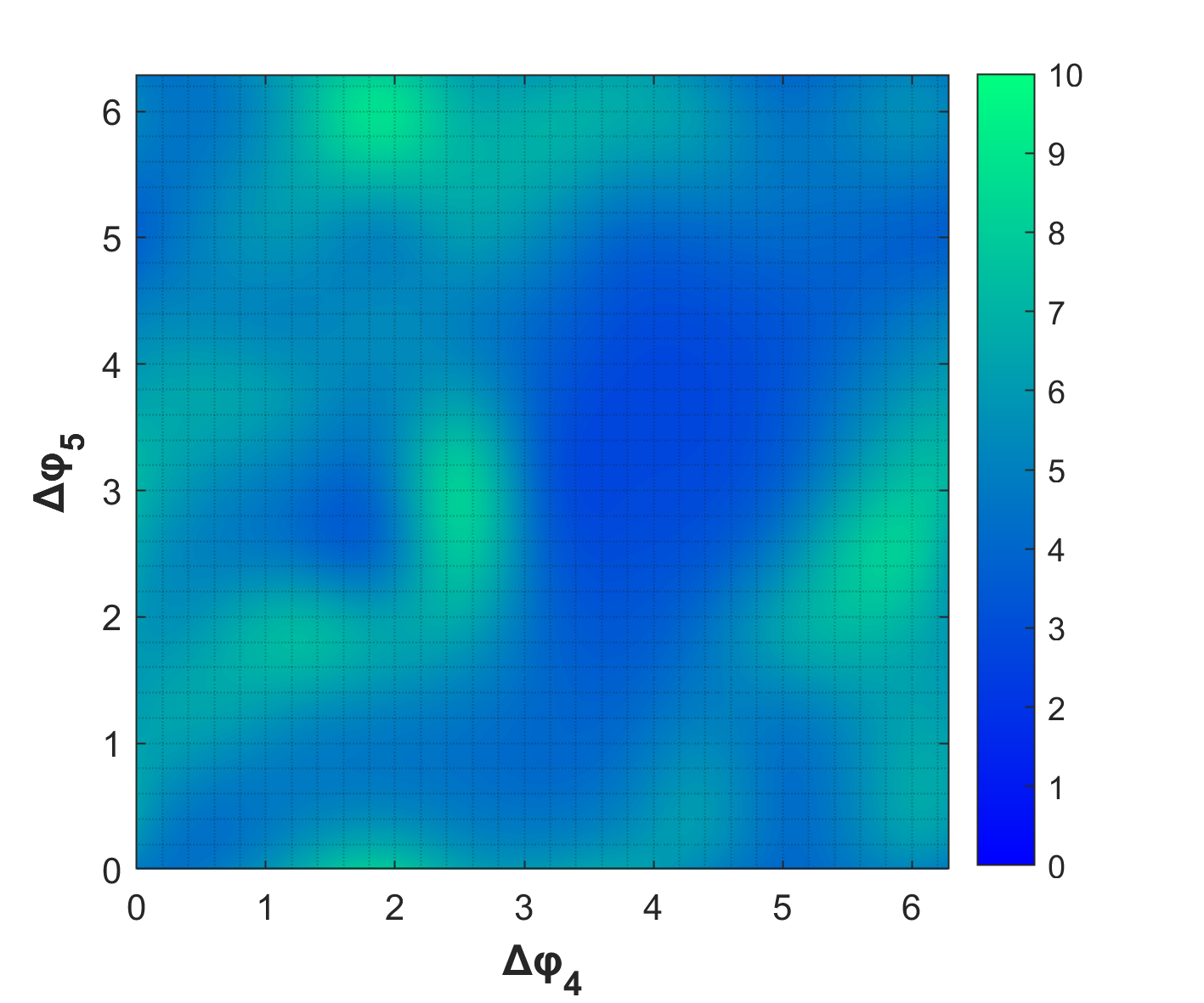}
\end{minipage}
}
\caption{Variation of $DAGDOP$ with respect to $\Delta\phi_4$ and $\Delta\phi_5$ in a 7-day arc}
\label{Result_long_Arc}
\end{figure}

 For long arcs, differences in the AOD accuracy arising from different design parameters under the same OD arc length become negligible \autocite{Liu2014}. Fig.~\ref{Result_long_Arc} demonstrates this by showing the variation of $DAGDOP$ over a 7-day arc for two representative cases. Apart from the arc length, all other parameters in Fig.~\ref{Result_long_Arc}(a) and (b) are identical to those used in simulations 3 and 4 respectively. The figure confirms that in the long-arc scenario, the $DAGDOP$ factor is largely insensitive to variations in the out-of-plane design parameters.

\section{Discussion}
\subsection{$DAGDOP$ and AOD accuracy}

\begin{figure}[h]
    \centering
    \includegraphics[width=6.5cm]{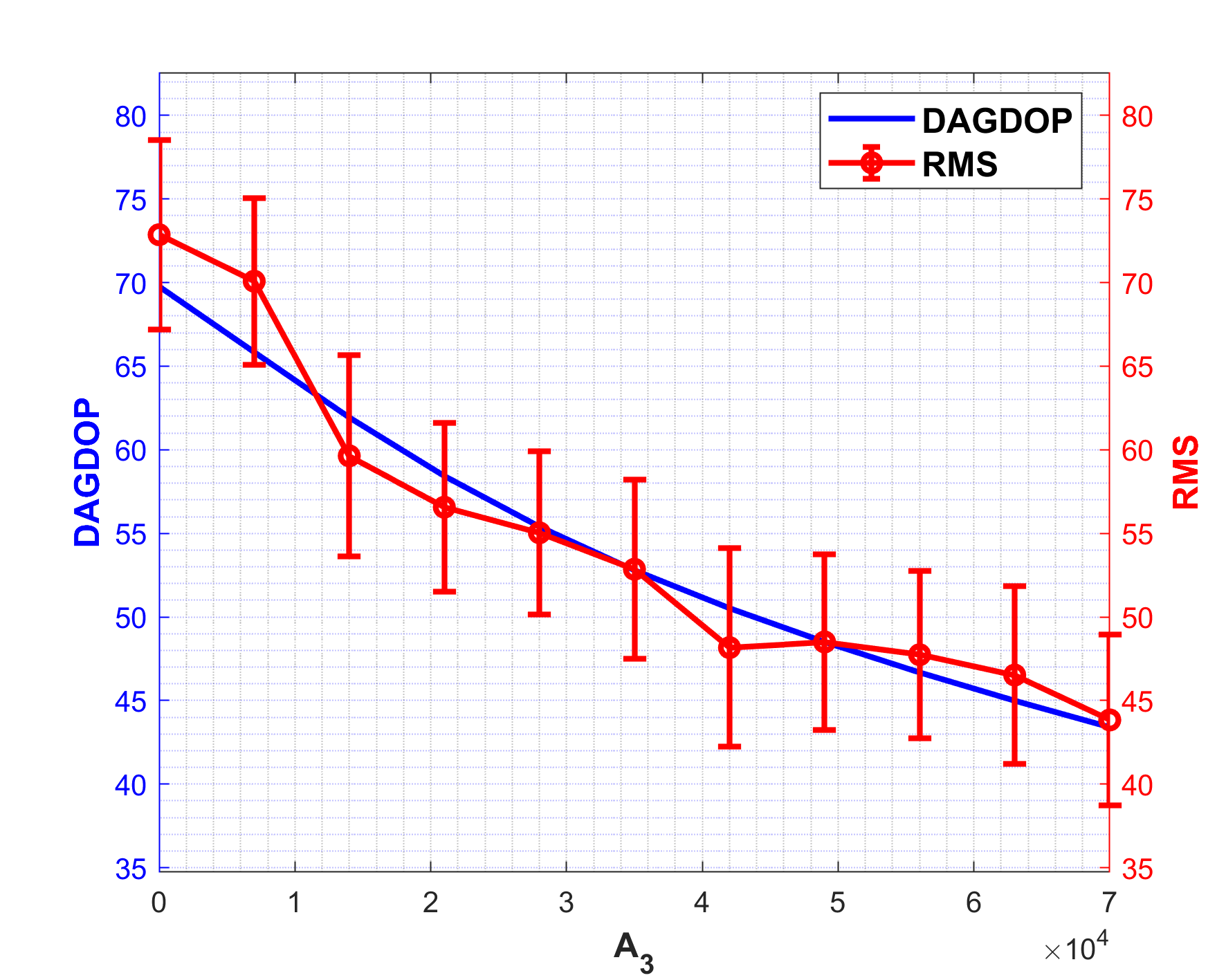}
    \caption{Variation of the $DAGDOP$ and the OD $RMS$ with respect to $A_3$}
    \label{DAGDOP_RMS}
\end{figure}

\par In the above analysis, we employ the $DAGDOP$ factor to reflect the AOD accuracy. In previous studies \autocite{Gao2022}, the agreement between the OD accuracy and this factor has already been demonstrated. Here we also demonstrate its validity in our study. The AOD accuracy of the navigation constellation can also be described by the $RMS$. For the $m^{th}$ satellite, its position $RMS^m$ is calculated by:
\begin{equation}
RMS^m = \sqrt{\sum_{k=1}^{N}\frac{\left(\Delta{x_k^m}\right)^2+\left(\Delta{y_k^m}\right)^2+\left(\Delta{z_k^m}\right)^2}{N}},
\nonumber
\end{equation}
where $N$ is the number of observations; $\Delta{x_k^m}$, $\Delta{y_k^m}$, and $\Delta{z_k^m}$ are the position error of the $k^{th}$ observation for the $m^{th}$ satellite. For a navigation  constellation with $M$ satellites, the overall $RMS$ can be expressed as:
\begin{equation}
RMS = \sqrt{\sum_{m=1}^{M}\left(RMS^m\right)^2}.
\nonumber
\end{equation}

\par Fig.~\ref{DAGDOP_RMS} shows the relationship between the $DAGDOP$ factor and the OD $RMS$ in simulation 1. The $RMS$ results at each sampled $A_3$ value are derived from 100 Monte Carlo AOD simulations. There is good agreement between the $DAGDOP$ factor and the AOD $RMS$, which justifies our above analysis using the $DAGDOP$ factor.

\subsection{Influence of the In-plane Parameters}

\par The above analysis identifies the out-of-plane parameters as dominate factors influencing the AOD accuracy of the constellation. In this section, we briefly show the influence of the remaining 8 parameters (the in-plane parameters) on the short-arc AOD accuracy, to demonstrate that they do have little influence on the AOD accuracy. In this subsection, $T_{arc}$ is still set as 3 days, and $\Delta{t}$ is set as 5 minutes.

\begin{itemize} 
\item \textbf{The in-plane parameters of L4 and L5 satellites}
\end{itemize} 
\par
\begin{figure}[h]
\centering 
\subfigure[Variation of $DAGDOP$ with respect to $A_4^p$ and $\phi_4^p$] 
{
\begin{minipage}[b]{.45\linewidth} 
\centering
\includegraphics[scale=0.6]{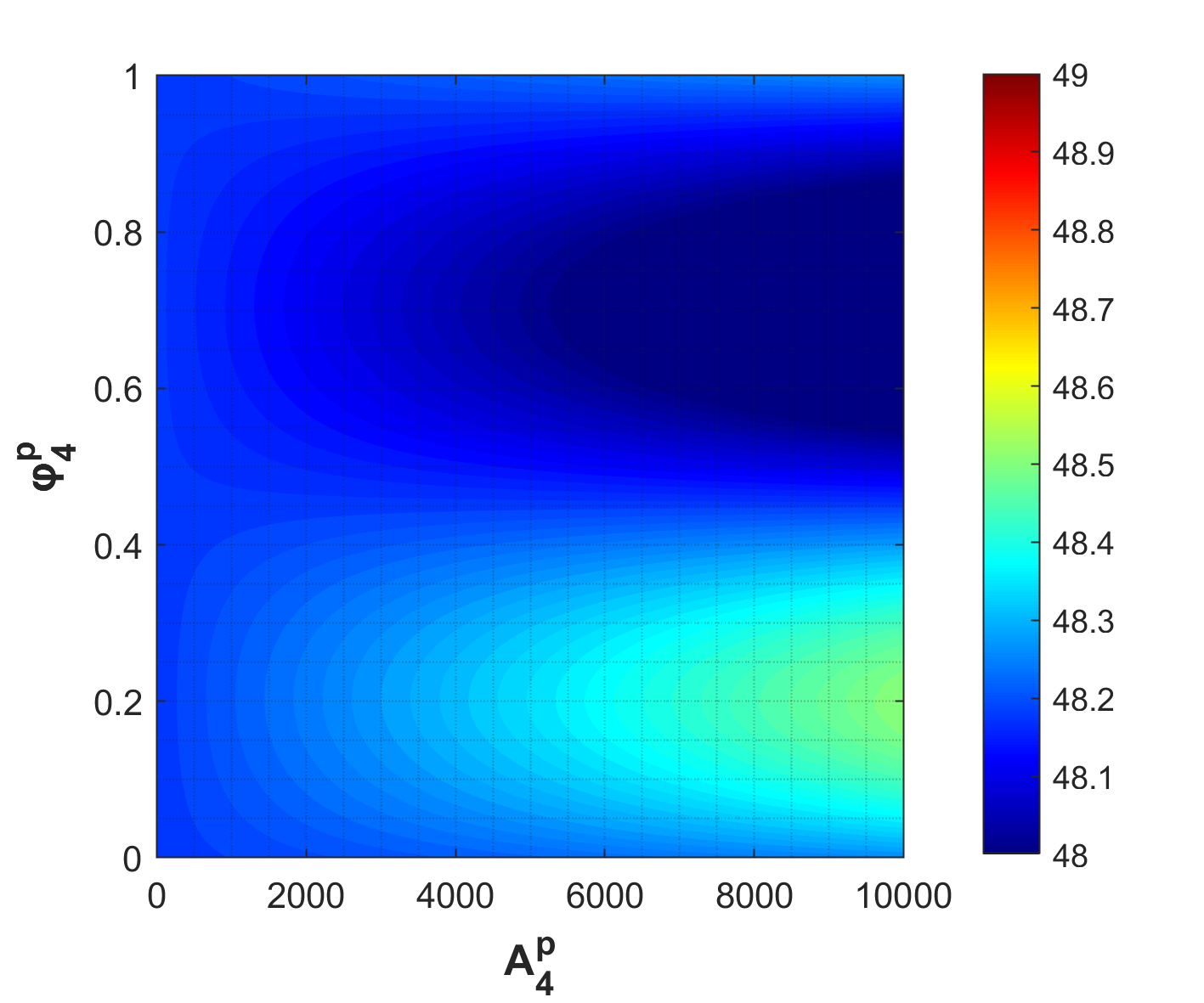}
\end{minipage}
}
\subfigure[Variation of $DAGDOP$ with respect to $A_5^p$ and $\phi_5^p$]
{
\begin{minipage}[b]{.45\linewidth}
\centering
\includegraphics[scale=0.6]{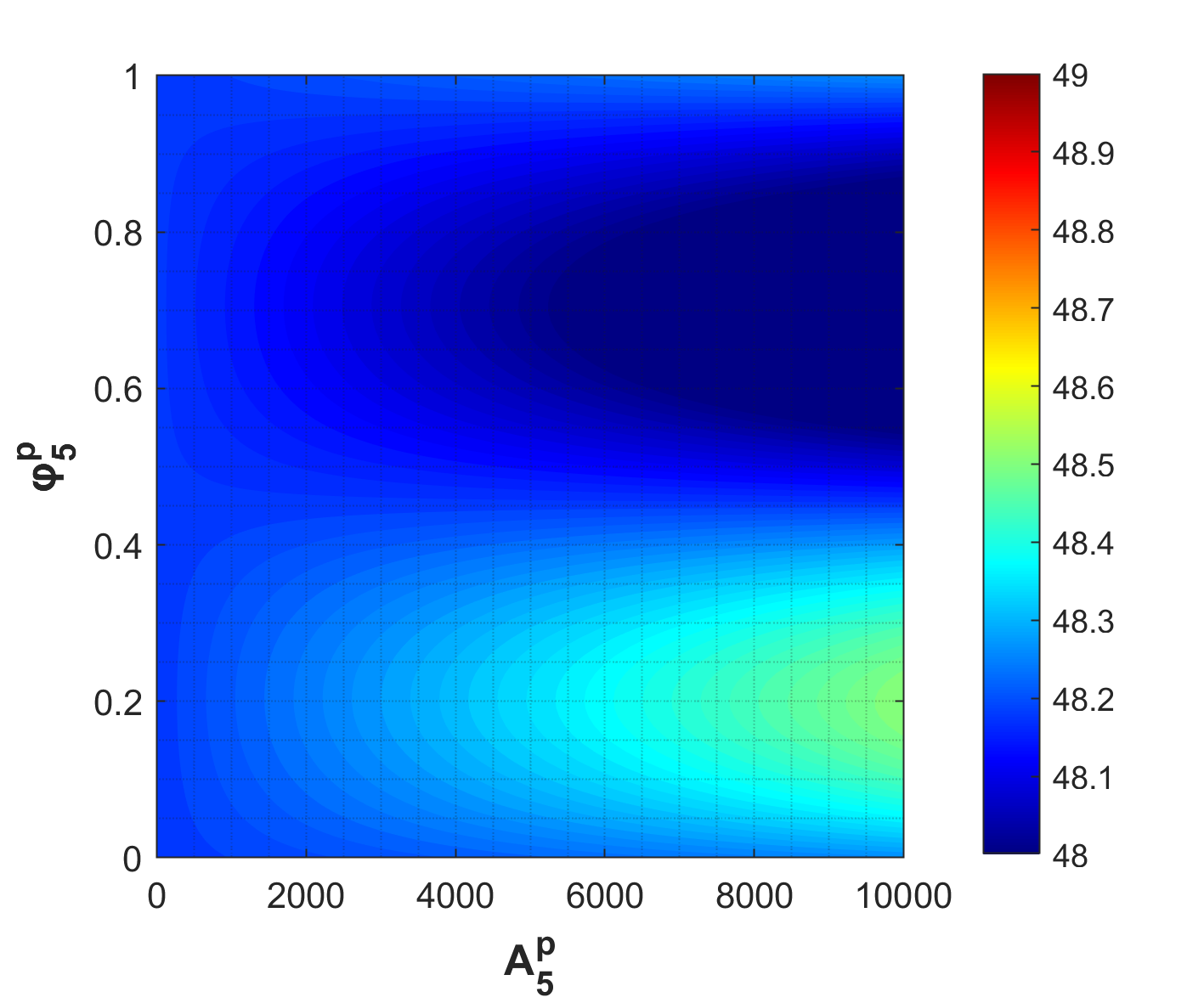}
\end{minipage}
}
\caption{Variation of $DAGDOP$ with respect to the in-plane parameters of the L4 and L5 satellites}
\label{DAGDOP_planar_L4L5}
\end{figure}

 As shown in Fig.~\ref{fig_constellation}, the free in-plane motions of L4 and L5 satellites are so small that they can be neglected. This observation is corroborated by the results presented in Fig.~\ref{DAGDOP_planar_L4L5}. The in-plane parameters of L4 satellites are varied over the range:
\begin{equation}
a_4^p\in[0,10000]~km,\quad \phi_4^p\in[0,2\pi].
\nonumber
\end{equation}
For L5 satellite, the range is:
\begin{equation}
a_5^p\in[0,10000]~km,\quad \phi_5^p\in[0,2\pi].
\nonumber
\end{equation}
Clearly, the in-plane parameters of L4 and L5 satellites have negligible influence on the short-arc AOD accuracy.

\begin{itemize} 
\item \textbf{The in-plane parameters of L3 satellite}
\end{itemize} 
\par
\begin{figure}[!h]
\centering 
\subfigure[Variation of $DAGDOP$ with respect to $A_3^p$ and $\phi_3^p$] 
{
\begin{minipage}[b]{.45\linewidth} 
\centering
\includegraphics[scale=0.6]{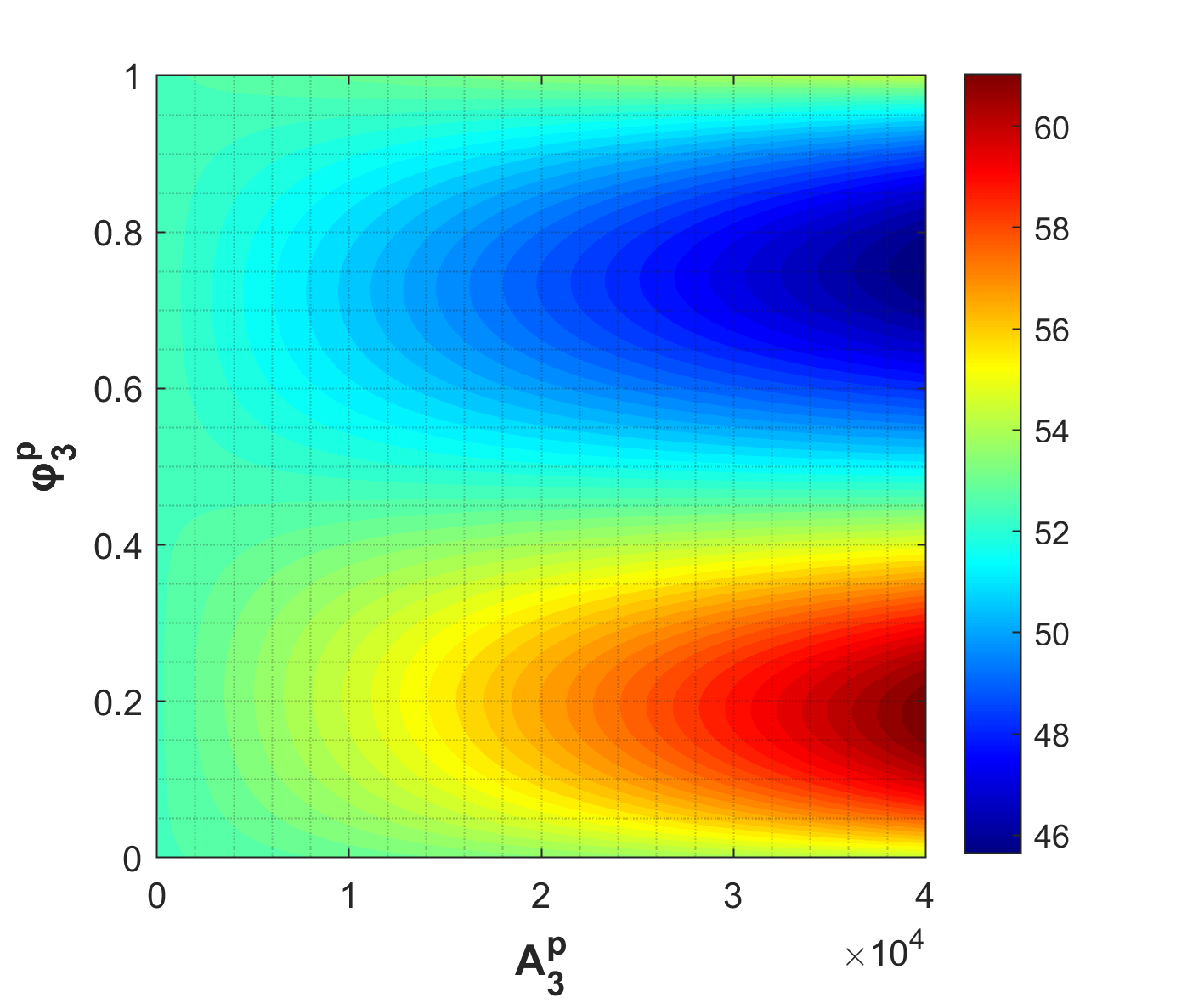}
\end{minipage}
}
\subfigure[Variation of $DAGDOP$ with respect to $A_{DRO}^p$ and $\phi_{DRO}^p$]
{
\begin{minipage}[b]{.45\linewidth}
\centering
\includegraphics[scale=0.6]{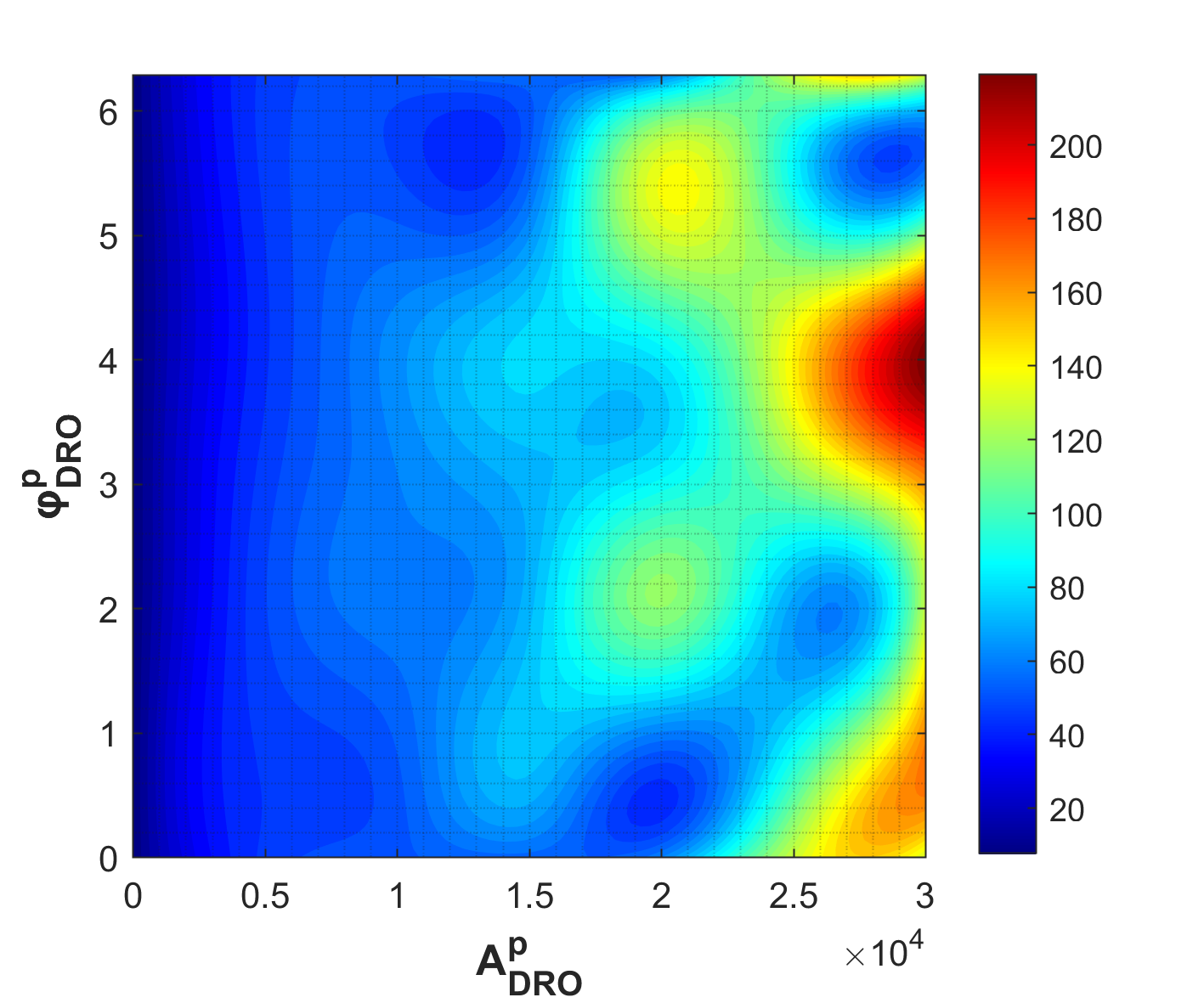}
\end{minipage}
}
\caption{Variation of $DAGDOP$ with respect to the in-plane parameters of L3 and DRO satellites}
\label{DAGDOP_planar_L3DRO}
\end{figure}

 As shown in Fig.~\ref{fig_constellation}, the in-plane motion amplitude of L3 satellite is larger than those of L4 and L5 satellites. We set the range of the in-plane parameters of L3 satellite as:
\begin{equation}
a_3^p\in[0,40000]~km,\quad \phi_3^p\in[0,2\pi].
\nonumber
\end{equation}
Fig.~\ref{DAGDOP_planar_L3DRO}(a) reveals the influence of the L3 satellite's in-plane motion parameters on $DAGDOP$. Two key trends are observed: (1) When $A_3^p$ is small (approximately smaller than 10000km), variations in $\phi_3^p$ have a negligible effect on $DAGDOP$. (2) As $A_3^p$ increases (approximately larger tham 10000km), the periodic variation of $\phi_3^p$ imparts a corresponding periodicity to $DAGDOP$. However, even in this regime, the average effect of the L3 in-plane motion on $DAGDOP$ remains insignificant.

\begin{itemize} 
\item \textbf{The parameters of DRO satellite}
\end{itemize}
\par Analysis of the in-plane amplitude of DRO satellite is more complex. In contrast to the L3, L4, and L5 satellites, the DRO's in-plane motions seem to have non-negligible effects on the AOD results. This complexity stems from the fact that the orbital period of the DRO is strongly influenced by its amplitude. For smaller size DROs, their period is shorter, which means a longer geometric arc is available over the same time length. As a result, better AOD accuracy can be expected for smaller size DROs, which agrees with the trend in Fig.~\ref{DAGDOP_planar_L3DRO}(b). Therefore, we do not aim to improve the AOD accuracy by optimizing the DRO parameters, but rather focus solely on comparing their influence. Fig.~\ref{DAGDOP_planar_L3DRO}(b) displays the results. The ranges of $a_{DRO}^p$ and $\phi_{DRO}^p$ are:
\begin{equation}
a_{DRO}^p\in[1000,30000]~km,\quad \phi_{DRO}^p\in[0,2\pi].
\nonumber
\end{equation}

As shown in Fig.~\ref{DAGDOP_planar_L3DRO}(b), the variation of $DAGDOP$ with respect to DRO parameters is complex. When $a_{DRO}^p$ is small (approximately less than 15000km), the variance of $DAGDOP$ is not obvious. In this case, the above analysis in sections 4 and 5 still holds. As $a_{DRO}^p$ increases, $DAGDOP$ exhibits a certain degree of growth. 

\subsection{Support from the GNSS}

\par
\begin{figure}[h]
\centering   
\subfigure[Variation of $DAGDOP$ with respect to $A_4$ and $A_5$] 
{
\begin{minipage}[b]{.45\linewidth} 
\centering
\includegraphics[scale=0.6]{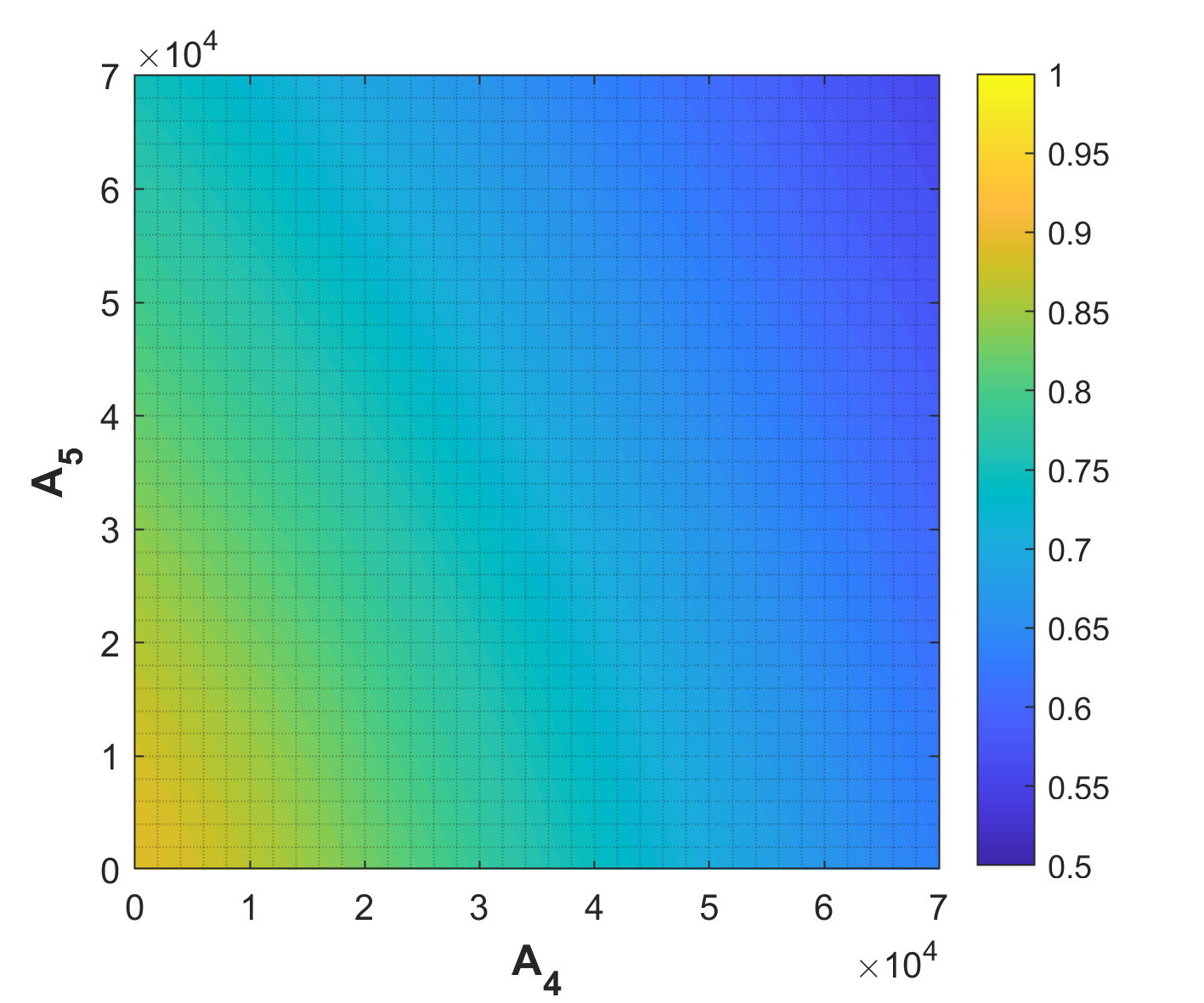}
\end{minipage}
}
\subfigure[Variation of $DAGDOP$ with respect to $\Delta\phi_4$ and $\Delta\phi_5$]
{
\begin{minipage}[b]{.45\linewidth}
\centering
\includegraphics[scale=0.6]{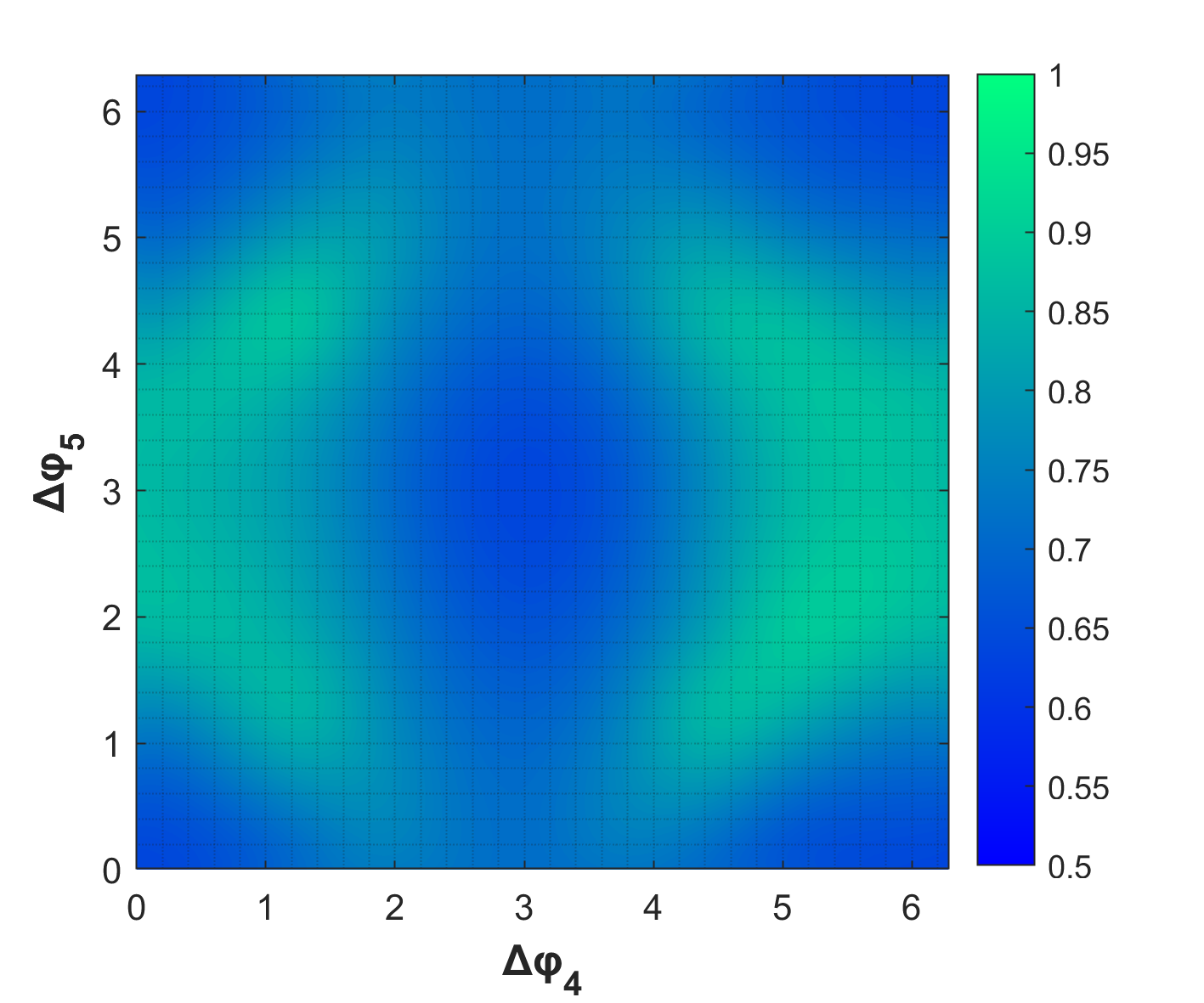}
\end{minipage}
}
\caption{Variation of $DAGDOP$ with the support of three GNSS satellites}\
\label{Result_GNSS_support}
\end{figure}

 Assuming support from ground stations or GNSS satellites is available, the addition of GNSS satellites can improve the OD accuracy significantly. Here we give two examples. We use three inclined geosynchronous orbit (IGSO) satellites as the supporting satellites. The initial states of the three supporting satellites are shown in Table~\ref{IGSOs}. Parameters of the navigation constellation for these two examples are the same as those in simulations 1 and 3. Figs.~\ref{Result_GNSS_support}(a) and \ref{Result_GNSS_support}(b) show the results, respectively. 

\begin{table}[htp]
    \renewcommand\arraystretch{1.2}
    \caption{\textbf{Initial positions and velocities of the IGSOs used in the numerical simulation}}
    \centering
    \begin{tabular}{|c|c|c|} \hline
    \textbf{Orbit IDs} & \textbf{Position (x,y,z) [m]} & \textbf{Velocity (x,y,z) [m$\cdot$s$^{-1}$]} \\ \hline
    IGSO-1 & 42164137.00 \quad 0.000 \quad 0.000 & 0.000\quad1763.55\quad2518.62 \\ \hline
    IGSO-2 & -21082068.50\quad-20944266.19\quad-29911512.01 & 2662.73\quad-881.778\quad-1259.31 \\ \hline
    IGSO-3 & -21082068.50\quad20944266.19\quad29911512.01 & -2662.73\quad-881.77\quad-1259.31 \\ \hline
    \end{tabular}
    \label{IGSOs}
\end{table}

\par Comparing Fig.~\ref{Result_GNSS_support} with Fig \ref{Result_P4P5_1}(b) and \ref{Result_P4P5_2}(b), the addition of GNSS satellites significantly decreases $DAGDOP$, which means improved OD accuracy compared with the case of without GNSS satellites. A careful comparison shows that the above conclusions still hold, although the differences between different values of $A_3, A_4$ (or $\Delta \phi_4, \Delta \phi_5$) are not pronounced. 

\subsection{The Choice of Satellite Maneuver Timing}
\par Navigation satellites, including the L4 and L5 TLPO satellites, require periodic station-keeping maneuvers throughout their operational lifetime \autocite{LI_2022_YF}. After each maneuver, the orbits must be re-determined as quickly as possible, which corresponds to the short-arc AOD scenario in our study. Consequently, the optimal time to execute a station-keeping maneuver better corresponds to the configuration in which the highest AOD accuracy can be obtained in a short-arc after maneuvering. According to the analysis in Section 4, this means that $P_\text{r}$ is maximized corresponding to the approximate minimization of $DAGDOP$, which occurs when the satellites' out-of-plane motion simultaneously reach their peak values. From the analysis in Section 5.1.3, the initial phase combination of $\Delta\phi_4 = \Delta\phi_5 = \pi$ is a favorable option, because it can achieve the highest accuracy at the aforementioned epochs.

\section{Conclusion}
\par In this work, we focus on the AOD accuracy of a four-satellite navigation constellation, comprising an L3 Lissajous satellite, an L4 TLPO satellite, an L5 TLPO satellite, and a satellite close to the Moon. After some comparisons, we choose the DRO over the NRHO as the orbit type close to the Moon. The $DAGDOP$ factor is adopted to quantify the OD performance of the constellation. Analysis reveals that the out-of-plane parameters exert a more significant influence on the constellation's AOD accuracy than the in-plane parameters. To characterize this, we propose the Relatively Planar Factor $P_\text{r}$ of the out-of-plane parameters. For short-arc AOD, $P_\text{r}$ exhibits an approximate negative correlation with the $DAGDOP$ factor and serves as a simple yet useful tool for qualitative analysis. Minimizing $DAGDOP$ is equivalent to maximizing $P_\text{r}$ which can be expressed analytically as a function of the out-of-plane parameters in the short-arc case. Our numerical experiments validate the negative correlation between $P_\text{r}$ and $DAGDOP$, thereby justifying our analytical analysis. Based on these findings, we draw the following conclusions for short-arc AOD:
\begin{itemize}
\item In short-arc AOD, $P_\text{r}$ and the $DAGDOP$ factor exhibit a clear negative correlation. Since $P_\text{r}$ can be directly computed from the constellation configuration, it is a useful indicator for measuring the short-arc AOD accuracy of the constellation.
\item In order to achieve high AOD performance in short-arc AOD, a cislunar navigation constellation should consist of at least four satellites to avoid co-planarity of the constellation satellites.
\item When the vertical amplitudes $A_3$, $A_4$ and $A_5$ increase, the AOD accuracy improves. This suggests that constellation satellites with larger out-of-plane amplitudes are recommended.
\item For optimal short-arc AOD performance in specific application scenarios, the initial phase combination $\Delta\phi_4 = \Delta\phi_5 = \pi $ is a favorable choice. When $\phi_3=0$ or $\phi_3=\pi$ (i.e., when the L3 satellite approaches its maximum out-of-plane amplitude), near-optimal AOD accuracy can be achieved. This phase configuration is also beneficial for the implementation of satellite maneuver operations.
\item Smaller DROs can improve the AOD performance. In addition, the in-plane motions of the L3, L4, and L5 satellites do not significantly affect the AOD performance.
\item With the support of ground stations or GNSS satellites, variation in OD accuracy due to different out-of-plane parameters is not pronounced. 
\end{itemize}
\par For long-arcs, the difference of AOD accuracy between different constellation configurations becomes insignificant. 

\newpage
\appendix
\section{Orbits used in Section 2.5} 
\begin{table}[htp]
    \renewcommand\arraystretch{1.2}
    \caption{\textbf{Initial positions and velocities of the used orbits in the E-BCRS}}
    \centering
    \begin{tabular}{|c|c|c|} \hline
    \textbf{Orbit Types} & \textbf{Position (x,y,z) [m]} & \textbf{Velocity (x,y,z) [m$\cdot$s$^{-1}$]} \\ \hline
    DRO & 339572669.531 \quad -222212655.559 \quad -79951716.175 & 106.376 \quad 132.254 \quad 74.429 \\ \hline
    NRHO & 303316550.277 \quad -229848639.549 \quad -1293636.435 & 617.294 \quad 776.339 \quad 392.035 \\ \hline
    L3-S & -328505851.060 \quad 175148440.299 \quad 126628539.389 & -547.093 \quad -729.962 \quad -374.320 \\ \hline
    L4-S & 343899832.327 \quad 143208815.657 \quad 145930435.583 & -451.442 \quad 830.156 \quad 240.487 \\ \hline
    L5-S & -25761504.233 \quad -379265898.199 \quad -138718454.487 & 959.790 \quad -122.654 \quad 134.609 \\ \hline
    L3-P & -326261997.066 \quad 206127730.342 \quad 73264071.000 & -554.031 \quad -712.159 \quad -399.330 \\ \hline
    L4-P & 348764554.993 \quad 160948729.595 \quad 109697582.331 & -471.969 \quad 782.137 \quad 344.911 \\ \hline
    L5-P & -11025708.411 \quad -337506228.591 \quad -223699342.590 & 970.880 \quad -79.075 \quad 58.023 \\ \hline
    \end{tabular}
    \label{Initial_orbit}
\end{table}

\begin{table}[htp]
    \caption{\textbf{Design parameters of the used orbits}}
    \renewcommand\arraystretch{1.2}
    \centering
    \begin{tabular}{|c|c|c|c|c|} \hline
    \multirow{2}{*}{\textbf{Orbit Types}} & \multicolumn{2}{c}{\textbf{Out-of-plane Parameters}} & \multicolumn{2}{|c|}{In-plane Parameters} \\ \cline{2-5}
    & Amplitude & initial phase & Amplitude & initial phase \\ \hline
    L3-S & $A_3=60000\text{km}$ & ${\phi_3=0}$ & $A_3^p=10000\text{km}$ & $\phi_3^p=0$  \\ \hline
    L4-S & ${A_4=60000\text{km}}$ & ${\phi_4=\pi/4}$ & $A_4^p=10000\text{km}$ & $\phi_4^{p}=0$ \\ \hline
    L5-S & ${A_5=60000\text{km}}$ & ${\phi_5=7\pi/4}$ & $A_5^p=10000\text{km}$ & $\phi_5^p=0$ \\ \hline
    L3-P & $A_3=0\text{km}$ & ${\phi_3=0}$ & $A_3^p=10000\text{km}$ & $\phi_3^p=0$  \\ \hline
    L4-P & ${A_4=0\text{km}}$ & ${\phi_4=0}$ & $A_4^p=10000\text{km}$ & $\phi_4^{p}=0$ \\ \hline
    L5-P & ${A_5=0\text{km}}$ & ${\phi_5=0}$ & $A_5^p=10000\text{km}$ & $\phi_5^p=0$ \\ \hline
    DRO & \multicolumn{2}{c|}{\diagbox{}} & $A_{DRO}^p=8000\text{km}$ & $\phi_{DRO}^p=\pi$ \\ \hline
    \end{tabular}
    \label{Fundamental Set}
\end{table}

\nocite{*}
\printbibliography[title=References]

\end{document}